\documentclass[12pt]{article}
\usepackage{amsfonts,amssymb,epsfig,amsmath,mathtools,xcolor}
\usepackage{color}

\usepackage{mathrsfs}
\usepackage{graphicx}
\usepackage{subcaption}
\usepackage{hyperref}

\renewcommand{\baselinestretch}{1.2}

\newcommand{\tr}{\textrm{tr}}
\newcommand{\nn}{\nonumber}

\begin{document}

\makeatletter \@addtoreset{equation}{section} \makeatother
\renewcommand{\theequation}{\thesection.\arabic{equation}}
\renewcommand{\thefootnote}{\alph{footnote}}

\begin{titlepage}

\begin{center}
\hfill {\tt SNUTP23-002}\\
\hfill {\tt KIAS-P23070}\\
\hfill {\tt LCTP-23-20}\\

\vspace{2cm}

{\Large\bf Finite $N$ black hole cohomologies}

\vspace{2cm}

\renewcommand{\thefootnote}{\alph{footnote}}

{\large Jaehyeok Choi$^1$, Sunjin Choi$^{2,3}$, Seok Kim$^1$, Jehyun Lee$^1$ and Siyul Lee$^4$}

\vspace{0.7cm}

\textit{$^1$Department of Physics and Astronomy \& Center for
Theoretical Physics,\\
Seoul National University, 1 Gwanak-ro, Gwanak-gu, Seoul 08826, Republic of Korea}\\

\vspace{0.2cm}

\textit{$^2$School of Physics, Korea Institute for Advanced Study,\\
85 Hoegi-ro, Dongdaemun-gu, Seoul 02455, Republic of Korea}\\

\vspace{0.2cm}

\textit{$^3$Kavli Institute for the Physics and Mathematics of the Universe (WPI),\\
The University of Tokyo Institutes for Advanced Study, The University of Tokyo,\\
Kashiwa, Chiba 277-8583, Japan}\\

\vspace{0.2cm}

\textit{$^4$Leinweber Center for Theoretical Physics, University of Michigan,\\ 
Ann Arbor, MI 48109, USA.}\\

\vspace{0.7cm}

E-mails: {\tt zaddere7jp@snu.ac.kr, sunjin.choi@ipmu.jp, \\
seokkimseok@gmail.com, ljs9125@snu.ac.kr, siyullee@umich.edu}

\end{center}

\vspace{1cm}

\begin{abstract}

We study new cohomologies for the BPS operators of the $\mathcal{N}=4$ Yang-Mills theory with 
$SU(3)$ and $SU(4)$ gauge groups, to better understand the black hole microstates. 
We first study the index of these black hole operators and 
identify their apparent threshold levels. For $SU(3)$, we find many towers of states 
and partial no-hair behaviors. We explicitly construct the threshold cohomology 
in the $SU(3)$ theory. We study throughout this paper a 
subsector of the field theory corresponding to the BMN matrix theory. 
We also argue that the BMN sector exhibits a black hole like entropy growth at large $N$.

\end{abstract}

\end{titlepage}

\renewcommand{\thefootnote}{\arabic{footnote}}

\setcounter{footnote}{0}

\renewcommand{\baselinestretch}{1}

\tableofcontents

\renewcommand{\baselinestretch}{1.2}

\section{Introduction and summary}

Microscopically counting the black hole entropy 
\cite{Strominger:1996sh} and better characterizing its microstates 
are one of the central problems in quantum gravity. These questions were also
naturally asked in AdS/CFT from its early days  
\cite{Witten:1998zw,Sundborg:1999ue,Aharony:2003sx,Kinney:2005ej}. 
The entropy of BPS black holes in AdS$_5$ has been counted from the CFT dual rather recently 
\cite{Cabo-Bizet:2018ehj,Choi:2018hmj,Benini:2018ywd} by studying 
the index \cite{Romelsberger:2005eg,Kinney:2005ej}. 
Motivated by this success, we wish to better characterize 
the BPS microstates which contribute to this entropy.
Explicitly constructing these quantum states at strong coupling is difficult, 
even for BPS states. However, a modest version of this program was 
posed in terms of the 1-loop BPS states and their cohomologies 
\cite{Kinney:2005ej,Berkooz:2006wc,minwalla,Janik:2007pm,Grant:2008sk,Chang:2013fba}. 
In this program, the 1-loop BPS states annihilated by a pair of classical 
supercharges $Q$, $Q^\dag$ are studied from $Q$-cohomologies,  
assuming that they remain BPS at strong coupling. See \cite{Chang:2022mjp} for 
a study of perturbative non-renormalization.
There has been progress in this program recently 
\cite{Chang:2022mjp,Choi:2022caq,Choi:2023znd,Chang:2023zqk,Budzik:2023vtr,Budzik:2023xbr}.
In particular, new cohomologies \cite{Chang:2022mjp,Choi:2022caq,Choi:2023znd} 
and 1-loop BPS states \cite{Chang:2023zqk,Budzik:2023vtr}
beyond the familiar `multi-graviton type' operators were constructed 
in the $SU(2)$ maximal super-Yang-Mills theory.

Although the $SU(2)$ results already shed interesting light on the black hole 
microstates, this theory is at best a highly quantum toy model of AdS/CFT. 
The ultimate goal is to study the $SU(N)$ theory at parametrically 
large $N$. With this in mind, in this paper, we find and study the new cohomologies 
for $N>2$. We shall detect and construct new cohomologies for the 
$SU(3)$ and $SU(4)$ theories, and discuss their physics.

Identifying and constructing these operators are computationally very demanding. 
To partly overcome this difficulty, we focus on a quantum mechanical subsector 
corresponding to the BMN matrix model \cite{Berenstein:2002jq,Kim:2003rza}.  
Among the full set of the BPS letters (elementary fields dressed by derivatives), this 
subsector keeps three scalars $\bar\phi^m$ ($m=1,2,3$), three fermions $\psi_{m+}$ 
and a component $f_{++}$ of the field strength only, without any derivatives. The computations 
are easier in this sector, though we should further pursue
efficient methods. As we shall explain in section 2.1, 
this sector exhibits a black hole like entropy 
growth at large $N$. Throughout this paper, we shall 
only study cohomologies within this sector.

In the rest of this section, we sketch the main technical/conceptual
advances in this paper. 

Completely finding all cohomologies is a very difficult calculation 
when the charges or $N$ are large. So far the comprehensive studies 
are made only in \cite{Chang:2022mjp} for $N=2,3,4$ till not-too-large charges. 
At fixed charges, the procedure consists of  
the following steps: 
\begin{enumerate}

\item Find all $Q$-closed operators.

\item Find a subset of operators found in step 1 which are not $Q$-exact.

\item Find a subset of cohomologies in step 2 which are not of multi-graviton type. 

\end{enumerate}
Each step demands substantial computational resources. 
One should also scan through all the charges by repeating these procedures.  
In this paper, we establish streamlined strategies, partly based on \cite{Choi:2023znd}, which allow faster detection and construction of new cohomologies. 

Before going through the $3$ steps above, 
we compute the index over the non-graviton cohomologies at finite $N$.
With this index computed, one can easily identify the charge sectors that contain non-graviton states and construct their cohomologies.
Of course one could miss pairs of non-graviton states which cancel in the index: 
we give up finding all cohomologies and search for those captured by the index. 
Exactly computing the full index in the BMN sector is easy at not too large $N$.  
More difficult is to count the finite $N$ gravitons to be subtracted,  
taking into account the trace relations of finite $N$ matrices. As we review in section 3, 
counting finite $N$ gravitons reduces to counting certain class of polynomials,
whose generators are subject to certain relations.
In principle, these relations can be systematically studied using the  Gr\"obner basis. 
In practice, finding Gr\"obner basis can be computationally very difficult. 
We used a hybrid method of the Gr\"obner basis (in a subsector in which this basis can be found easily) and a more brutal counting of independent polynomials by computer, order by order in the charges. 

Even after identifying the charge sectors to study from the index, the steps $1$, $2$, and $3$ are still quite hard. 
We have established the following procedures which make steps 1 and 2 somewhat easier, and trivialize step 3 within our setup. 
See p.23 for how our implementation trivializes step 3.

As for step 1, we present a class of ansatz for the $Q$-closed operators. 
In order for the final cohomology not to be of graviton type, $Q$ acting on the operator should vanish by trace relations.  
Systematically finding useful trace relations is difficult.
We find a method of constructing a class of operators which become $Q$-closed only after imposing trace relations. 
Our ansatz uses the trace relations of the graviton cohomologies that we detected while computing the index.
Trace relations of gravitons mean that certain polynomials of single-graviton cohomologies are $Q$-exact. 
These relations satisfy `relations of relations', i.e. certain linear combinations of trace relations (with the coefficients being graviton cohomologies) are identically zero. 
In other words, relations of relations are linear combinations of $Q$-exact terms which vanish. So they provide operators which become $Q$-closed thanks to the trace relations, which are our ans\"atze for the non-graviton cohomologies. 
See section 4.1 for more details.

Some $Q$-closed operators mentioned in the previous paragraph 
are not $Q$-exact, providing new cohomologies, while others are $Q$-exact. 
Determining whether our ans\"atze are $Q$-exact or not, step 2, is very hard. 
We developed a numerics-assisted approach to make this step 
affordable on the computer, by ordering the Grassmann variables and then inserting many random integers to the matrix elements. 
See section 4.2 for the details.

The new cohomology thus constructed, and often just the index for the non-graviton cohomologies, provides insights on the structures of the BPS spectrum. 

First, for $SU(3)$ and $SU(4)$, we address the threshold (lightest) non-graviton cohomologies as seen by the index. 
Let us denote the $SO(6)$ R-charges and spins by $R_1,R_2,R_3$ and $J_1,J_2$ respectively. 
For $SU(3)$, the index detects one fermionic threshold cohomology at $R_1=R_2=R_3$, $J_1=J_2$ ($\equiv J$) and $j\equiv 2(R_1+R_2+R_3)+6J=24$. 
($j$ is a combination which can grade the index.) For $SU(4)$, the index 
detects six fermionic threshold cohomologies at $J_1=J_2\equiv J$ and 
$j=28$ in the $[2,0]$ representation of $SU(3)\subset SO(6)$. 
Combining with the results known in the $SU(2)$ theory \cite{Chang:2022mjp,Choi:2022caq}, the threshold levels for $j$ at $N=2,3,4$ form a non-decreasing sequence $24,24,28$. 
These are apparent thresholds for $N=3,4$. 
They would be true thresholds if there are no lower cohomologies canceling in the index or outside the BMN sector.

For $SU(3)$, we construct the threshold cohomology. 
The threshold cohomology is fermionic and has charges $R_1=R_2=R_3=\frac{5}{2}$ ($\equiv R$) and $J=\frac{3}{2}$, thus the scaling dimension $E=3R+2J=\frac{21}{2}$. 
The scaling dimension is larger than $E=\frac{19}{2}$ of the $SU(2)$ threshold operator at $j=24$. 
So the energy threshold for non-graviton operators increases in $N$. 
Such an increase is also expected at large $N$ because the graviton description should be reliable till $E=N$.  

The non-graviton index itself reveals various organized 
patterns which allow us to guess some underlying structures of the spectrum. 
For $SU(4)$, we barely managed to detect the threshold level and did not  
go to higher orders to find richer structures. 
However, we computed the $SU(3)$ non-graviton index up to $j\leq 54$, much higher than the threshold $j=24$, and observed various intriguing structures. 
First of all, like the $SU(2)$ non-graviton index, we find organized towers of states. 
In $SU(2)$ BMN case, we found just one tower of states starting from the threshold operator. 
However, for $SU(3)$, there are more than one towers starting at different levels. 
We also find that some towers are related to others by developing hierarchies. 
We constrain their charge structures and compare them with the expected 
asymptotic entropy of the BMN sector. Possibly, these towers will continue to 
exist at larger $N$'s and may be related to a particular subset of the  
giant graviton towers \cite{Imamura:2021ytr,Gaiotto:2021xce,Murthy:2022ien,Lee:2022vig}.

Another important aspect of the $SU(3)$ non-graviton index is the partial no-hair behavior. 
Namely, the index does not see many product cohomologies made of core black hole cohomologies multiplied by graviton cohomologies. 
In $SU(2)$, many of these product states not seen by the index were shown to be $Q$-exact, thus absent in the BPS spectrum \cite{Choi:2023znd}. 
So the black hole 
cohomologies do not want to be dressed by certain gravitons, reminiscent of the black hole no-hair theorem. 
It was also pointed out, both from $SU(2)$ QFT and the large $N$ gravity dual, that the BPS no-hair theorem holds only partially. 
We find an empirical signal that the $SU(3)$ no-hair theorem may also be partial, in that assuming particular graviton hairs makes the remaining core spectrum better organized. 
See section 3.1 for the details.  

The remaining part of this paper is organized as follows. 
In section 2, we explain the cohomology problem, the BMN sector and its entropy, and the graviton cohomologies. 
In section 3, we present our strategies for computing the non-graviton index and present the results for $N=3,4$. 
Qualitative discussions are also made for the $SU(3)$ index. In section 4, we present the new $SU(3)$ cohomology at the threshold level.  
In section 5, we discuss future directions. 
In the appendix, we list the graviton trace relations for the $SU(3)$ theory.

\section{The cohomology problem}

The $\mathcal{N}=4$ Yang-Mills theory with $SU(N)$ gauge group 
has six real scalars $\Phi_{ij}=-\Phi_{ji}$ (subject to 
${\overline{\Phi}}^{ij}\sim \frac{1}{2}\epsilon^{ijkl}\Phi_{kl}$), 
fermions $\Psi_{i\alpha},{\overline{\Psi}}^i_{\dot\alpha}$ and the gauge field 
$A_\mu\sim A_{\alpha\dot\beta}$, all in the $SU(N)$ 
adjoint representation. (Here, $i,j\cdots=1,\cdots,4$, $\alpha=\pm$, 
$\dot\beta=\dot{\pm}$, $\mu=1,\cdots,4$.) 
For later convenience, we arrange these fields into 
$\mathcal{N}=1$ supermultiplets as follows, with manifest 
covariance only for the $SU(3)\subset SU(4)$ part of R-symmetry,  
\begin{eqnarray}
  \textrm{vector multiplet}&:&A_{\alpha\dot\beta}\ ,\ 
  \lambda_\alpha=\Psi_{4\alpha}\ ,\ \bar\lambda_{\dot\alpha}=\overline{\Psi}^4_{\dot\alpha}~,\\
  3\textrm{ chiral multiplets}&:&\phi_m=\Phi_{4m}\ ,\ \bar\phi^m=\overline{\Phi}^{4m}\ ,\ 
  \psi_{m\alpha}=-i\Psi_{m\alpha}\ ,\ \bar\psi^m_{\dot\alpha}=i\overline{\Psi}^m_{\dot\alpha}
  \nonumber\ ,
\end{eqnarray}
where $m=1,2,3$. In this paper, we consider the Euclidean CFT on $\mathbb{R}^4$. This  
is related to the Lorentzian CFT on $S^3\times\mathbb{R}$ by radial quantization, 
which regards the radius of $\mathbb{R}^4$ as the exponential of the Euclidean time $\tau$
and makes a Wick rotation $\tau=it$. Here we note the operator-state map, 
in which the local operators at the origin of $\mathbb{R}^4$ map to the states propagating 
in $S^3\times\mathbb{R}$. The CFT carries a marginal coupling constant $g_{\rm YM}$.

The CFT is invariant under $32$ supersymmetries, represented by the 
$16$ Poincare supercharges $Q^i_\alpha$, $\overline{Q}_{i\dot\alpha}$ and the 
$16$ conformal supercharges $S_{i\alpha}$, $\overline{S}^i_{\dot\alpha}$. 
In the radially quantized theory, $S$'s are Hermitian conjugates of $Q$'s:
$S_i^\alpha=(Q^i_\alpha)^\dag$, $\overline{S}^{i\dot\alpha}=(\overline{Q}_{i\dot\alpha})^\dag$.
Together with other symmetry generators, these supercharges 
form the $PSU(2,2|4)$ superconformal algebra. The most important part of 
the algebra for this paper is \cite{Kinney:2005ej}
\begin{equation}\label{QS-algebra}
  \{Q^i_\alpha,S_j^\beta\}={\textstyle \frac{1}{2}}H\delta^i_j\delta_\alpha^\beta+
  {R^i}_j\delta_\alpha^\beta+{J_\alpha}^\beta\delta^i_j~,
\end{equation}
where $H$ is the dilatation operator (or the Hamiltonian on
$S^3\times\mathbb{R}$ multiplied by the radius of $S^3$), ${R^i}_j$ is the $SU(4)$ R-charges, and ${J_\alpha}^\beta$ is the left
$SU(2) \subset SO(4)$ angular momenta. The BPS states/operators of our interest preserve 
$2$ Hermitian supercharges $Q\equiv Q^4_-$ and $Q^\dag\equiv S_4^-$, 
thus called $\frac{1}{16}$-BPS states/operators. These two supercharges satisfy 
$Q^2=0$, $(Q^\dag)^2=0$, and from (\ref{QS-algebra}) one obtains  
\begin{equation}
  2\{Q,Q^\dag\}=H-(R_1+R_2+R_3+J_1+J_2)\ .
\end{equation}
On the right hand side, we expressed $2{R^4}_4=-R_1-R_2-R_3$ and 
$2{J_-}^-=-J_1-J_2$ in terms of the five charges which rotate the mutually orthogonal 
2-planes on $\mathbb{R}^6\supset S^5$ and $\mathbb{R}^4\supset S^3$, respectively, 
all normalized to have $\pm\frac{1}{2}$ values for spinors. The BPS 
operators of our interest saturate the bound 
$E\geq R_1+R_2+R_3+J_1+J_2$. The charges $R_I$, $J_i$ on the right hand side 
are part of the non-Abelian charges and cannot depend on the coupling $g_{\rm YM}$.
However, $E$ is in general a function of $g_{\rm YM}$, so that 
a BPS state may become anomalous as $g_{\rm YM}$ changes.

The gauge-invariant BPS operators are easily identified (and counted) in 
the free theory limit, $g_{\rm YM}\rightarrow 0$. They are given by 
any gauge-invariant operators made of the following fields
\begin{equation}
  \bar\phi^m\ ,\ \psi_{m+}\ ,\ f_{++}\ ,\ \bar\lambda_{\dot\alpha}
\end{equation}
as well as the derivatives $\partial_{+\dot\alpha}$ acting on these fields, 
subject to the free equation of motion constraint 
$\partial_{+\dot\alpha}\bar\lambda^{\dot\alpha}=0$. See, for instance, 
\cite{Kinney:2005ej,Choi:2023znd} for a more detailed explanation.
We want to study how many of these operators remain  BPS 
at the 1-loop level. The dilatation operator $H(g_{\rm YM})$ 
can be expanded in $g_{\rm YM}^2$, $H(g_{\rm YM})=\sum_{L=0}^\infty
g_{\rm YM}^{2L} H_{(L)}$. At least in perturbation theory, this operator can be 
diagonalized within the subspace of free BPS operators.\footnote{More precisely, 
for the gauge invariance in the interacting theory, the subsector 
is defined at $g_{\rm YM}\neq 0$ by promoting the derivatives $\partial_{+\dot\alpha}$ 
appearing in the operators to the covariant derivatives 
$D_{+\dot\alpha}\equiv\partial_{+\dot\alpha}-i[A_{+\dot\alpha},\ ]$.} 
Within this subspace, $H_{(0)}$ is zero. 
We want to find the subset of free BPS operators 
which are annihilated by $H_{(1)}$. Within the free BPS sector, 
one finds that
\begin{equation}\label{QS-algebra-expand}
  \{Q(g_{\rm YM}),Q^\dag(g_{\rm YM})\}=H(g_{\rm YM})-\sum_I R_I-\sum_i J_i=
  \sum_{L=1}^\infty g_{\rm YM}^{2L}H_{(L)}\ .
\end{equation}
$Q$ and $Q^\dag$ also depend on $g_{\rm YM}$. 
Since the free BPS fields are annihilated by them at the leading $\mathcal{O}(g_{\rm YM}^0)$ order, their coupling expansions start from the $\mathcal{O}(g_{\rm YM}^1)$ `half-loop' order.
Therefore, the leading 1-loop Hamiltonian $H_{(1)}$ in (\ref{QS-algebra-expand}) is given by the anticommutator of $Q$ and $Q^\dag$ at the half-loop order.
In particular, $Q_{(\frac{1}{2})}$ at $\mathcal{O}(g_{\rm YM}^1)$ is precisely the supercharge of the classical interacting field theory. 
So the 1-loop BPS operators should be annihilated by both $Q$ and $Q^\dag$ at the classical half-loop order.

Due to the nilpotency of $Q$ and $Q^\dag$, the operators annihilated by $Q$ and $Q^\dag$ are in 1-1 map with the cohomologies of $Q$, which are $Q$-closed operators with identifications of operators which differ by $Q$-exact operators. 
We shall construct and study the representatives of the cohomologies of the classical half-loop supercharge $Q$, which maps to the 1-loop BPS operators. 
The actions of classical (half-loop) $Q$ on the free BPS fields are given by 
\begin{equation}\label{Q-classical}
  Q\bar\phi^m=0\ ,\ \ Q\bar\lambda_{\dot\alpha}=0
  \ ,\ Q\psi_{m+}=-{\textstyle \frac{i}{2}}\epsilon_{mnp}[\bar\phi^n,\bar\phi^p]\ ,\ 
  Qf=-i[\bar\phi^m,\psi_{m+}]\ ,\ [Q,D_{+\dot\alpha}]=-i[\lambda_{\dot\alpha},\ \ \}\ ,
\end{equation}
where we absorbed the $g_{\rm YM}$ factors on the right hand sides into the normalization of fields.

It is well known that there are fewer BPS states at the 1-loop level than in the free theory. 
It has been conjectured (for instance, explicitly in \cite{minwalla}) 
that the 1-loop BPS states remain BPS at general non-zero coupling. 
Some perturbative evidence of this conjecture was discussed in \cite{Chang:2022mjp}. 
Various discussions in this paper will assume this conjecture.

In our studies, the index over these cohomologies will be important. It is defined by
\begin{equation}
  Z(\Delta_I,\omega_i)={\rm Tr}\left[(-1)^F e^{-\Delta_1 R_1-\Delta_2 R_2-\Delta_3 R_3
  -\omega_1 J_!-\omega_2 J_2}\right]
\end{equation}
with the constraint $\Delta_1+\Delta_2+\Delta_3=\omega_1+\omega_2$ (mod $4\pi i$). The trace is taken over the BPS states, or equivalently over the cohomologies. 
We may regard it as the index over the cohomologies of $Q$ given by (\ref{Q-classical}). 
A matrix integral formula for this index is given in 
\cite{Kinney:2005ej}.

\subsection{The BMN sector}

The cohomology problem has a consistent truncation to the so-called BMN matrix model. 
This is the sector in which only the following three letters are used to 
construct the operators,
\begin{equation}
  \bar\phi^m\ ,\ \ \psi_{m+}\ ,\ \ f_{++}\ ,
\end{equation}
without any derivatives acting on them. In this paper, to simplify the calculations, 
we shall study cohomologies only in this sector. To simplify notations, 
from now on we call them
\begin{equation}
  (\bar\phi^m,\psi_{m+},f_{++})\ \rightarrow \ (\phi^m,\psi_m,f)\ .
\end{equation}
This truncation is consistent only in the classical interacting field theory, 
i.e. till 1-loop BPS states. The truncation forbids 
all the modes in the classical field theory with $J_1\neq J_2$. In general quantum 
theories, the letters carrying nonzero $J_1-J_2$ charges may mix with the fields 
in the BMN sector. So the 1-loop BPS operators
in the BMN sector may mix with non-BMN operators at higher orders in $g_{\rm YM}^2$. 
But if the conjecture that the spectrum of 1-loop BPS states is isomorphic to the 
BPS states at general coupling is true, the spectrum computed in the BMN  
sector will remain unchanged. 

Since the truncation is made for the modes of elementary fields, 
the same truncation can be implemented to compute the index 
over the cohomologies in the BMN sector, 
\begin{equation}
  Z(\Delta_I)={\rm Tr}_{\rm BMN}[(-1)^Fe^{-\sum_{I=1}^3\Delta_I(R_I+J)}]\ ,
\end{equation}
where $J=\frac{J_1+J_2}{2}$. 
Its matrix integral expression is given by \cite{Choi:2023znd}  
\begin{eqnarray}\label{BMN-index-integral}
  Z(\Delta_I)&=&\frac{1}{N!}\int_0^{2\pi} \frac{d^N\alpha}{(2\pi)^N}
  \frac{\prod_{a\neq b}(1-e^{i\alpha_{ab}})\prod_{a,b=1}^N\prod_{I<J}
  (1-e^{-\Delta_I-\Delta_J}e^{i\alpha_{ab}})}
  {\prod_{a,b=1}^N\left[(1-e^{-(\Delta_1+\Delta_2+\Delta_3)}e^{i\alpha_{ab}})
  \prod_{I=1}^3(1-e^{-\Delta_I}e^{i\alpha_{ab}})\right]}\nonumber\\
  &&\hspace{2.7cm}
  \times\ \frac{(1-e^{-(\Delta_1+\Delta_2+\Delta_3)})\prod_{I=1}^3(1-e^{-\Delta_I})}
  {\prod_{I<J}(1-e^{-\Delta_I-\Delta_J})}~,
\end{eqnarray}
where the second line (inverse of the $U(1)$ index) is multiplied 
to make it an $SU(N)$ index rather than $U(N)$.
This integral can be computed either exactly using the residue sum 
or in a series expansion in $t$ defined by 
$(e^{-\Delta_1},e^{-\Delta_2},e^{-\Delta_3})=t^2(x,y^{-1},x^{-1}y)$.

The entropy of BMN cohomologies will be smaller than the entropy of all cohomologies. 
Despite, the large $N$ BMN entropy will still exhibit the black hole like growth.
Taking $j$ (schematically) to be the charges, 
the black hole like entropy growth is 
\begin{equation}\label{entropy-bh-scaling}
  S(j,N)=N^2f({\textstyle \frac{j}{N^2}})~,
\end{equation}
for a function $f(x)$ which does not explicitly depend on $N$, 
where $N\gg 1$, $j\gg 1$ and the ratio $\epsilon\equiv\frac{j}{N^2}$ does not scale in $N$. 
Roughly, the scaled charge parameter 
$\epsilon$ measures the size of the black hole in the AdS unit. In 
this subsection, we show that the BMN entropy scales like (\ref{entropy-bh-scaling}) 
when $\epsilon$ is parametrically small (but not scaling in $N$). 
We expect the same to be true at general $\epsilon$, although we do not prove it. 

We first compute the large $N$ entropy in the `small black hole' regime: 
$j\gg 1$, $N\gg 1$ and $\epsilon\equiv\frac{j}{N^2}$ fixed and much smaller 
than $1$ (but not scaling in $N$).\footnote{The term `small black hole' 
has at least three different meanings in the literature. It sometimes denotes 
string scale black holes, for which 2-derivative gravity description breaks down 
near the horizon. 
In our example, since $\epsilon$ does not scale in $N$, the 2-derivative 
gravity is reliable everywhere. Also, small black holes
sometimes mean AdS black holes with negative specific heat or susceptibility. 
What we call `small black holes' belong to this class, but are more specific. 
Our notion is precisely the same as \cite{Bhattacharyya:2010yg,Choi:2021lbk}.} 
This regime is reached by taking all $\Delta_I$'s to be small. 
The approximate large $N$ calculation of the entropy 
can be done by following all the calculations in section 5.3 of \cite{Choi:2021lbk} 
with minor changes in the setup. In particular, the calculations from  (5.88) 
to (5.91) there can be repeated by simply replacing all 
$2-(-e^\gamma)^n-(-e^{\gamma})^{-n}$ by $1$ (which are the denominators 
of the letter indices in the two setups) and remembering 
that $\beta_I$ there are $\frac{\Delta_I}{2}$ here. The resulting 
eigenvalue distribution is along the interval $\alpha\in (-\pi,\pi)$ on the real axis 
(the gap closes in the small black hole limit), with the distribution function
\begin{equation}
  \rho(\alpha)=\frac{3}{4\pi^3}(\pi^2-\alpha^2)\ .
\end{equation}
The free energy $\log Z$ 
of this saddle point is given by
\begin{equation}
  \log Z=-\frac{3N^2}{2\pi^2}\Delta_1\Delta_2\Delta_3\ .
\end{equation}
(For small black holes with negative susceptibility, the grand canonical index is not 
well defined. Whenever we address $\log Z$, a Laplace transformation to the micro-canonical ensemble is assumed.)
The entropy at given charges $q_I\equiv R_I+J$ is given by extremizing 
\begin{equation}
  S_{\rm BMN}(q_I;\Delta_I)=\log Z+\sum_I q_I\Delta_I~,
\end{equation}
in $\Delta_I$'s, which is given by
\begin{equation}
  S_{\rm BMN}(q_I)=2\pi\sqrt{\frac{2q_1q_2q_3}{3N^2}}\ .
\end{equation}
This expression is valid when $q_I=N^2\epsilon_I$ with $\epsilon_I\ll 1$. 
The entropy $\sim N^2\sqrt{\epsilon_1\epsilon_2\epsilon_3}\propto N^2$ 
exhibits a black hole like scaling (\ref{entropy-bh-scaling}).
$S_{\rm BMN}$ is smaller than the full entropy 
$S(q_I)=2\pi\sqrt{\frac{2q_1q_2q_3}{N^2}}$ 
in the small black hole regime \cite{Choi:2021lbk} by a factor of $\frac{1}{\sqrt{3}}$.
This is natural since the BMN truncation loses cohomologies. 
However, the fact that $S_{\rm BMN}(q_I)$ scales like $S(q_I)$ implies 
that the truncation provides a good simplified model for black holes, 
at least for $\epsilon_I\ll 1$.

At large charges, we have not computed the asymptotic entropies at large $N$ and $q_I$.
Instead, we have computed the large charge entropies at $N=2,3,4$.
This was done by first computing 
the exact index $Z(t)$ by a residue sum, and then extracting the coefficient
$\Omega_j$ of the expansion $Z(t)=\sum_j \Omega_jt^j$ for large 
$j=2(q_1+q_2+q_3)$. This calculation was done both by saddle point approximation at 
large $j$, and also by expanding $Z(t)$ in $t$ by computer until very high order. 
The asymptotic BMN entropies $S_{\rm BMN}(j,N)$ at very large $j\gg 1$ are given by
\begin{equation}
  S_{\rm BMN}(j,2)\sim 3\log j\ ,\ 
  S_{\rm BMN}(j,3)\sim 2\log j\ ,\ S_{\rm BMN}(j,4)\sim 8 \log j\ .
\end{equation}
Even at small $N$'s, these are much slower growths 
than the full entropy, which is
\cite{Choi:2018hmj,Honda:2019cio,ArabiArdehali:2019tdm}
\begin{equation}\label{cardy-entropy}
  S\propto (N^2\!-\!1)^{\frac{1}{3}} j^{\frac{2}{3}}\ .
\end{equation}
The large discrepancy $S_{\rm BMN}/S\ll 1$ is natural since truncating a 4d QFT to 
quantum mechanics loses almost all cohomologies at higher energies. 
This is because the dominant part of the entropy (\ref{cardy-entropy}) 
is supposed to come from the infinitely many letters dressed by derivatives. 
If $n$ unconstrained derivatives can appear in the operators, 
one expects an entropy $S\propto j^{\frac{n}{n+1}}$. 
The case with $n=2$ BPS derivatives in 4d indices yields (\ref{cardy-entropy}),
while the quantum mechanical case with 
$n=0$ replaces $j^{\frac{0}{0+1}}$ by $\log j$.

So we expect the BMN subsector to give us good lessons on 
the small black holes for sufficiently large $N$. 
We hope that the studies at finite $N=3,4$ in this paper will start to
unveil important structures which will continue at larger $N$'s.

\subsection{The graviton cohomologies}

Some cohomologies are very well known, which are the multi-graviton cohomologies.  
We want to exclude them in our discussions. In this subsection, we review the notion of
graviton cohomologies, especially at finite $N$, and also explain how to
list and count them. 

The multi-graviton cohomologies are `defined' to be the polynomials of single-trace cohomologies. 
(This definition naturally yields the familiar large $N$ cohomologies for the supergravitons.) 
Single-trace cohomologies are completely understood \cite{Kinney:2005ej,Janik:2007pm,Chang:2013fba}, as we shall review in a moment. 
Single-trace cohomologies are nontrivial cohomologies at 
arbitrary $N$ by definition, since no trace relations apply
within the single-trace sector. Polynomials of these single-trace cohomologies define the
multi-trace cohomologies. Some polynomials may be trivial, 
i.e. $Q$-exact at finite $N$. However, they are $Q$-closed at arbitrary $N$ without 
using any trace relations. 
This will be in contrast to the black hole cohomologies, which should become 
$Q$-closed only after applying trace relations at particular $N$.

When $N$ is larger than the energy, the multi-graviton operators 
defined above are all nontrivial cohomologies since no trace relations can 
be applied to make them $Q$-exact. So in this setup, the `graviton cohomologies' 
defined abstractly in the previous paragraph actually map to the familiar 
$\frac{1}{16}$-BPS graviton states in $AdS_5\times S^5$. Trace number of the operator 
is regarded as the particle number.

At finite $N$, all the multi-trace operators mentioned in the previous paragraphs 
are still $Q$-closed. However, some of their linear combinations may be zero or 
$Q$-exact  when their energies are larger than $N$,
due to the trace relations. So the independent graviton cohomologies 
reduce at finite $N$. Such reductions of states are a well known finite $N$ effect in 
the gravity dual. It is called the stringy exclusion principle \cite{Maldacena:1998bw}, 
which happens because gravitons polarize into D-brane giant gravitons 
\cite{McGreevy:2000cw,Grisaru:2000zn,Hashimoto:2000zp}. 
The reduction/exclusion mechanism is the same for any $N$ 
in QFT, making it natural to call them `finite $N$ gravitons' at general finite $N$.

Now we concretely explain the list of the graviton cohomologies.
One starts by listing the single-trace graviton cohomologies. These are completely found 
and collected into supermultiplets. The relevant algebra for these multiplets is the
$PSU(1,2|3)$ subset of the superconformal symmetry $PSU(2,2|4)$ that commutes 
with $Q,Q^\dag$. The multiplets for single-trace graviton cohomologies  
are called $S_n$ with $n=2,3,\cdots$ \cite{Kinney:2005ej}. $S_n$ is obtained by 
acting the Poincare supercharges $Q^m_+$, $\overline{Q}_{m\dot\alpha}$ and the 
translations $P_{+\dot\alpha}$ in $PSU(1,2|3)$ on 
the following primary operators
\begin{equation}\label{Sn-primary}
  u^{i_1i_2\cdots i_n}={\rm tr}(\bar\phi^{(i_1}\bar\phi^{i_2}\cdots\bar\phi^{i_n)})\ .
\end{equation}
See \cite{Kinney:2005ej} for more details. At large $N$, 
multiplying the operators in $S_n$'s yields independent multi-trace cohomologies.
At finite $N$, trace relations reduce the independent single-trace 
and multi-trace operators. Following \cite{Choi:2023znd}, we first identify 
the dependent single-trace operators as follows. Using the 
Cayley-Hamilton identity, one can show that all single-trace operators in $S_{n\geq N+1}$ 
can be expressed as polynomials of operators in $S_{n\leq N}$ \cite{Choi:2023znd}. So
it suffices to use only the operators in $S_{n\leq N}$ to 
generate graviton cohomologies. The remaining single-trace generators in $S_{n\leq N}$ 
are not independent when we multiply 
them. In other words, there are further trace relations for gravitons within $S_{n\leq N}$.
The last trace relations are not systematically understood, to the best of our 
knowledge.

To simplify the discussions, let us consider the BMN sector only from now. 
The subset of $PSU(2,2|4)$ that acts within the BMN sector is $SU(2|4)$. 
The subset $SU(1|3)\subset SU(2|4)$ commutes with $Q,Q^\dag$ and generates 
the supermultiplets of BMN cohomologies.
In each $S_n$, there is a finite number of single-trace generators in the BMN sector. 
They are given by 
\begin{eqnarray}\label{BMN-mesons}
  (u_n)^{i_1\cdots i_n}&=&{\rm tr}(\phi^{(i_1}\cdots \phi^{i_n)})
  \ \ \ 
  \\
  {(v_n)^{i_1\cdots i_{n\!-\!1}}}_j&=&{\rm tr}(\phi^{(i_1}\cdots\phi^{i_{n\!-\!1})}\psi_j)
  -\textrm{`trace'}
  \ \ \ \ \nonumber\\
  (w_n)^{i_1\cdots i_{n\!-\!1}}&=&{\rm tr}(\phi^{(i_1\cdots i_{n\!-\!1})}f+
  {\textstyle \frac{1}{2}}\epsilon^{jk(i_p}{\textstyle \sum_{p=1}^{n\!-\!1}}
  \phi^{i_1}\cdots \phi^{i_{p\!-\!1}}\psi_j\phi^{i_{p\!+\!1}}\cdots\phi^{i_{n\!-\!1})}\psi_k)\ .
  \ \ \ \ \nonumber 
\end{eqnarray}
Here, `trace' denotes the terms to be subtracted to ensure that 
the contractions of the upper/lower $SU(3)$ indices are zero.
The BMN multi-graviton cohomologies are 
polynomials of $u_n$, $v_n$, $w_n$. These polynomials are subject to trace relations. 
These trace relations hold up to $Q$-exact 
terms.\footnote{In principle there might be relations which hold  
without any $Q$-exact terms. In practice, with extensive studies of the $SU(2)$ and 
$SU(3)$ graviton operators in the BMN sector, all trace relations of this 
sort that we found have nontrivial $Q$-exact terms.} 
For instance, the lowest trace relations for $N=2$ are 
\begin{equation}\label{relation-example}
  R_{ij}\equiv\epsilon_{ikm}\epsilon_{jln}(u_2)^{kl}(u_2)^{mn}
  =Q\left[-i\epsilon_{a_1a_2(i}{\rm tr}(\psi_{j)}\phi^{a_1}\phi^{a_2})\right]\ .
\end{equation}
More concretely, some components of these relations are
\begin{equation}
  {\rm tr}(X^2){\rm tr}(Y^2)-[{\rm tr}(XY)]^2\sim 0\ ,\ 
  {\rm tr}(XY){\rm tr}(XZ)-{\rm tr}(X^2){\rm tr}(YZ)\sim 0~,
\end{equation}
where $\sim$ hold up to $Q$-exact terms.
Such $Q$-exact combinations are zeros in 
cohomology. Of course multiplying gravitons to such relations yields further 
relations. Trace relations cannot be seen if one does not know that the `meson' or
`glueball' operators $u_n,v_n,w_n$ are made of `gluons' $\phi,\psi,f$.
To enumerate graviton cohomologies without overcounting, 
we first consider the Fock space made by the operators $\{u_n,v_n,w_n\}$ with
$n=2,\cdots, N$ and then take care of the trace relations to eliminate the dependent 
states.

It is important to find all fundamental trace relations of the polynomials of $u_n,v_n,w_n$, 
which cannot be decomposed into linear combinations of smaller relations. 
Let us denote by $R_a(\{u_n,v_n,w_n\})$ the fundamental trace relations, 
with $a$ being the label. 
Non-fundamental trace relations are obtained by linear combinations of $R_a$'s,
\begin{equation}\label{non-fundamental-relation}
  \sum_a f_a(\{u_n,v_n,w_n\})R_a(\{u_n,v_n,w_n\})\ .
\end{equation}
In general, (\ref{non-fundamental-relation}) is nonzero and $Q$-exact. 
However, for some choices of $f_a$'s, the combination (\ref{non-fundamental-relation}) 
may be exactly zero. If (\ref{non-fundamental-relation}) exactly vanishes, 
this yields a `relation of relations.' 
In terms of the mesonic variables $u_n,v_n,w_n$, they are trivial expressions, meaning 
that various terms just cancel to zero. They just represent the ways 
in which fundamental relations $R_a$ can be redundant at higher orders. 
For example, consider the relations $R_{ij}$ of (\ref{relation-example}) in the 
$SU(2)$ gauge theory. 
Some relations of these relations are given by
\begin{equation}
  u^{ik}R_{jk}(u_2)-{\textstyle \frac{1}{3}}\delta^i_ju^{kl}R_{kl}(u_2)=0~,
\end{equation}
in the $[1,1]$ representation. For instance, one can immediately see for $i=1$, $j=2$ 
that 
\begin{equation}\label{rel-of-rel-example1}
  u^{1i}R_{2i}=u^{11}[u^{23}u^{13}-u^{12}u^{33}]
  +u^{12}[u^{33}u^{11}-(u^{13})^2]+u^{13}[u^{12}u^{13}-u^{11}u^{23}]=0\ .
\end{equation}
This is a trivial identity if expanded in mesons.  
$u^{11}R_{21}$ and $-u^{12}R_{22}-u^{13}R_{23}$ represent same constraint 
$u^{11}(u^{23}u^{13}-u^{12}u^{33})=Q[\cdots]$, 
implying that $R_{ij}$'s are not independent. 

Interestingly, trace relations described so far will be used in section 4
to construct the ansatz for the non-graviton cohomologies.
In the meantime, we shall exploit a more practical way of enumerating the
graviton cohomologies, as we explain in section 3.

\section{The index}

In this section, we explain how to enumerate the finite $N$ graviton cohomologies
described in section 2.2. Then it is straightforward to compute the index
over graviton cohomologies, and subtract from the full index to obtain the index
over non-graviton cohomologies.
The results for $SU(3)$ and $SU(4)$ are presented in respective subsections.

Recall that graviton operators in the BMN sector are the polynomials of 
mesons (\ref{BMN-mesons}).
We wish to enumerate linearly independent operators among these, 
i.e. we wish to mod out by linear relations between them.
There are two main strategies that we exploit to ease this computation:
the eigenvalue counting and the Gr\"obner basis.

Let us explain the first idea, the eigenvalue counting.
We first review how the multi-gravitons 
made of the chiral primaries $u_n$ of (\ref{Sn-primary}) are enumerated. 
Based on rather physical arguments, \cite{Kinney:2005ej} proposed to count 
them by taking all three scalars $\phi^m$ to be diagonal matrices.\footnote{The argument 
is often dubbed `quantizing the moduli space' of the QFT. For exact quantum states, 
it relies on the protection of the moduli space against  
quantum corrections. At the level of classical cohomologies, its proof 
should be elementary, although we do not pursue it here.}
With this restriction, the problem of enumerating gauge-invariant 
operators reduces to enumerating Weyl-invariant polynomials of the eigenvalues. 
Now our interest is in counting the finite $N$ graviton cohomologies 
involving all the descendants (\ref{BMN-mesons}) in $S_n$, not only the chiral primaries $u_n$.
The descendants are obtained from $u_n$ by acting 
the supercharges in $PSU(1,2|3)$. Since the single-graviton states belong to absolutely
protected multiplets $S_n$, and since their multiplications trivially remain 
in cohomology both for free and 1-loop calculations, we can generate 
the descendants by acting the supercharges of the strictly free theory \cite{Choi:2023znd}. 
The actions of free supercharges are linear so that 
diagonal $\phi^m$'s transform to diagonal $\psi_m$ and $f$. (In the BMN 
sector, the supercharges $Q^m_+$ in $SU(1|3)$ act linearly even in the classical 
interacting theory.)
Therefore, $\psi_m$ and $f$ that appear in descendants can be taken to be diagonal matrices
as well, for the purpose of enumerating graviton operators.

So the counting of graviton operators is reduced to the counting 
of certain polynomials of the eigenvalues. We have $N-1$ eigenvalues 
for each field $\phi^m$, $\psi_m$ and $f$, so in total $7(N-1)$ variables
are needed to describe graviton operators in the BMN sector.
Let us denote these eigenvalues collectively as $\lambda_I$.
Let us also denote the `mesonic generators' $\{u_n,v_n,w_n\}$ (\ref{BMN-mesons}) with $n=2,\cdots,N$,
collectively as $g_i$'s.
These are now regarded as polynomials $g_i(\lambda_I)$ of the eigenvalues $\lambda_I$.
Then, we want to count the polynomials $p(g_i)$ of the mesons $g_i$,
which can be written as polynomials $p(g_i(\lambda_I))$ of eigenvalues $\lambda_I$.

These polynomials are not all independent because certain polynomials $p(g_i)$ of $g_i$'s
may be zero when written as polynomials $p(g_i(\lambda_I))$ of $\lambda_I$.
Such polynomials can be thought of as constraints on the space of polynomials.
These are remnants of the trace relations 
of the $N\times N$ matrices. Had we been keeping all the $N\times N$ matrix elements, 
trace relation would have been zero up to a $Q$-exact term. Since the action of $Q$ 
yields a commutator, the $Q$-exact term vanishes when the fields are diagonal.
So general trace relations up to $Q$-exact terms reduce to exact polynomial constraints.

Counting constrained polynomials is a classic mathematical problem,
with known solution.
This brings us to the second strategy that we exploit: the Gr\"obner basis. 
See e.g. \cite{Cox_2015}.
Let us briefly explain a flavor of its properties and 
how it is used to solve the enumeration problem.

Recall that the multi-graviton operators are
given by the set of all polynomials $p(g_i)$ of $g_i$'s.
However, this set is overcomplete and therefore not suitable for the counting purpose, because of the constraints.
That is, some of the polynomials are zero and consequently some of the polynomials are equivalent to each other. 

We want to better understand the constraints, i.e. polynomials of $g_i$ that are zero.
The constraints appear because each meson $g_i$ 
is not an independent variable but instead made of the gluons $\lambda_I$,
i.e. $g_i=g_i(\lambda_I)$ where the right hand side is a polynomial of $\lambda_I$ that corresponds to the meson $g_i$.
All constraints are derived from the fact that
\begin{equation}\label{meson-def}
 G_i(g_i, \lambda_I) \equiv g_i-g_i(\lambda_I) = 0\ ,
\end{equation}
for each meson labeled by $i$.
Therefore, the set of all polynomials of the mesons $g_i$ and the eigenvalues $\lambda_I$ 
that are zero (also known as the ideal) is \emph{generated} by (\ref{meson-def}), 
in the sense that any element of this set can be written as 
\begin{equation}\label{zero-basis} 
  \sum_i q_i(g_i,\lambda_I) G_i (g_i,\lambda_I)~,
\end{equation}
where $q_i(g_i,\lambda)$ are polynomials of $g_i$ and $\lambda_I$.
If we restrict to elements of this `set of zeroes’ that only involve $g_i$ but not $\lambda_I$, those will be precisely the constraints that mod out the set of all polynomials $p(g_i)$.

Although (\ref{meson-def}) is the most intuitive basis that generates the set of zeroes 
like (\ref{zero-basis}), it is often not the most convenient basis. 
The same set of zeroes can be generated by many different choices of the basis,
possibly with different numbers of generators.
Gr\"obner basis is one of these choices with the following special property.
Let $\{G_a(g_i,\lambda_I)\}$ be a basis of the set of zero polynomials of $(g_i,\lambda_I)$.
Then, for any polynomial $p (g_i,\lambda_I)$,
suppose one tries to `divide’ this polynomial by the basis $\{G_a (g_i,\lambda_I)\}$.
This is a process of writing the polynomial as
\begin{equation}\label{poly-div}
  p (g_i,\lambda_I) =\sum_a q_a (g_i,\lambda_I) G_a (g_i,\lambda_I)  +r (g_i,\lambda_I)\ ,
\end{equation}
where $r (g_i,\lambda_I)$ can no longer be `divided by’ $\{G_a (g_i,\lambda_I)\}$, which can be well-defined by setting an ordering scheme between variables and their monomials.
Naturally, $r (g_i,\lambda_I)$ can be thought of as the remainder of the division.
In general, there can be multiple ways --- with different $q_a$ and $r$ --- to write $p (g_i,\lambda_I)$ as (\ref{poly-div}).
The special property of the Gr\"obner basis is that if $\{G_a (g_i,\lambda_I)\}$
were the Gr\"obner basis of the set of zeroes, then the remainder
$r (g_i,\lambda_I)$ is unique for each given $p (g_i,\lambda_I)$.
Note that since $\{G_a (g_i,\lambda_I)\}$ generates the set of zeroes,
(\ref{poly-div}) implies that the polynomial $p (g_i,\lambda_I)$ is equivalent to its remainder $r (g_i,\lambda_I)$.
It follows that the set of all polynomials $p (g_i,\lambda_I)$ is identical to the set of 
all possible remainders $r (g_i,\lambda_I)$ under division by the Gr\"obner basis.
However, unlike in the set of all polynomials $p (g_i,\lambda_I)$,
there are no polynomials in the set of all remainders that are equivalent due to the constraints, because otherwise one of them should have been divided once more to yield the other as the remainder.
Therefore, the set of remainders can be used to count the number of independent polynomials of $(g_i,\lambda_I)$ under constraints.

There is a canonical procedure to find the Gr\"obner basis of the set of zeroes
given one choice of basis \eqref{meson-def},
known as \emph{Buchberger's algorithm}.
Many computer algebra softwares implement this algorithm or its improved versions.
The Gr\"obner basis depends wildly on the ordering scheme between
variables and monomials, so it is important to choose
a nice ordering scheme which eases the calculations. 
This ordering is difficult to know in advance,
so some amount of trials and errors is involved in finding the Gr\"obner basis.

By setting an appropriate ordering scheme,
it is possible to consistently truncate the Gr\"obner basis for zero polynomials
of $(g_i,\lambda_I)$, into that for zero polynomials of $g_i$ only.
Then, the set of all possible remainders $r(g_i)$ under division
by the truncated Gr\"obner basis form a faithful --- complete but not overcomplete ---
set of all independent polynomials of $g_i$, and therefore the set of
all independent graviton operators.
Moreover, one can easily construct a monomial basis for this set of remainders,
from which it is straightforward to compute both the partition function and the index over graviton operators.

Employing the two strategies --- the eigenvalue counting and the Gr\"obner basis --- explained so far,
we have obtained a closed-form expression for the graviton index for the $SU(2)$ theory,
\begin{equation}\label{BMNgravindex-SU2}
\hspace*{-.3cm}Z_{\rm grav}^{SU(2)} = \frac{1\!+\!3t^4\!-\!8t^6\!-\!6t^{10}\!+\!10t^{12}\!+
\!9t^{14}\!-\!9t^{16}\!+\!16t^{18}\!
-\!18t^{20}\!-\!3t^{22}\!+\!t^{24}\!-\!3t^{26}\!+\!9t^{28}\!-\!2t^{30}\!+\!3t^{32}\!-\!3t^{34}}
{(1-t^4)^3 (1-t^8)^3}~.
\end{equation}
This result was obtained in \cite{Choi:2023znd} where the eigenvalue counting
strategy was used but the Gr\"obner basis was not.
Here, we have reproduced this result by finding a Gr\"obner basis of
relations between $SU(2)$ BMN gravitons that consists of 66 generators (after truncation),
and counting the set of all possible remainders under division by those.

Unfortunately, the computation of the Gr\"obner basis quickly becomes very cumbersome if 
the generators of the constraints $\{ g_i - g_i(\lambda_I)\}$ are numerous and complicated. 
For relations between a subset of $SU(3)$ BMN gravitons that do not involve $f$,
i.e. $u_n$ and $v_n$ in \eqref{BMN-mesons},
we found the Gr\"obner basis with $1170$ generators (after truncation) after several hours
of computation on a computer.
For the complete set of $SU(3)$ BMN gravitons including $w_n$,
we were unable to find the Gr\"obner basis due to lack of computing resources:
it takes months at least and it is tricky to parallelize.
Therefore, we have devised a hybrid method to take maximal advantage of the
Gr\"obner basis obtained for the non-$f$ subsector as we now describe.

We first list the complete and independent monomial basis of graviton operators,
i.e. set of monomials of the mesons $g_i$,
that consist of $u_n,v_n$ but not of $w_n$ ($n=2,3$),
up to the charge order $j=54$.
This can be done for any order $j$ because the Gr\"obner basis for the non-$f$ subsector
has been obtained.
Then, one can construct an overcomplete set of all graviton operators by multiplying
each basis from the previous step by arbitrary numbers of $w_2$ and $w_3$,
again up to $j=54$.
Note that $w_2$ and $w_3$ include 3 and 6 different species of single-graviton operators,
respectively, so the size of the overcomplete set grows quickly.

It is helpful to fragment the problem by classifying the operators according to their charges.
Namely, each charge sector is specified by 4 non-negative integers
$2J$ and $q_I = R_I + J$ (where $I=1,2,3$).
Note that the overall order $j$ used for grading the operators is $j=2(q_1+q_2+q_3)$
and therefore it is always even in the BMN sector.
This classification is useful because all single-graviton operators $u_n,v_n,w_n$
and therefore all multi-graviton operators have definite charges,
and operators with different sets of charges can never have a linear relation between them.
Moreover, different charge sectors with merely permuted charges $(q_1, q_2, q_3)$
should contain the same number of independent graviton operators.
Therefore, we separately consider the overcomplete basis of gravitons in each charge sector
with $q_1 \leq q_2 \leq q_3$.

In order to count linearly independent operators among the overcomplete set in any charge sector,
we rewrite each operator as a polynomial of the eigenvalues.
This is done by substituting the mesons with corresponding eigenvalue polynomials
$u_n(\lambda_I)$, $v_n(\lambda_I)$ and $w_n(\lambda_I)$,
which are obtained by writing the gluons in terms of their eigenvalues.
For the eigenvalues of the $SU(3)$ traceless elementary fields, we use the convention
\begin{eqnarray}\label{SU3diag}
f = \begin{pmatrix} f_1 & 0 & 0 \\ 0 & f_2 & 0 \\ 0 & 0 & -f_1-f_2 \end{pmatrix}~,
\end{eqnarray}
and likes.

The number of independent polynomials within each charge sector
is determined as the rank of their coefficient matrix.
We have used the software \texttt{Singular} \cite{DGPS} for finding the Gr\"obner basis,
writing each operator as an eigenvalue polynomial, and extracting the coefficient matrix within each charge sector,
and \texttt{numpy} for computing the rank of the matrix.

The computation of indices for the $SU(3)$ theory have been performed up to $j=54$
on personal computers.
For example, the computation for the charge sector $(2J, q_1, q_2, q_3) = (7,9,9,9)$,
which turns out to be the largest, 
the coefficient matrix was $31026 \times 20940$ with rank 3242.

For the counting of $SU(4)$ BMN gravitons, we take a similar hybrid approach.
Separation into charge sectors works identically to the $SU(3)$ theory.
However, computation of the Gr\"obner basis is even more heavy, both time-wise and memory-wise,
so we were only able to obtain the Gr\"obner basis for a subsector of $SU(4)$
BMN gravitons involving $u_n$ ($n=2,3,4$), i.e. the chiral primaries.
We first list the complete and independent monomial basis of the chiral primaries
$u_n$ using the Gr\"obner basis, up to the order $j=30$.
Then we construct an overcomplete set of all multi-graviton operators
within each charge sector by multiplying each independent basis
by appropriate numbers of $v_2$, $v_3$, $v_4$, $w_2$, $w_3$ and $w_4$, again up to $j=30$.

We write each operator in the overcomplete basis as a polynomial of the eigenvalues.
For the traceless elementary fields in the $SU(4)$ theory,
we used the following convention for the diagonal entries:
\begin{eqnarray}\label{SU4diag}
f = \begin{pmatrix} f_1 & 0 & 0 & 0 \\ 0 & f_2 & 0 & 0 \\ 0 & 0 & f_3-f_1 & 0 \\
0 & 0 & 0 & - f_2 - f_3 \end{pmatrix}~,
\end{eqnarray}
which slightly simplifies the polynomials compared to the more canonical convention
$f = {\rm diag}(f_1,~ f_2,~ f_3,~ -f_1-f_2-f_3)$.

The computation of indices for the $SU(4)$ theory have been performed up to $j=30$
on personal computers.
For example, the computation for the charge sector $(2J, q_1, q_2, q_3) = (3,5,5,5)$,
which turns out to be the largest, 
the coefficient matrix was $12079 \times 116042$ with rank 3788.

\subsection{$SU(3)$}

Following the computational procedures explained above, we have computed 
the $SU(3)$ graviton index $Z_{\rm grav}$ until $t^{54}$ order.
We write the difference $Z-Z_{\rm grav}$ with the full index $Z$,
which is the index over non-graviton cohomologies or the `black hole' cohomologies, as
\begin{equation}\label{difference-factored}
  Z-Z_{\rm grav}=Z_{\rm core}(\Delta_I)\cdot 
  \prod_{I=1}^3\frac{1}{1-e^{-\Delta_I}e^{-\Delta_1-\Delta_2-\Delta_3}}
  \cdot \prod_{I=1}^3(1-e^{-\Delta_I})\ .
\end{equation}
The factors that dress the index over \emph{core} non-graviton cohomologies
will be explained shortly.
$Z_{\rm core}(\Delta_I)\equiv f(t,x,y)$ with 
$e^{-\Delta_1}=t^2x$, $e^{-\Delta_2}=t^2y^{-1}$, $e^{-\Delta_3}=t^2x^{-1}y$ 
can be expanded as
\begin{equation}
  f(t,x,y)=\sum_{j=0}^{54}\sum_{{\bf R}_j}
  (-1)^{F({\bf R}_j)}\chi_{{\bf R}_j}(x,y)t^{j}+\mathcal{O}(t^{56})\ ,
\end{equation}
where ${\bf R}_j$ runs over the $SU(3)$ irreducible representations which 
appear at $t^j$ order ($j$ is even in the BMN sector), $\chi_{{\bf R}_j}(x,y)$ is 
its character, and 
$F({\bf R}_j)$ is its fermion number. 
The representations ${\bf R}_j$ appearing in the expansion of $f$, together 
with their bosonic/fermionic natures, are shown in Table \ref{tower}.
We have classified the representations into several groups, i.e. what we suspect 
to be the fermionic towers $F_0,...,F_4$, the bosonic towers $B_1,...,B_3$, 
and the remainders $F_{\rm exc}$, $B_{\rm exc}$ for which we 
do not see particular patterns (thus named `exceptional').

\begin{table}[t]
\begin{center}
\begin{tabular}{c||c|c|c|c|c|c||c|c|c|c}
	\hline
    $j$ & $F_0$ & $F_1$  & $F_2$ & $F_3$ & $F_4$ & $F_{\rm exc}$ & $B_1$ & $B_2$ 
    & $B_3$ & $B_{\rm exc}$ \\
	\hline 
    $24$ & $[0,0]$ &&&&&&&&& \\
    $26$ &&&&&&&&&& \\
    $28$ &&&&&&&&&& \\
    $30$ & $[0,0]$ & $[3,0]$ &&&&&&&& \\
    $32$ && $[4,0]$ &&&&&&&& \\
    $34$ && $[5,0]$ &&&&& $[3,1]$ &&& \\
    $36$ & $[0,0]$ & $[6,0]$ &&&&& $[4,1]$ &&& $[3,0]$ \\
    $38$ && $[7,0]$ &&&& $[1,0]$ & $[5,1]$ &&& \\
    $40$ && $[8,0]$ & $\color{cyan} [5,0]$ && $\color{cyan}[3,1]$ && $[6,1]$ &&& \\
    $42$ & $[0,0]$ & $[9,0]$ & $\color{cyan}[6,0]$ && $\color{cyan}[4,1]$ && $[7,1]$ &&& $[1,1]$ \\
    $44$ && $[10,0]$ & $\color{cyan}[7,0]$ && $\color{lightgray} [5,1]$ && $[8,1]$ & $\color{lightgray} [5,1]$ && \\
    $46$ && $[11,0]$ & $\color{cyan}[8,0]$ && $\color{lightgray} [6,1]$ & $[2,0]$ & $[9,1]$ & $\color{lightgray} [6,1]$ && $[5,0]$ \\
    $48$ & $[0,0]$ & $[12,0]$ & $\color{cyan}[9,0]$ && $\color{lightgray} [7,1]$ & $[3,0]$ & $[10,1]$ & $\color{lightgray} [7,1]$ && $[4,1]$ \\
    $50$ && $[13,0]$ & $\color{cyan}[10,0]$ & $\color{cyan}[7,0]$ & $\color{lightgray} [8,1]$ && $[11,1]$ & $\color{lightgray} [8,1]$ && $[4,0]$ \\
    $52$ & & [14,0] & \color{cyan} [11,0] & \color{cyan} [8,0] & \color{lightgray} [9,1] & 
    [2,0] & $[12,1]$ & \color{lightgray} [9,1] && [3,1]\\ 
    $54$ & & [15,0] & \color{cyan} [12,0] & \color{cyan} [9,0] & \color{lightgray} [10,1] & 
    [4,1] & $[13,1]$ & \color{lightgray} [10,1] & \color{cyan} [7,1] & \\
    \hline
\end{tabular}
\end{center}
\caption{$SU(3)$ Dynkin labels of fermionic/bosonic black hole cohomologies 
after factoring out the descendants and the conjectured graviton hairs of $w_2$. 
The towers in cyan/gray colors may be related to $F_1$ or $B_1$ tower
by partial hairs of $w_3$ gravitons.
The representations colored in gray are not detected by the index, but may
exist in cancelling pairs in $F_4$/$B_2$.}\label{tower}
\end{table}

We comment on the factors which we have taken out in (\ref{difference-factored}).
The factor $\prod_{I}(1-e^{-\Delta_I})$ accounts for $SU(1|3)$ descendants.
For each non-graviton cohomology in ${\bf R}_j$ that contributes to $Z_{\rm core}$,
the entire $SU(1|3)$ multiplet obtained by acting the three fermionic generators $Q_+^m$
must also be non-graviton cohomologies.
Every such multiplet is a long multiplet of the $SU(1|3)$,
so the corresponding character is simply the contribution from the primary
times the factor $\prod_{I}(1-e^{-\Delta_I})$.
This fact can be argued using the embedding supergroup
$PSU(2,2|4)$ of the 4d $\mathcal{N}=4$ theory.
For any of the three generators $Q_+^m$ to annihilate the $SU(1|3)$ primary,
the primary of a bigger representation of $PSU(2,2|4)$ that includes the $SU(1|3)$ multiplet
must be annihilated by $Q_+^4$ \emph{and} by the $SU(4)_R$ lowering operator
that is not part of the $SU(3) \subset SU(4)_R$.
The only $PSU(2,2|4)$ representations that satisfy this property are
$B_1 \bar{B}_1 [0;0]^{(0,n,0)}$, namely the graviton operators, or the identity.
For details on the relevant representation theory, we refer to
\cite{Cordova:2016emh}, particularly its section 2.2.4,
or to appendix B of \cite{Choi:2023znd}.

The second factor of (\ref{difference-factored}) was taken out for an empirical 
reason, with an expectation that they come from the graviton hairs of $w_2$'s 
in (\ref{BMN-mesons}). 
Namely, we conjecture that $w_2$ gravitons multiplying 
the core black hole cohomologies represented by $Z_{\rm core}$ provide 
nontrivial product cohomologies.
Although we have little logical justification of the last claim (except that similar 
hairs are allowed in the $SU(2)$ theory), we think that the phenomenological evidence 
of this claim is compelling since various simple patterns in 
Table \ref{tower} are clear only after factoring it out.

Now we comment on various structures that we observe from Table \ref{tower}.

We first discuss the possible product cohomologies obtained by multiplying 
black hole cohomologies and gravitons. When a product is $Q$-exact, it is
interpreted \cite{Choi:2023znd} 
as a finite $N$ generalization of the black hole no-hair theorem in the BPS
sector.\footnote{This does not mean that the product states are absent: 
the $Q$-exact products acquire anomalous dimensions and become non-BPS. 
Similar structures are known in the gravity dual: 
a charged bulk field develops hairs around AdS black holes 
in the non-BPS regime, which become pathological/singular in the BPS limit 
\cite{Markeviciute:2018yal,Markeviciute:2018cqs,Choi:2023znd}.} 
We cannot conclude just from the index whether these product states exist 
or not. However, if possible product states do not appear in the index, it is 
suggestive that the corresponding hair is not allowed in the cohomology. In the $SU(2)$
theory, several simple product cohomologies which do not appear in the index were 
explicitly shown to be $Q$-exact. Similar calculations are  
much more difficult for $SU(3)$, which we did not attempt.

First of all, as explained above, all possible product cohomologies obtained by 
multiplying $w_2$ are factored out. Mainly appealing to the simpler spectral structures 
of the resulting $Z_{\rm core}$ as shown in Table \ref{tower}, we conjecture
that the $w_2$ hairs are (at least mostly) allowed. Apart from $w_2$, 
we discuss below the possibilities of other graviton hairs. We divide 
the discussions into the possible hairs of $w_3$ (which are the only gravitons 
heavier than $w_2$) and the rest.

Table \ref{tower} does not seem to signal hairs from
gravitons which are no heavier than $w_2$. The gravitons with $j$ charges 
no larger than $w_2$ are: $u_2$ in $[2,0]_4$, $v_2$ in $[1,1]_6$, 
$u_3$ in $[3,0]_{6}$, $v_3$ in $[2,1]_8$, where the subscript denotes $j$. 
Some products clearly do not appear in Table \ref{tower}. 
For instance, $[0,0]_{24}^{F_0}$ in Table \ref{tower} times $u_2$ does not 
appear in the index, because we find no states at $[2,0]_{28}$ with fermionic 
statistics in the table. There are many more products which similarly do not appear 
in the table. Of course, just by matching charges and 
representations, there are several possibilities in which an entry of Table \ref{tower} 
can be accounted for by a tensor product of another entry and these gravitons. 
For instance, $u_2$ may  
multiply the state $[p,q]_j$ in $F_1,...,F_4$ or $B_1,...,B_3$ (with either 
$q=0,1$) to yield $[p+2,q]_{j+4}$ in the same tower. However, 
we feel that this seems quite unlikely. For instance, in $F_1$,
this may explain $[5,0]_{34},[7,0]_{38},\cdots$ from $[3,0]_{30}$, but not 
$[4,0]_{32},[6,0]_{36},\cdots$. In other words, all these towers are more naturally 
generated by adding one scalar at a time (as an adjoint letter) rather than 
two scalars (in the gauge-invariant form of $u_2$). Similarly, $u_3$ 
may multiply $[p,q]_j$ in a tower to yield $[p+3,q]_{j+6}$ in the 
same tower, but again for the same reason we feel 
this unappealing. Apart from these, sporadically, there exist several group theoretic possibilities 
of light graviton hairs in Table \ref{tower}, in particular accounting for 
some entries in $F_{\rm exc}$ or $B_{\rm exc}$ as product cohomologies. 
We shall not list all the possibilities here. It may be worthwhile 
to think about whether some of the `exceptional' cohomologies in 
$F_{\rm exc}$ or $B_{\rm exc}$ can be accounted for by such sporadic graviton hairs. 

Now we consider possible hairs obtained by multiplying $w_3$ of (\ref{BMN-mesons}). 
Again, this is just a possibility from matching the charges/representations. 
It is group theoretically possible 
that the towers $F_2$, $F_3$ are the hairy cohomologies obtained by 
multiplying $w_3$'s to the $F_1$ tower. This is 
because the product of $[n,0]_j$ in $F_1$ and $[2,0]_{10}$ 
decomposes to
\begin{equation}\label{hair-w3-decompose}
  [n,0]_{j}\otimes[2,0]_{10}=[n+2,0]_{j+10}
  \oplus [n,1]_{j+10}\oplus[n-2,2]_{j+10}\ ,
\end{equation}
containing 
$[n+2,0]_{j+10}$ which is in $F_2$. Similarly, multiplying two such gravitons, 
one obtains states in $[n+4,0]_{j+20}$ which is in $F_3$. 
If $F_4$ also forms a tower, with partial cancellations with $B_2$, 
then $F_4$ may also be product cohomologies of $F_1$ and these gravitons 
because the right hand side of (\ref{hair-w3-decompose}) contains $[n,1]_{j+10}$. 
Similarly, if the tower $B_2$ exists (by cancelling partly with 
$F_4$), $B_2$ and $B_3$ could be product cohomologies of $B_1$ and $w_3$'s 
because $[n,1]_j\otimes[2,0]_{10}$ contains $[n+2,1]_{j+10}$.
The possibilities discussed above suggest that the colored towers (cyan or gray)
in Table \ref{tower} may be product cohomologies containing $w_3$.
If these colored towers are indeed obtained by multiplying $w_3$ to 
$F_1$ or $B_1$, note that the last possible tower of $[n-2,2]_{j+10}$ 
in (\ref{hair-w3-decompose}) does not exist in the index, which could mean 
that this part of the product is $Q$-exact.

If the possibilities raised in the previous paragraph are indeed true, it implies 
that the allowed hair structure of the $w_3$ gravitons in $[2,0]_{10}$ is more 
delicate than that of $w_2$. For instance, our conjecture on  
$w_2$ is that their hairs are universally allowed, irrespective of 
the core black hole cohomology chosen.
On the other hand, if the scenario of the previous paragraph is true, 
the allowed/disallowed combinations of the $w_3$ hair are 
determined after entangling the gravitons with the core black hole states. 
This is not the familiar form of the no-hair theorem of semiclassical black hole 
physics. That is, the no-hair theorem as well as its violation is stated for a given 
black hole background, which represents the whole ensemble of states. 
It will be interesting to see if the allowed hairs exhibit subtle 
dependence on the fine-grained information within  the black hole ensemble or not, 
and if they do, whether it may have implications to the fuzzball paradigm 
of black holes \cite{Mathur:2005zp}. We emphasize that all the scenarios discussed 
above can be straightforwardly confirmed/disproved once the cohomologies of 
$F_1,B_1$ are constructed, by checking whether the product cohomologies 
discussed above are $Q$-exact or not.

Now we discuss the possible charge structures and the field contents 
of the towers. We first make general considerations. Suppose 
that $n_\phi$ scalars, $n_\psi$ fermions and $n_f$ field strengths are used to 
make the operator in $SU(3)$ representation $[p,q]$. The operator is not necessarily 
made by just one choice of $n_\phi,n_\psi,n_f$ but superposes many different terms in general. 
So we are in fact studying the structures of a given term in the operator.
Let us also introduce 
the following non-negative integers: 
$l$ pairs of contractions are made between the $SU(3)$ indices of $\phi$'s and $\psi$'s; 
$m_\phi$ pairs of $\phi$'s are contracted with $\epsilon_{abc}$ to yield lower indices;
$m_\psi$ pairs of $\psi$'s are contracted with $\epsilon^{abc}$ to yield upper indices;
$b_\phi$ threesomes of $\phi$'s are contracted with $\epsilon_{abc}$;
$b_\psi$ threesomes of $\psi$'s are contracted with $\epsilon^{abc}$. 
Once the above contractions are made, the remaining upper/lower indices are 
respectively symmetrized to yield the $[p,q]$ representation (after subtracting 
certain terms to ensure that the indices are traceless). 
So one obtains 
\begin{eqnarray}\label{pq-express}
  p&=&n_\phi-l-2m_\phi-3b_\phi+m_\psi~, \\
  q&=&n_\psi-l-2m_\psi-3b_\psi+m_\phi\ .
  \nonumber
\end{eqnarray}
Since $\phi$, $\psi$ and $f$ carry $(R,J)=(\frac{1}{3},0)$, $(\frac{1}{6},\frac{1}{2})$ 
and $(0,1)$ respectively, the charges are given by
\begin{eqnarray}\label{charge-SU3-rep}
  R&=&{\textstyle \frac{n_\phi}{3}+\frac{n_\psi}{6}}=
  {\textstyle \frac{p}{3}+\frac{q}{6}+\frac{l}{2}+\frac{m_\phi}{2}+b_\phi+\frac{b_\psi}{2}}~,\\
  J&=&{\textstyle \frac{n_\psi}{2}}+n_f={\textstyle \frac{q}{2}+n_f
  +\frac{l}{2}+m_\psi+\frac{3b_\psi}{2}-\frac{m_\phi}{2}}~,
  \nonumber\\
  j&=&6(R+J)=2p+4q+6n_f+6l+6m_\psi+6b_\phi+12b_\psi\ .
  \nonumber
\end{eqnarray}

Now we apply these results to the tower $F_1$, for the states with 
$[p,q]_j=[n+3,0]_{30+2n}$, where $n=0,1,2,\cdots$. (Similar studies can 
be made to other towers in Table \ref{tower} except $F_0$.) 
Inserting these expressions to (\ref{charge-SU3-rep}), one obtains
\begin{eqnarray}\label{F1-charges}
  4&=&n_f+l+m_\psi+b_\phi+2b_\psi~,\\
  J&=&n_f+{\textstyle \frac{l}{2}+m_\psi+\frac{3b_\psi}{2}-\frac{m_\phi}{2}}~,
  \nonumber\\
  R&=&{\textstyle \frac{n}{3}+1+\frac{l}{2}+\frac{m_\phi}{2}+b_\phi+\frac{b_\psi}{2}}
  \geq {\textstyle \frac{n}{3}+1}\ .
  \nonumber
\end{eqnarray}
The last inequality is saturated if and only if 
$l=m_\phi=b_\phi=b_\psi=0$. On the other hand, from 
$j=2n+30$ and the lower bound of $R$, one obtains
\begin{equation}
  J={\textstyle \frac{j}{6}-R}={\textstyle 5+\frac{n}{3}-R}\leq 4\ .
\end{equation}
This implies that $F_1$ is a tower in which $R$ increases 
indefinitely with $J\leq 4$ bound. 
From the first equation of (\ref{F1-charges}), $n_f,l,m_\psi,b_\phi,b_\psi$ 
are all bounded from above. Also, from the second equation of (\ref{pq-express}), 
one finds $n_\psi=l+2m_\psi+3b_\psi-m_\phi$. The first three terms on the right 
hand side are bounded, so $n_\psi$ is bounded from above. Furthermore, for the non-negativity 
of $n_\psi$, $m_\phi$ is also bounded from above. The only unbound non-negative
integer is $n_\phi$. So this tower has increasing numbers 
of scalars. In other words, this
is a Kaluza-Klein tower, carrying increasing momentum charges on $S^5$, 
rather than a higher-spin tower.

We do not know which type of tower $F_0$ is, between higher-spin and Kaluza-Klein. 
Apparently, the states in $F_0$ appear when $j$ increases by $6$, except 
at $j=54$ where we found no states in $[0,0]$ representation. 
It is unclear which of the following is true (if any): 
(i) the absence of the tower $F_0$ beyond this 
charge; (ii) the multiple tower structure within $F_0$ with different 
periods; (iii) existence of an exceptional bosonic cohomology at this order 
which cancels with the tower at $j=54$. 
Related to this tower, 
we comment that the $SU(2)$ cohomologies 
in the BMN sector showed the following index \cite{Choi:2023znd}:
\begin{eqnarray}
  Z-Z_{\rm grav}&=&Z_{\rm core}(\Delta_I)\cdot 
  \prod_{I=1}^3\frac{1}{1-e^{-\Delta_I}e^{-\Delta_1-\Delta_2-\Delta_3}}
  \cdot \prod_{I=1}^3(1-e^{-\Delta_I})~,\nonumber\\
  Z_{\rm core}&=&=-\sum_{n=0}^\infty t^{24+12n}\ .
\end{eqnarray}
The whole tower of cohomologies for $Z_{\rm core}$ was also constructed
at arbitrarily large $j$, with $R$ fixed and $J$ increasing.
So this is a higher spin tower of core black hole primaries.
This suggests that the $SU(2)$ tower may be protected by certain symmetries, perhaps of 
the sort discussed in 
\cite{Budzik:2023xbr}. It is not clear whether the $F_0$ tower in Table \ref{tower} 
can be understood in a similar way.

The Kaluza-Klein towers are apparently not related to any symmetries and thus 
may not be protected at large energies or large $j$. 
In particular, it will be interesting to see if these towers are 
related to the towers of giant gravitons appearing in 
the so-called `giant graviton expansion' of the index  
\cite{Imamura:2021ytr,Gaiotto:2021xce,Murthy:2022ien,Lee:2022vig}. 
This expansion recasts the index as an auxiliary summation 
over a tower, with increasing giant graviton number. More precisely, 
the giant graviton tower includes both finite $N$ gravitons (or the finite $N$ 
trace relations to be subtracted) as well as new black hole states formed by bound states 
of D-branes and open string excitations. In this framework, in certain charge regimes, 
the black hole entropy is obtained by first computing the entropy $S(j,n)$ with 
fixed giant graviton number $n$ and then maximizing $S(j,n)$ in $n$ at fixed $j$
\cite{Kinney:2005ej,Choi:2022ovw}. (See \cite{Choi:2022ovw,Beccaria:2023hip} for how 
one may generalize this calculation.) Indeed there is a signal 
that the giant graviton tower loses its meaning at higher $n\gg N$ \cite{Choi:2022ovw}.
With this interpretation in mind, it will be interesting to compute $Z-Z_{\rm grav}$ either 
till higher orders in $j$ or exactly, to see whether the KK towers 
survive at arbitrary high $j$ or not.

\subsection{$SU(4)$}

In the $SU(4)$ case, we computed $Z_{\rm grav}$ till $j=30$ level. 
The index $Z-Z_{\rm grav}$ over non-graviton cohomologies is given by
\begin{equation}
  Z-Z_{\rm grav}=\left[-\chi_{[2,0]}(x,y)t^{28}
  -\chi_{[3,0]}(x,y)t^{30}+\mathcal{O}(t^{32})
  \right]\cdot \prod_{I=1}^3(1-e^{-\Delta_I})\ .
\end{equation}
The second factor generates the Fock space of each $SU(1|3)$ multiplet, 
while the first factor in the square parenthesis represents the non-graviton primaries.
One finds that the BMN index predicts an apparent threshold of non-graviton 
cohomologies at $j=6(R+J)=28$. Again, conservatively, this is an upper bound 
for the threshold for two different reasons: first because the index may miss a 
pair of canceling threshold cohomologies at lower charges, and also because the true 
threshold might lie outside the BMN sector (carrying nonzero $SU(2)_r$ spin $J_1-J_2$).
Anyway, the above apparent threshold is higher than the $SU(3)$ threshold. 
So it is natural to expect that it was an exception that 
the $SU(2)$ and $SU(3)$ thresholds were the same: the (apparent) 
thresholds for $j=6(R+J)$ are $24,24,28,\cdots$ for $N=2,3,4\cdots$. 
To obtain the threshold level in terms of energy $E=3R+2J$, 
one should construct the actual cohomologies which account for the 
$t^{28}$ term. This will not be done in this paper.

\section{Constructing cohomologies}

The cohomologies we would like to construct should be $Q$-closed and not $Q$-exact. 
Unlike gravitons, the $Q$-closedness of the black hole 
cohomologies should be ensured by the trace relations. (Otherwise, if it is a
cohomology at given energy and at arbitrary values of $N$,  
it is a graviton cohomology.) So it is important to know 
what kind of nontrivial trace relations are available for $N\times N$ matrices 
when the number of fields is larger than $N$.

It seems to be widely believed that all $SU(N)$ trace relations are derived 
from the Cayley-Hamilton identity. For instance, see \cite{Ebertshauser:2001nj} 
(p.7, below eqn.(19)) and \cite{Dempsey:2022uie}.
But in practice it is inefficient to search for the trace 
relations that we need just from this identity. Fortunately, we already 
know many trace relations from the calculations reported in section 3.
Namely, when enumerating finite $N$ gravitons, 
we have counted them subject to various trace relations of the generators $g_i$. 
So one can take advantage of these trace relations to construct black hole cohomologies. 
This leads to our `ansatz' for black hole cohomologies, which we 
explain now.

We can motivate the ideas with a simple example in the $SU(2)$ theory \cite{Chang:2022mjp,Choi:2022caq,Choi:2023znd}. 
A representative of the threshold non-graviton cohomology in $SU(2)$ is given by
\begin{equation}\label{SU2-threshold}
 O_0\equiv\epsilon^{abc}{(v_2)^m}_a{(v_2)^n}_b{\rm tr}(\psi_{(c}\psi_{m}\psi_{n)})
\end{equation}
where $v_2$ is the graviton operator in the $S_2$ multiplet. 
Let us see how this operator becomes $Q$-closed. Acting $Q$ on $O_0$, 
$Q$ acts only on ${\rm tr}(\psi_{(c}\psi_m\psi_{n)})$ since $v_2$ is $Q$-closed.
One obtains
\begin{equation}\label{trace-relation-example}
  Q{\rm tr}(\psi_{(c}\psi_m\psi_{n)})\propto \epsilon_{ab(c}{(v_2)^a}_m{(v_2)^b}_{n)}
  \equiv R(v_2)_{cmn}
\end{equation}
after using $SU(2)$ trace relations. 
Plugging this into $QO_0$, one obtains
\begin{equation}\label{syzygy-example}
  QO_0\propto \epsilon^{abc}{(v_2)^m}_a{(v_2)^n}_bR(v_2)_{cmn}=0\ .
\end{equation}
At the last step, one can show that the quartic mesonic polynomial 
$\epsilon^{abc}{(v_2)^m}_a{(v_2)^n}_bR(v_2)_{cmn}$ 
is identically zero \cite{Choi:2023znd}.
From the viewpoint of section 3, (\ref{trace-relation-example}) are 
graviton trace relations and the last step of (\ref{syzygy-example}) is 
a relation of relations. So the operator $O_0$ is shown to be $Q$-closed by 
using the trace relations and a relation of relations of the finite $N$ graviton operators.

This idea can be extended to construct operators which become $Q$-closed 
only after using trace relations. Namely, for each relation of relations 
such as (\ref{syzygy-example}),
we can construct a $Q$-closed operator such as (\ref{SU2-threshold}).
We still need to check that they are not $Q$-exact for them to represent 
nontrivial $Q$-cohomologies.
Also, there are non-graviton cohomologies which are not constructed 
this way \cite{Choi:2023znd}.
For these reasons, the $Q$-closed operators constructed in this way are
mere ans\"atze for the non-graviton cohomologies.
In the appendix, we have collected 
all $SU(3)$ fundamental trace relations including $u_n,v_n$ only, 
and manifestly wrote them in $Q$-exact forms. We have also found trace relations 
involving $u_n,v_n,w_n$ till $j=20$ order.
We have also found all relations between the fundamental graviton 
trace relations at $j=24$ and some of them at $j=30$ orders in the $SU(3)\subset SO(6)_R$ 
singlet sector where the index predicts non-graviton cohomologies (see Table \ref{tower}). 
In other charge sectors, one can immediately write down $Q$-closed operators 
if one finds new relations of the fundamental trace relations.

When we write a fundamental trace relation $R_a$ in a $Q$-exact form, 
$R_a\sim Qr_a$, there is an ambiguity in $r_a$ by addition of 
arbitrary $Q$-closed operators. We partly fix it so that 
$r_a$ vanishes when all the letters are restricted to diagonal matrices. Since the $Q$-closed 
operators constructed from relations of relations are linear combinations of $r_a$'s, they 
vanish with diagonal letters. This makes it impossible for our ansatz to be gravitons. 
So our ansatz either yields $Q$-exact operators or non-graviton cohomologies.

Based on the ansatz, in subsection 4.1, we construct a number of gauge-invariant $Q$-closed
non-graviton operators at $j=24$ order that are singlets of $SU(3)$ global symmetry.
Only one of them is not $Q$-exact, representing 
the non-graviton cohomology predicted by the index.
In subsection 4.2, we sketch how we checked the (non-)$Q$-exactness of these operators,
while also showing that there are no other non-graviton cohomologies
in the $j=24$, $SU(3)$ singlet sector.

\subsection{$SU(3)$ threshold cohomology from ans\"atze}

In this subsection, we present the explicit form of the black hole cohomology at the threshold level $j=24$ which is singlet under $SU(3) \subset SU(4)_R$ global symmetry, 
in the BMN sector of the $SU(3)$ gauge theory. We first list the non-graviton $Q$-closed operators from our ansatz. We find one non-$Q$-exact operator among them,  
which is the threshold cohomology.

At $j\equiv 6(R+J)=24$, operators are further distinguished by R-charges 
$R \equiv \frac{R_1+R_2+R_3}{3}$. The BMN operators which are $SU(3) \subset SU(4)_R$ singlets 
satisfy $R_1=R_2=R_3$ and $J_1=J_2$. Then the possible charges of the operators are $(R,J) = (\frac{n}{2},\frac{8-n}{2})$ where $n=0,\cdots , 8$. In each charge sector, the number of letters 
is fixed to $n+4$. However, our ansatz further restricts the charges since acting $Q$ on our ansatz should become a polynomial of 
$u_{2,3}, v_{2,3}, w_{2,3}$. As a result, there exist total 7 possible charge sectors within our ansatz: $(R,J) = (\frac{n}{2},\frac{8-n}{2})$ where $n=1,\cdots , 7$.

When $(R,J) = (\frac{1}{2}, \frac{7}{2})$ or $(1,3)$, there are no $Q$-closed 
operators within our ansatz using the trace relations in the appendix.
One can understand it heuristically as follows. At these charges, 
$R$ is so small that only a small number of scalars is admitted.  
As the graviton generators contain at least one scalar field, only few types of graviton polynomials 
exist in these sectors, which are not enough to host relations 
of relations. Therefore, these charge sectors are incompatible with our ansatz. 
The other $5$ charge sectors host $Q$-closed operators in our ansatz, 
whose explicit forms will be presented below.

We now present the $Q$-closed non-graviton 
operators in each of the five charge sectors,
$(R,J) = (\frac{n}{2},\frac{8-n}{2})$ where $n=3,\cdots ,7$. For convenience, we 
rewrite here the definition of the single-trace generators of the $SU(3)$ 
BMN gravitons $u_{2,3}, v_{2,3}, w_{2,3}$:
\begin{equation}
    \begin{aligned}
        u^{ij} \equiv & \; \textrm{tr}  \left(\phi^{(i} \phi^{j)}\right)\ ,\ \   
        u^{ijk} \equiv \; \textrm{tr} \left(\phi^{(i} \phi^{j} \phi^{k)}\right)\ , \\
        {v^{i}}_j \equiv & \; \textrm{tr} \left( \phi^i \psi_{j}\right) - {\textstyle \frac{1}{3} }
        \delta^i_j\textrm{tr}\left(\phi^a \psi_{a}\right)\ , \ \ 
        {v^{ij}}_k \equiv \; \textrm{tr} \left(\phi^{(i} \phi^{j)} \psi_{k} \right) 
        - {\textstyle \frac{1}{4}} \delta^{i}_{k}  
        \textrm{tr} \left( \phi^{(j} \phi^{a)} \psi_{a} \right) 
        - {\textstyle \frac{1}{4}} \delta^{j}_{k}  \textrm{tr} \left( \phi^{(i} \phi^{a)} \psi_{a} \right)\ , \\
        w^i \equiv & \; \textrm{tr} \left(f \phi^i + {\textstyle \frac{1}{2}} \epsilon^{ia_1a_2} \psi_{a_1} \psi_{a_2}\right)\ ,\ \ 
        w^{ij} \equiv \; \textrm{tr}\left(f \phi^{(i} \phi^{j)} +\epsilon^{a_1a_2(i}\phi^{j)} \psi_{a_1} \psi_{a_2}\right)\ .
    \end{aligned}
\end{equation}

\paragraph{i) $(R,J) = (\frac{3}{2}, \frac{5}{2})$} The operators in this sector are made of 7 letters.
The possible numbers $(n_\phi,n_\psi,n_f)$ of scalars, fermions and $f$ in each term are $(n_\phi, n_\psi, n_f) = (4,1,2), (3,3,1),(2,5,0)$.
We find 1 $Q$-closed operator in this sector from the trace relations and a relation of 
relations in appendix A. This $Q$-closed operator is given by
\begin{equation}\label{O1} 
        O^{(2,1)} \equiv  65u^{ij}  (r_{20}^{(2,1)})_{ij} -39w^{ij}   (r_{14}^{(1,1)})_{ij} +5w^i   (r_{16}^{(1,1)})_i  
        +312{v^{jk}}_i   (r_{16}^{(1,2)})^i_{jk} +26{v^j}_i    (r_{18}^{(1,2)})^i_j +6w^i   (r_{16}^{(0,3)})_i\ .
\end{equation}
The superscripts denote $(n_f, n_\psi)$ of the terms with maximal $n_f$ in the operator. 
$r_j^{(n_f,n_\psi)}$'s are given in \eqref{tr-r}, \eqref{tr-r-f} where $R_j^{(n_f,n_\psi-1)} \equiv i \, Q\, r_j^{(n_f,n_\psi)}$'s are 
the fundamental trace relations.
The $Q$-closed operator \eqref{O1} turns out to be $Q$-exact.
In fact, \eqref{O1} is even under the parity transformation of \cite{Beisert:2004ry}. It is already known that 
all such even operators in this charge sector are $Q$-exact for all $N\geq 3$
\cite{Budzik:2023vtr}.

\paragraph{ii) $(R,J) = (2, 2)$} The operators in this sector are made of 8 letters. Allowed $(n_\phi, n_\psi, n_f)$ are $(6,0,2), (5,2,1),(4,4,0)$. We find 4 $Q$-closed operators in this sector given by
\begin{equation}\label{O2}
    \begin{aligned}
        O_1^{(1,2)} \equiv & -3{v^{(j}}_{i} w^{k)}   (r_{10}^{(0,1)})^i_{jk} -3u^{(ij}w^{k)}   (r_{12}^{(0,2)})_{ijk} +\epsilon_{a_1a_2i} u^{a_1j} w^{a_2}  (r_{12}^{(0,2)})^i_j\ , \\
        O_2^{(1,2)} \equiv & -9u^{a(i} {v^{j)}}_{a}   (r_{14}^{(1,1)})_{ij} +10\epsilon_{a_1a_2(i}u^{a_1k} {v^{a_2}}_{j)}   (r_{14}^{(1,1)})^{ij}_k + 30 {v^{(j}}_{i} w^{k)}   (r_{10}^{(0,1)})^i_{jk}
        +60u^{(jk} {v^{l)}}_i   (r_{14}^{(0,3)})^i_{jkl}\ , \\
        O_3^{(1,2)} \equiv & -3u^{a(i} {v^{j)}}_{a}   (r_{14}^{(1,1)})_{ij} 
        +6\epsilon_{a_1a_2(i}u^{a_1k} {v^{a_2}}_{j)}   (r_{14}^{(1,1)})^{ij}_k
        +4u^{ijk}   (r_{18}^{(1,2)})_{ijk} 
        +14{v^{(j}}_{i} w^{k)}   (r_{10}^{(0,1)})^i_{jk} \\
        & -6w^{ij}   (r_{14}^{(0,2)})_{ij} 
        -12\epsilon^{a_1a_2(i} {v^{j}}_{a_1} {v^{k)}}_{a_2}   (r_{12}^{(0,2)})_{ijk} 
        -4{v^j}_a {v^a}_i   (r_{12}^{(0,2)})^i_j\ , \\
        O_4^{(1,2)} \equiv & -3u^{a(i} {v^{j)}}_{a}   (r_{14}^{(1,1)})_{ij} 
        +14\epsilon_{a_1a_2(i}u^{a_1k} {v^{a_2}}_{j)}   (r_{14}^{(1,1)})^{ij}_k
        +8{v^{jk}}_{i}   (r_{16}^{(1,1)})^i_{jk} 
        +42{v^{(j}}_{i} w^{k)}   (r_{10}^{(0,1)})^i_{jk} \\
        &+12 u^{(ij}w^{k)}   (r_{12}^{(0,2)})_{ijk} 
        -24 w^{ij}   (r_{14}^{(0,2)})_{ij} 
        -36\epsilon^{a_1a_2(i} {v^{j}}_{a_1} {v^{k)}}_{a_2}   (r_{12}^{(0,2)})_{ijk} 
        - 8 {v^{jk}}_i   (r_{16}^{(0,3)})^i_{jk}\ .
    \end{aligned}
\end{equation}
All operators in \eqref{O2} are $Q$-exact.

\paragraph{iii) $(R,J) = (\frac{5}{2}, \frac{3}{2})$} The operators in this sector are made of 9 letters. Allowed $(n_\phi, n_\psi, n_f)$ are $(7,1,1),(6,3,0)$. We find 13 $Q$-closed operators in this sector given by
\begin{equation}\label{O3}
    \begin{aligned}
    &O_1^{(1,1)} \equiv \epsilon_{a_1a_2 i } u^{a_1 (j} w^{k) a_2}   (r_{10}^{(0,1)})^i_{jk} \ , \\
    &O_2^{(1,1)} \equiv \epsilon_{a_1a_2 i } u^{a_1 jk} w^{a_2}   (r_{10}^{(0,1)})^i_{jk} \ , \\
    &O_3^{(1,1)} \equiv \epsilon_{a_1a_2i}\epsilon_{b_1b_2j} u^{a_1b_1}u^{a_2b_2k}   (r_{14}^{(1,1)})^{ij}_k 
    +5{v^a}_i {v^{jk}}_a (r_{10}^{(0,1)})^i_{jk} -2{v^{(j}}_a {v^{k)a}}_i  (r_{10}^{(0,1)})^i_{jk} \ , \\
    &O_1^{(0,3)}  = -\epsilon_{i a_1 a_2} \left( 4 u^{a_1 b} {v^{j a_2}}_{b} + 3u^{j a_1b} {v^{a_2}}_{b}\right)   (r_{12}^{(0,2)})^i_j = \frac{1}{2} i\, Q ((r_{12}^{(0,2)})^i_j (r_{12}^{(0,2)})^j_i) \ , \\
    &O_2^{(0,3)}  = -\epsilon_{a_1 a_2 (i}\left(u^{a_1(k}{v^{l)a_2}}_{j)}+
        u^{kl a_1}{v^{a_2}}_{j)}\right)   (r_{12}^{(0,2)})^{ij}_{kl} = \frac{1}{2} i\, Q ((r_{12}^{(0,2)})^{kl}_{ij} (r_{12}^{(0,2)})^{ij}_{kl})\ , \\
    &O_3^{(0,3)} \equiv -u^{a(i}{v^{jk)}}_a   (r_{12}^{(0,2)})_{ijk} \ , \\
    &O_4^{(0,3)} \equiv -\epsilon_{a_1 a_2 i} u^{a_1 b} {v^{a_2}}_b   (r_{14}^{(0,2)})^i\ , \\
    &O_5^{(0,3)} \equiv 6{v^a}_i {v^{jk}}_a  (r_{10}^{(0,1)})^i_{jk}
    +6u^{a(ij}{v^{k)}}_a   (r_{12}^{(0,2)})_{ijk} 
    +\epsilon_{a_1a_2 i} u^{a_1 bj}{v^{a_2}}_{b}   (r_{12}^{(0,2)})^i_j\ ,\\
    &O_6^{(0,3)} \equiv 24 {v^{(j}}_a {v^{k)a}}_i  (r_{10}^{(0,1)})^i_{jk} 
    +6 u^{a(i} {v^{j)}}_{a}   (r_{14}^{(0,2)})_{ij} 
    - \epsilon_{a_1a_2(i} u^{a_1 k} {v^{a_2}}_{j)}   (r_{14}^{(0,2)})^{ij}_k \ ,\\
    &O_7^{(0,3)} \equiv {v^a}_i {v^{jk}}_a  (r_{10}^{(0,1)})^i_{jk} 
    -10{v^{(j}}_a {v^{k)a}}_i  (r_{10}^{(0,1)})^i_{jk} 
    +6u^{a(ij}{v^{k)}}_a   (r_{12}^{(0,2)})_{ijk}  
    +10 \epsilon_{a_1a_2(i} u^{a_1 kl}{v^{a_2}}_{j)}   (r_{12}^{(0,2)})^{ij}_{kl}  \ ,\\
    &O_8^{(0,3)} \equiv  5{v^a}_i {v^{jk}}_a  (r_{10}^{(0,1)})^i_{jk} 
    -2{v^{(j}}_a {v^{k)a}}_i  (r_{10}^{(0,1)})^i_{jk} 
    +9u^{a(ij}{v^{k)}}_a   (r_{12}^{(0,2)})_{ijk}  
    +6\epsilon_{a_1 a_2 i} u^{a_1 (j}u^{kl) a_2}   (r_{14}^{(0,3)})^{i}_{jkl}\ ,\\
    &O_9^{(0,3)} \equiv 6{v^a}_i {v^{jk}}_a  (r_{10}^{(0,1)})^i_{jk} 
    +12{v^{(j}}_a {v^{k)a}}_i  (r_{10}^{(0,1)})^i_{jk} 
    +18u^{a(ij}{v^{k)}}_a  (r_{12}^{(0,2)})_{ijk}  
    -\epsilon_{a_1a_2(i} u^{a_1 k} {v^{a_2}}_{j)}   (r_{14}^{(0,2)})^{ij}_k \ ,\\
    &O_{10}^{(0,3)} \equiv 38{v^a}_i {v^{jk}}_a  (r_{10}^{(0,1)})^i_{jk} 
    +4{v^{(j}}_a {v^{k)a}}_i  (r_{10}^{(0,1)})^i_{jk} 
    +24u^{a(ij}{v^{k)}}_a  (r_{12}^{(0,2)})_{ijk}  
    +5u^{(jk}{v^{l)}}_i   (r_{14}^{(0,2)})^i_{jkl}  \ .
    \end{aligned}
\end{equation}
All except for $O^{(0,3)}_6$ in \eqref{O3}  are $Q$-exact. 
Therefore, a representative of the cohomology in this sector can be written as\footnote{{\bf v3:}
the last line of \eqref{Q-coho} differs from that in v2, due to the correction of $(r_{14}^{(0,2)})_{ij}$ in \eqref{tr-rel}.
The representative as written in v2 was $Q$-exact, whereas the corrected one is not.
It has been shown in \cite{Kim:2026} that \eqref{Q-coho} is cohomologous to a non-Coulomb graviton, which was assumed to be absent in this paper.}
\begin{eqnarray}\label{Q-coho}
    O & \equiv & -6O_6^{(3)} \\
    &=& \; 288 {v^{(j}}_a {v^{k)a}}_i\times \epsilon_{c_1c_2(j} \tr \left( \bar\phi^{c_1}\bar\phi^{c_2}\bar\phi^{i}\psi_{k+)} \right) -72 {v^{a}}_b {v^{bk}}_a\times \epsilon_{c_1c_2(k} \tr \left( \bar\phi^{c_1}\bar\phi^{c_2}\bar\phi^{d}\psi_{d+)} \right) \nn\\ 
    &&+36\epsilon_{a_1a_2(i} u^{a_1 k} {v^{a_2}}_{j)} \times \left[ 2\tr\left(\phi^{(i} \phi^{c} \phi^{j)} \psi_{(c} \psi_{k)}\right)+2\tr\left(\phi^{(i|} \phi^{c} \phi^{|j)} \psi_{(c} \psi_{k)}\right) \right. \nn\\
    &&\qquad \qquad \qquad \qquad \quad \left. +9 \tr\left( \phi^{(i} \phi^{j} \psi_{(c} \phi^{c)} \psi_{k)} \right)-6 \tr\left( \phi^{(i} \phi^{j)} \psi_{(c} \phi^c \psi_{k)} \right)\right] \nn\\
    &&-9\epsilon_{a_1a_2j} u^{a_1 b} {v^{a_2}}_{b} \times \left[ 2\tr\left(\phi^{(j} \phi^{c} \phi^{d)} \psi_{(c} \psi_{d)}\right)+2\tr\left(\phi^{(j|} \phi^{c} \phi^{|d)} \psi_{(c} \psi_{d)}\right) \right.\nn\\
    &&\qquad \qquad \qquad \quad \;\;\, \left. \, +9 \tr\left( \phi^{(j} \phi^{d} \psi_{(c} \phi^{c)} \psi_{d)} \right)-6 \tr\left( \phi^{(j} \phi^{d)} \psi_{(c} \phi^c \psi_{d)} \right)\right] \nn\\
    &&-20 u^{a(i} {v^{j)}}_{a} \times \epsilon_{a_1a_2a_3}\left[2 \tr\left(\psi_{(i}\psi_{j)} \phi^{a_1} \phi^{a_2}\phi^{a_3}\right) + \tr\left(\psi_{(i} \phi^{a_1}\psi_{j)}\phi^{a_2}\phi^{a_3}\right)\right] \nn \\
    &&-36 u^{a(i} {v^{j)}}_{a} \times \epsilon_{a_1a_2(i}\left[\tr\left(\psi_{j)} \psi_{a_3}\phi^{a_1}\phi^{a_2}\phi^{a_3}\right)+\tr\left(\psi_{j)} \psi_{a_3}\phi^{a_1}\phi^{a_3}\phi^{a_2}\right)+\tr\left(\psi_{j)} \psi_{a_3}\phi^{a_3}\phi^{a_1}\phi^{a_2}\right)\right] \nn\\
    &&-36 u^{a(i} {v^{j)}}_{a} \times \epsilon_{a_1a_2(i}\left[\tr\left(\psi_{j)} \phi^{a_1}\psi_{a_3}\phi^{a_2}\phi^{a_3}\right)+\tr\left(\psi_{j)} \phi^{a_1}\psi_{a_3}\phi^{a_3}\phi^{a_2}\right)+\tr\left(\psi_{j)} \phi^{a_3}\psi_{a_3}\phi^{a_1}\phi^{a_2}\right)\right] \nn\\
    &&-36 u^{a(i} {v^{j)}}_{a} \times \epsilon_{a_1a_2(i}\left[\tr\left(\psi_{j)} \phi^{a_1}\phi^{a_2}\psi_{a_3}\phi^{a_3}\right)+\tr\left(\psi_{j)} \phi^{a_1}\phi^{a_3}\psi_{a_3}\phi^{a_2}\right)+\tr\left(\psi_{j)} \phi^{a_3}\phi^{a_1}\psi_{a_3}\phi^{a_2}\right)\right] \nn\\
    &&-36 u^{a(i} {v^{j)}}_{a} \times \epsilon_{a_1a_2(i}\left[\tr\left(\psi_{j)} \phi^{a_1}\phi^{a_2}\phi^{a_3}\psi_{a_3}\right)+\tr\left(\psi_{j)} \phi^{a_1}\phi^{a_3}\phi^{a_2}\psi_{a_3}\right)+\tr\left(\psi_{j)} \phi^{a_3}\phi^{a_1}\phi^{a_2}\psi_{a_3}\right)\right] \nn\\
    &&+60 u^{a(i} {v^{j)}}_{a} \times \epsilon_{a_1a_2(i}\left[\tr\left(\psi_{j)}\phi^{a_1}\phi^{a_2}\right)\tr\left(\psi_{a_3}\phi^{a_3}\right)+\tr\left(\psi_{j)}\phi^{a_1}\phi^{a_3}\right)\tr\left(\psi_{a_3}\phi^{a_2}\right) \right. \nn\\ 
    &&\qquad \qquad \qquad \qquad \;\;\, \left. +\tr\left(\psi_{j)}\phi^{a_3}\phi^{a_1}\right)\tr\left(\psi_{a_3}\phi^{a_2}\right)\right]\ . \nn
\end{eqnarray}
The scaling dimension of this cohomology $O$ is $E=3R+2J=\frac{21}{2}$. Note that the 
representative found above does not contain the letter $f$.

\paragraph{iv) $(R,J) = (3, 1)$} The operators in this sector are made of 10 letters. Allowed $(n_\phi, n_\psi, n_f)$ are $(9,0,1),(8,2,0)$. We find 6 $Q$-closed operators in this sector given by
\begin{equation}\label{O4}
    \begin{aligned}
        &O^{(0,2)}_1 \equiv - \epsilon_{a_1a_2i} u^{a_1 b} u^{jk} {v^{a_2}}_{b}   (r_{10}^{(0,1)})^i_{jk}
        +2\epsilon_{a_1a_2 i}u^{a_1 b} u^{a_2 (j} {v^{k)}}_{b}   (r_{10}^{(0,1)})^i_{jk} \ , \\
        &O^{(0,2)}_2 \equiv -6\epsilon_{a_1a_2i} u^{a_1b(j} {v^{k)a_2}}_{b}   (r_{10}^{(0,1)})^i_{jk} 
        - \epsilon_{a_1a_2(i} u^{a_1(k} {v^{l)a_2}}_{j)}   (r_{12}^{(0,1)})^{ij}_{kl} \ , \\
        &O^{(0,2)}_3 \equiv -\epsilon_{a_1a_2i} u^{a_1 b} u^{jk} {v^{a_2}}_{b}   (r_{10}^{(0,1)})^i_{jk} 
        - \epsilon_{a_1a_2(i} u^{a_1kl} {v^{a_2}}_{j)}   (r_{12}^{(0,1)})^{ij}_{kl} \ , \\
        &O^{(0,2)}_4 \equiv -\epsilon_{a_1a_2i} u^{a_1 b} u^{jk} {v^{a_2}}_{b}   (r_{10}^{(0,1)})^i_{jk} 
        + \epsilon_{a_1a_2(i} \epsilon_{j)b_1b_2} u^{a_1b_1} u^{a_2b_2} u^{kl}   (r_{12}^{(0,2)})^{ij}_{kl} \ , \\
        &O^{(0,2)}_5 \equiv -4\epsilon_{a_1a_2i} u^{a_1 b} u^{jk} {v^{a_2}}_{b}   (r_{10}^{(0,1)})^i_{jk} 
        - 24\epsilon_{a_1a_2i} u^{a_1b(j} {v^{k)a_2}}_{b}   (r_{10}^{(0,1)})^i_{jk} 
        - \epsilon_{a_1a_2(i} \epsilon_{j)b_1b_2} u^{a_1b_1} u^{a_2b_2k}   (r_{14}^{(0,2)})^{ij}_k  \ , \\
        &O^{(0,2)}_6 \equiv -\epsilon_{a_1a_2i} u^{a_1 b} u^{jk} {v^{a_2}}_{b}   (r_{10}^{(0,1)})^i_{jk} 
        + 12\epsilon_{a_1a_2i} u^{a_1b(j} {v^{k)a_2}}_{b}   (r_{10}^{(0,1)})^i_{jk} 
        + 3\epsilon_{a_1a_2i} u^{a_1(j} u^{kl)a_2}   (r_{14}^{(0,2)})^i_{jkl} \ .
    \end{aligned}
\end{equation}
All the operators in \eqref{O4} are $Q$-exact.

\paragraph{v) $(R,J) = (\frac{7}{2}, \frac{1}{2})$} The operators in this sector are made of 11 letters. Allowed $(n_\phi, n_\psi, n_f)$ is $(9,1,0)$. We find 1 $Q$-closed operator in this sector given by
\begin{eqnarray}\label{O5}
    O^{(0,1)} & \equiv& 36 \epsilon_{a_1a_2a_3} \epsilon_{b_1b_2 i} u^{a_1b_1} u^{a_2b_2} u^{a_3jk}   (r_{10}^{(0,1)})^i_{jk} 
    +5 \epsilon_{a_1 a_2 a_3} \epsilon_{b_1 b_2 b_3} u^{a_1 b_1}u^{a_2 b_2}u^{a_3 b_3}   r_{12}^{(0,1)} 
    \nonumber\\
    &&-6 \epsilon_{a_1a_2(i} \epsilon_{j)b_1b_2} u^{a_1b_1} u^{a_2b_2} u^{kl}   (r_{12}^{(0,1)})^{ij}_{kl}\ . 
\end{eqnarray}
The operator \eqref{O5} is $Q$-exact.

In summary, we have found 1 fermionic black hole cohomology using our ansatz, which is a singlet under $SU(3) \subset SU(4)_R$, at $j=24$ whose representative is given by (\ref{Q-coho}).
Its charges and scaling dimension are given by $(R,J,E) = \left(\frac{5}{2},\frac{3}{2},\frac{21}{2}\right)$.

\subsection{$Q$-exactness checks and ansatz-independent studies}

In this subsection, we sketch how to determine $Q$-exactness of various
$Q$-closed operators introduced in the previous subsection.
We also show that \eqref{Q-coho} is the only non-graviton cohomology
at $j=24$ in the $SU(3)$ singlet sector most generally, without imposing the ansatz. 

To check whether a given operator is $Q$-exact or not, especially to check 
non-$Q$-exactness, one has to 
rule out all possible ways of writing the operator as $Q$ of `something'.
That being said, one needs to construct all possible operators
that can participate in `something' (the meaning of which will be made clear shortly)
and show that the \emph{target} operator is linearly independent of $Q$-actions of them.
More specifically, we divide the check of $Q$-exactness into 4 steps,
that we summarize as follows.
\begin{enumerate}
    \item Construct all gauge-invariant operators whose $Q$-action may participate
    in reproducing the \emph{target}.
    \item Count the number of linearly independent operators from step 1,
    and extract the maximal subset of linearly independent operators.
    This is called the \emph{basis}.
    \item Act $Q$ on the basis operators, then again count and extract
    the maximal subset of linearly independent ones between them.
    \item Check if the target is linearly independent of the result of step 3.
\end{enumerate}

Now we explain what operators `may participate in reproducing the target' in step 1.
This consists of two criteria: the charges and the parity under permutation.

First, the charges of the target operator constrain the charges,
thus the letter contents of the basis operators.
Note that the action of $Q$ increases $R_{I=1,2,3}$ by $\frac12$
and decreases $J = J_1 = J_2$ by $\frac12$.
Therefore, the basis operators must have the set of charges
that differ by the corresponding amount from the target,
otherwise their $Q$-actions are disjoint from the target.
Note that all of our targets are $SU(3)$ singlets, so we always have $R=R_1=R_2=R_3$.

Second, all of our targets being singlets under the $SU(3)$ subgroup
of the $SU(4)$ R-symmetry group,
imposes a stronger constraint than just restricting to the
charge sectors with $R_1=R_2=R_3$.
Each basis operator must be invariant under cyclic permutation
$\phi^{i} \to \phi^{i+1}$ and simultaneously $\psi_{i} \to \psi_{i+1}$,
where $i=1,2,3$ mod 3.
Moreover, if there are even/odd number of $\phi$'s and $\psi$'s combined,
which carries one $SU(3)$ index each,
it requires even/odd number of Levi-Civita symbols to write the operator
covariantly while contracting all indices.
Therefore, we may restrict to
i) operators with even number of $\phi$'s and $\psi$'s combined,
that are even under all $3!$ permutations of $SU(3)$ indices, and
ii) operators with odd number of $\phi$'s and $\psi$'s combined,
that are even under even (cyclic) permutations of $SU(3)$ indices
and odd under odd (swap) permutations of $SU(3)$ indices.
Also note that this permutation property commutes with the action of $Q$,
so that $Q$ of a non-trivial operator satisfies this property
if and only if the original operator does.
This permutation property is necessary but not sufficient for an operator
to be an $SU(3)$ singlet.
However, we impose this property on the basis instead of requiring $SU(3)$ singlets, because the latter requires many sums over dummy indices
and thus the former is computationally more efficient.
Our conclusions on the singlet sector will be valid despite.

For example, suppose that the target operator is \eqref{Q-coho},
which has charges $(R,J) = (\frac52,\frac32)$.
Operators whose $Q$-action may reproduce this target operator
must then have $(R,J) = (2,2)$.
Possible choices of letter contents are
$(n_\phi, n_\psi, n_f) = (6,0,2), (5,2,1)$, or $(4,4,0)$,
and numbers of $\phi^i$ minus numbers of $\psi_i$ must be equal between $i=1,2,3$.
Further taking into account the permutation property,
the basis operators whose $Q$-action `may participate in reproducing the target'
\eqref{Q-coho} can be classified into the following 7 subsectors.
($(-1)^\epsilon$ in subsectors 5 and 6 indicates minus sign for odd permutations,
because there are odd number of $\phi$'s and $\psi$'s in those subsectors.)
\begin{itemize}
    \item Subsector 1: $(\phi^1)^4(\psi_1)^4 + $ (permutations)
    \item Subsector 2: $(\phi^1)^3(\phi^2)^1(\psi_1)^3(\psi_2)^1 +$ (permutations) 
    \item Subsector 3: $(\phi^1)^2(\phi^2)^2(\psi_1)^2(\psi_2)^2 +$ (permutations)
    \item Subsector 4: $(\phi^1)^2(\phi^2)^1(\phi^3)^1(\psi_1)^2(\psi_2)^1(\psi_3)^1 +$ (permutations) 
    \item Subsector 5: $(\phi^1)^3(\phi^2)^1(\phi^3)^1(\psi_1)^2f^1 + (-1)^\epsilon$ (permutations)
    \item Subsector 6: $(\phi^1)^2(\phi^2)^2(\phi^3)^1(\psi_1)^1(\psi_2)^1f^1+ (-1)^\epsilon$ (permutations)
    \item Subsector 7: $(\phi^1)^2(\phi^2)^2(\phi^3)^2f^2 +$ (permutations) 
\end{itemize}
Appropriate sums over permutations of single- and multi-trace operators
in each of these subsectors are the result of step 1,
some of which we write down below to help visualize:
\begin{eqnarray}\label{QEx-basis-eg}
    &&\hspace{-1cm} {\rm tr}(\phi^1\phi^1\psi_1\phi^1\psi_1\psi_1\phi^1\psi_1)
    + {\rm tr}(\phi^2\phi^2\psi_2\phi^2\psi_2\psi_2\phi^2\psi_2)
    + {\rm tr}(\phi^3\phi^3\psi_3\phi^3\psi_3\psi_3\phi^3\psi_3)~, \nonumber \\
    &&\hspace{-1cm} 
    {\rm tr}(\phi^1\phi^1\phi^2\psi_2){\rm tr}(\psi_1\psi_2){\rm tr}(\phi^2\psi_1) +
    {\rm tr}(\phi^2\phi^2\phi^3\psi_3){\rm tr}(\psi_2\psi_3){\rm tr}(\phi^3\psi_2) +
    {\rm tr}(\phi^3\phi^3\phi^1\psi_1){\rm tr}(\psi_3\psi_1){\rm tr}(\phi^1\psi_3) \nonumber \\
    &&\hspace{-1cm}
    ~~+ {\rm tr}(\phi^3\phi^3\phi^2\psi_2){\rm tr}(\psi_3\psi_2){\rm tr}(\phi^2\psi_3) +
    {\rm tr}(\phi^1\phi^1\phi^3\psi_3){\rm tr}(\psi_1\psi_3){\rm tr}(\phi^3\psi_1) +
    {\rm tr}(\phi^2\phi^2\phi^1\psi_1){\rm tr}(\psi_2\psi_1){\rm tr}(\phi^1\psi_2)~, \nonumber \\
    &&\hspace{-1cm} {\rm tr}(\phi^2\phi^2\psi_1\psi_2\phi^3){\rm tr}(f\phi^1\phi^1) +
    {\rm tr}(\phi^3\phi^3\psi_2\psi_3\phi^1){\rm tr}(f\phi^2\phi^2) +
    {\rm tr}(\phi^1\phi^1\psi_3\psi_1\phi^2){\rm tr}(f\phi^3\phi^3) \nonumber \\
    &&\hspace{-1cm}~~- {\rm tr}(\phi^3\phi^3\psi_1\psi_3\phi^2){\rm tr}(f\phi^1\phi^1) -
    {\rm tr}(\phi^1\phi^1\psi_2\psi_1\phi^3){\rm tr}(f\phi^2\phi^2) -
    {\rm tr}(\phi^2\phi^2\psi_3\psi_2\phi^1){\rm tr}(f\phi^3\phi^3)~.
\end{eqnarray}

Given the operators from step 1, the rest is relatively straightforward, at least conceptually.
There are non-trivial trace relations between operators from step 1,
so in step 2 we 
extract linearly independent basis operators.
Then in step 3, we consider $Q$-actions of the basis operators,
and again count the number of linearly independent ones among them.
These should form a complete basis of all $Q$-exact operators in the
target charge sector and with the aforementioned permutation property.
Therefore, the target operator is $Q$-exact if and only if it is a linear combination
of the $Q$-actions of the basis operators.
More generally, if there are multiple target operators,
the number of cohomologies among them would be equal to
the number of linearly independent ones among the basis \emph{and} all target operators,
minus the number of linearly independent ones among the basis only.

Each of step 2-4 involves counting and/or finding linearly independent operators
among a given set of gauge-invariant operators.
Each operator is a sum over single- and multi-trace operators
written in terms of seven species of fields $\phi^m$, $\psi_m$ and $f$.
To completely account for trace relations between them,
we first convert the operators written in terms of adjoint fields
into polynomials of their matrix elements, by substituting
\begin{eqnarray}
f &=& \begin{pmatrix} f_1 & f_2 & f_3 \\ f_4 & f_5 & f_6 \\ f_7 & f_8 & -f_1-f_5 \end{pmatrix}~,
\end{eqnarray}
and likes for 6 other fields. In this way, every operator is now written as
a polynomial of $8 \times 7 = 56$ variables, 24 of which are Grassmannian.
So the problem boils down to finding linear dependence between a set of polynomials.
Although this is the same problem that was encountered while computing the
graviton index in section 3, the same method of extracting the
coefficient matrix is extremely unpractical here.
It is because there are four times as many variables (recall that for
counting gravitons, we substituted each field with a diagonal matrix),
and therefore exponentially larger number of monomials appear in polynomials.
As a result, the coefficient matrix will have a huge number of columns
that is not viable for computers.

For this reason, we have devised a numerics-assisted approach to find
linear dependence between the polynomials with large number of variables.
The approach stems from the basic fact that if some linear combination
of certain polynomials vanishes,
it must also be zero if we attribute any specific number to each variable.
So let us represent each polynomial by an array of numbers, i.e. a row vector,
by substituting each variable with a set of randomly chosen integers.
Then we examine the linear dependence between vectors, instead of polynomials.

The substitution can be repeated for arbitrarily many sets of integers,
so the row vector can be made arbitrarily long.
Obviously, the length of the row vectors, i.e. number of columns,
must be at least as many as there are independent polynomials.
Otherwise, it will be always possible to find a relation between the row vectors
even if the polynomials they represent are independent.
On the other hand, the length of the row vectors need not be much more than
the number of independent polynomials, as we will explain shortly.

This makes it clear why this method is efficient.
It naturally realizes the basic principle that in order to distinguish
$n$ different entities, one needs at least $n$ data for each entity,
whereas extracting the coefficient matrix for the polynomials with so many variables
will equivalently convert each polynomial into an unnecessarily long vector.

There are two issues with this approach that we need to address.
The first is that 24 of 56 variables are Grassmannian,
which cannot be properly substituted with c-numbers.
The second is that randomness is involved in this approach,
and it may lead to errors albeit unlikely.

The issue with Grassmann variables can be easily addressed by ordering them in a
definite manner within each monomial.
That is, we fully expand each polynomial (which includes eliminating squares of Grassmann variables),
and let variables be multiplied only in a certain order within each monomial.
During this process the coefficients may flip signs, but the result of this process
is unique for each polynomial.
Once we have done this, none of the Grassmann properties will be used
when finding linear relations between the polynomials,
because each monomial is now compared verbatim with monomials in other polynomials.
Therefore, it is now safe to substitute Grassmann variables with c-numbers.
This principle was also implied while extracting the coefficient matrix of
graviton operators in section 3.

As for the randomness, first note that substituting (sufficiently many sets of)
random integers never miss the true dependence between polynomials.
If there is a true linear dependence between polynomials,
i.e. a linear combination that vanishes,
the same linear combination must be zero for whatever numbers are put in,
so the row vectors corresponding to the polynomials must be linearly dependent.
Note that all polynomials have rational coefficients and we put in random integers,
so there is no issue with machine precision.

However, the converse is possible: this method may find false linear dependence between polynomials.
This is simply because a non-vanishing polynomial may evaluate to zero when
certain values are put into variables.
That is, the randomly chosen values could miraculously be the roots of the polynomial.
This type of error can be made arbitrarily more unlikely by increasing
the number of columns, i.e. number of sets of random integers that are put in.
Let us roughly estimate the unlikelihood.

\begin{figure}
    \centering
    \includegraphics[width=0.45\textwidth]{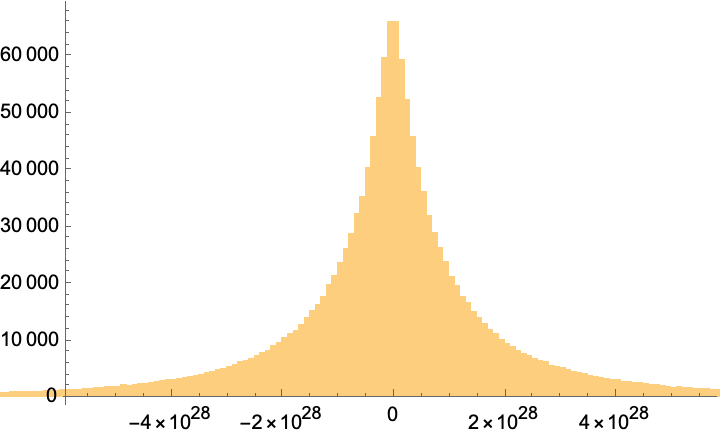}
    \caption{An example distribution of 1.5$\times 10^6$ evaluated values
    of $Q$-action of basis polynomials in paragraph {\bf iii)}.
    Width of each bin is $10^{27}$.}
    \label{fig:egdistribution}
\end{figure}

Suppose that the number of columns is $m+n$ where $m$ is the true number of independent polynomials.
For this method to find a false dependence, both of the followings must happen:
i) there exists a non-trivial linear combination of the polynomials
that vanishes for the first $m$ sets of random integers, and
ii) this polynomial further vanishes for the additional $n$ sets of random integers.
The probability of i) is relatively difficult to estimate, since it involves
intricate tuning of $m-1$ coefficients in a linear combination of the polynomials.
Therefore we only estimate the probability of ii) as follows.
A typical basis polynomial such as $Q$-action of those in (\ref{QEx-basis-eg})
\begin{footnote}{These are used in step 3 and 4 of determining $Q$-exactness
of $Q$-closed operators in the charge sector $(R,J)=(\frac52,\frac32)$,
of which one is the non-graviton cohomology \eqref{Q-coho}}\end{footnote}
evaluates to $\sim 10^{28}$ when a random integer between $1$ and $1000$
are substituted into each variable.
(See Fig. \ref{fig:egdistribution}. for an example.)
This is a natural scale considering that the typical polynomial is a sum over
$\sim 10^{6}$ monomials (with both signs) that each consists of 9 letters,
so for example $10^{6} \times (10^{2.5})^{9} \sim 10^{28}$.
This value is far smaller than the number of all possible random choices
--- which is $(10^3)^{56}$ if all 7 gluons, thus $7\times (3^2-1)$ variables,
are involved ---
so each integer value within magnitude $\sim 10^{28}$ will be sufficiently populated.
Furthermore, since a typical polynomial consists of many$\sim 10^{6}$ monomials,
we assume that the evaluation of the polynomial is like a random walk with
sufficient iterations, and thus the factorization property of integers is blurred.
For these reasons, let us assume that the distribution of the evaluated values
is continuous.
Then, the probability that this value falls within $O(1)$ is estimated to be
$\sim 10^{-28}$, even accounting for the shape of the distribution.
For ii), this must happen for $n$ independent sets of random variables,
so the probability of ii) is estimated to be $10^{-28n}$.
In step 3, $n$ was taken to be 175, so the estimated probability of ii)
is $10^{-4900}$.

This method of detecting linear dependence was used between numerous sets
of polynomials while determining $Q$-exactness of various operators
in different charge sectors.
Numbers that appeared in the previous paragraph slightly differ between occasions.
Typical values of the polynomials differ because they consist of different numbers
of letters, and $n$ is inevitably different because the number of columns are set
before we know the number of true independent polynomials.
However, in any case, we use at least $n \geq 30$ and the estimated probability
of ii) has order of magnitude of a few negative hundreds at the worst.
Furthermore, when a $Q$-closed operator is determined to be $Q$-exact,
we checked analytically the relation between the target and basis polynomials
to further eradicate the margin for error.

Employing the method explained so far, we have constructed the basis operators
in each and all charge sectors with $R_1 = R_2 = R_3$ at $j=24$ order,
with the aforementioned permutation property.
We have also evaluated the $Q$-actions of the bases,
that should form the basis of $Q$-exact operators.
Then we have determined $Q$-exactness of all $Q$-closed
non-graviton operators obtained from our ansatz in the previous subsection.
The result is that all operators except for the fermionic \eqref{Q-coho} are $Q$-exact.

\begin{table}
\begin{center}
\begin{tabular}{|c|c|c|c|c|c|c|c|c|}
     \hline
     $R$ & $J$ & \#letters & \#basis & \#closed & \#exact & \#coh. & \#gravitons & \#BH coh. \\
     \hline
     0 & 4 & 4 & 1 & 0 & 0 & 0 & 0 & 0 \\
     $\frac12$ & $\frac72$ & 5 & 9 & 1 & 1 & 0 & 0 & 0 \\
     1 & 3 & 6 & 91 & 8 & 8 & 0 & 0 & 0 \\
     $\frac32$ & $\frac52$ & 7 & 511 & 85 & 83 & 2 & 2 & 0 \\
     2 & 2 & 8 & 1369 & 445 & 426 & 19 & 19 & 0 \\
     $\frac52$ & $\frac32$ & 9 & 1898 & 953 & 924 & 29 & 28 & 1 \\
     3 & 1 & 10 & 1456 & 961 & 945 & 16 & 16 & 0 \\
     $\frac72$ & $\frac12$ & 11 & 633 & 505 & 495 & 10 & 10 & 0 \\
     4 & 0 & 12 & 136 & 136 & 128 & 8 & 8 & 0 \\
     \hline
\end{tabular}
\caption{ 
For each charge sector $R=R_1=R_2=R_3$ and $J$ at level $j=24$,
we present the numbers of letters, of independent basis operators with the permutation property,
of $Q$-closed operators, of $Q$-exact operators, and therefore of $Q$-cohomologies.
Then we compare with the number of gravitons with the same permutation property
to determine the number of non-graviton (black hole) cohomologies in each charge sector.}
\label{tab:countall}
\end{center}
\end{table}

From the fact that we have constructed and counted all
operators and their $Q$-actions in the $R_1 = R_2 = R_3$ charge sectors
at $j=24$ order with the permutation property,
we can also prove the non-existence of any other $SU(3)$ singlet non-graviton
cohomology at $j=24$ order.
Recall that the result of step 2 is a complete basis of all operators,
in a given charge sector $(R,J)$ and with permutation property that is
designed to include all $SU(3)$ singlets.
There are further linear relations between $Q$-actions of these basis operators, 
reducing the number of independent $Q$-exact operators at charge sector 
$(R+\frac{1}{2},J-\frac{1}{2})$ in step 3.
The reduced operators correspond to the $Q$-closed operators at charge sector $(R,J)$:
$$
\text{(\#closed)}_{(R,J)} = \text{(\#basis)}_{(R,J)} - \text{(\#exact)}_{(R+\frac12,J-\frac12)}~.
$$
Then the number of $Q$-cohomologies is given by
$$
\text{(\#coh.)}_{(R,J)} = \text{(\#closed)}_{(R,J)}-\text{(\#exact)}_{(R,J)}~.
$$
Meanwhile, we can also count the number of independent graviton cohomologies
in these charge sectors and with the same permutation property,
like we counted the full set of gravitons in subsection 3.1.
The number of non-graviton cohomologies is given by
$$
\text{(\#BH coh.)}_{(R,J)} = \text{(\#coh.)}_{(R,J)}-\text{(\#gravitons)}_{(R,J)}~.
$$
We present all the numbers mentioned in this paragraph in Table \ref{tab:countall}.
We find only one non-graviton cohomology in the $(R,J) = (\frac52,\frac32)$ sector,
which is the fermionic cohomology presented in \eqref{Q-coho}.
Since the operators with the permutation property in the $R_1=R_2=R_3$ charge sectors
include all $SU(3)$ singlets,
we conclude that \eqref{Q-coho} is the only non-graviton cohomology
that is an $SU(3)$ singlet at order $j=24$.

The computation presented in this subsection, of constructing the basis operators
and counting independent ones between them and their $Q$-actions,
is essentially the sort of computation that was performed in \cite{Chang:2022mjp},
although we find our numerics-assisted approach to be more efficient.
Moreover, we have only performed this computation in the $R_1=R_2=R_3$ charge sectors
at $j=24$ order in the BMN sector, and further restricted to operators with certain permutation property.
This is because we focused on the $SU(3)$ singlet sector at order $j=24$,
where the non-graviton index indicated the existence of a non-graviton cohomology.

\section{Future directions}

In this paper we studied the $Q$-cohomologies of 4d maximal super-Yang-Mills theory 
for the local BPS operators at 1-loop level. We detected new non-graviton cohomologies 
from the index for the $SU(3)$ and $SU(4)$ gauge theories, and constructed 
the apparent $SU(3)$ threshold cohomology. A goal 
of this program is to identify and characterize the microstates of BPS black holes 
in the $SU(N)$ theory with parametrically large $N$. Although we are currently very 
far from this goal, several novel structures are observed for $SU(3)$ and $SU(4)$ 
theories which we hopefully think will shed light on the large $N$ black hole physics.

Constructing new cohomologies (non-graviton type, or black hole type) requires us 
to find operators which become $Q$-closed only after using the trace relations of 
finite-sized matrices. Although the basic principle for the trace relations should 
be simple (e.g. repeated uses of the Cayley-Hamilton identities), it is hard in 
general to analytically generate useful trace relations. In this paper, 
we have developed a semi-systematic way of constructing $Q$-closed operators 
using certain trace relations. Within this framework, 
the main technical bottleneck is proving that the constructed $Q$-closed 
operator is not $Q$-exact. This demands us to check if the target 
$Q$-closed operators are $Q$-exact or not after using all possible trace relations. 
We developed numerics-assisted computational strategies 
to check this on a computer.

In \cite{Choi:2023znd}, $Q$-exactness was easy to show for a particular class of operators 
in the BMN sector. Namely, if a BMN operator contains a term without any scalars, this 
operator cannot be $Q$-exact. This is because $Q$ acting on the BMN fields always generate 
one of more scalars, so that $Q$ cannot generate a term without scalars.
An infinite number of cohomologies with this property was found in \cite{Choi:2023znd}, 
whose checks of $Q$-nonexactness were trivial.
These cohomologies are beyond our ansatz in this paper. 
This implies that our ansatz is far from sufficient to generate all the non-graviton 
cohomologies, even within the BMN sector. It would be highly desirable, if possible, 
to combine the analytic insights learned in \cite{Choi:2023znd} and in this paper.

Merely knowing the index $Z-Z_{\rm grav}$ over non-graviton cohomologies is very useful to 
learn their novel spectral structures. Section 3.1 has extensively discussed 
the $SU(3)$ non-graviton index up to charge $j=54$, finding the hints of 
novel partial no-hair behaviors and the tower structures. Technically, the full BMN 
index $Z$ is relatively easy to compute at not too large $N$, by computing the residue 
sum for the integral formula (\ref{BMN-index-integral}). 
The harder part is to compute 
the finite $N$ graviton index $Z_{\rm grav}$ by taking into account the trace relations. 
This counting problem reduces to counting independent 
polynomials of $4(N-1)$ bosonic and $3(N-1)$ fermionic variables subject to constraints. 
In principle this counting problem can be solved completely by knowing the so-called 
Gr\"obner basis of constraints. In practice, constructing the Gr\"obner bases is very 
cumbersome, even on computer. We have partially obtained the Gr\"obner bases for the 
$SU(3)$ BMN gravitons in the subsector not containing the letter $f$. To expand these 
results to the full BMN sector including all letters, we did a rather brutal 
computations on computer order by order in the charges. For $SU(4)$, the uses of 
the Gr\"obner bases were more limited to obtain results of section 4.2.
It will be very desirable to compute the full $Z_{\rm grav}$, and thus 
$Z-Z_{\rm grav}$, exactly. There are several features that we would like to 
check with these exact results. For instance, many towers of states were 
observed in $Z-Z_{\rm grav}$ in the $SU(3)$ theory till $j=54$, and to us it is 
unclear whether these towers continue indefinitely or not.

One may wonder if the analysis of spectrum will be simpler at large $N$, 
ignoring $\frac{1}{N}$ corrections in the computations. However, whether a state is 
BPS or not is an exact property so that $\frac{1}{N}$ corrections may affect 
the answer. Nevertheless, there should be large $N$ simplifications for studying 
the near-BPS operators with small anomalous dimensions, 
say, at order $\frac{1}{S}\sim \frac{1}{N^2}$ \cite{Boruch:2022tno}. 
Some studies in this direction were made in \cite{Chang:2023zqk,Budzik:2023vtr}.

\vskip 0.5cm

\hspace*{-0.8cm} {\bf\large Acknowledgements}
\vskip 0.2cm

\hspace*{-0.75cm} 
We thank Minkyoo Kim, Seunggyu Kim, Eunwoo Lee, Masaki Shigemori, Jaewon Song and especially 
Shiraz Minwalla for helpful discussions and comments. 
We also thank Goojin Kwon for independently deriving some results and Chi-Ming Chang for pointing out an error in section 2.1.
This work is supported in part by the NRF grant 2021R1A2C2012350 (JC, SK, JL), 
a KIAS Individual Grant PG081602 at Korea Institute for Advanced Study (SC),
World Premier International Research Center Initiative (WPI), MEXT, Japan (SC),
the DoE grant DE-SC0007859 (SL) and Rackham Predoctoral Fellowship (SL). 
This research was supported in part through computational resources and services provided by Advanced Research Computing (ARC), a division of Information and Technology Services (ITS) at the University of Michigan, Ann Arbor.

\appendix

\section{Graviton trace relations}

In this appendix, we first list the trace relations between the graviton cohomologies in the BMN sector of the $SU(3)$ theory. Then we construct the 
relations of relations at $j=24$ and $j=30$ which are singlets under 
the $SU(3) \subset SU(4)_R$ global symmetry. These are the two sectors in which 
the index predicted fermionic cohomologies in the $SU(3)$ singlet. The results at $j=24$ 
are used in section 4.1 to construct the threshold cohomology. The results at $j=30$ provide 
ans\"atze for the $Q$-closed operators, whose $Q$-exactness are not checked in this paper.

The trace relations are the linear dependence between the multi-trace operators, up to $Q$-exact operators, due to the finite size of the matrices.
In this appendix, we shall only consider the trace relations between gravitons.
Let us first arrange the trace relations by their level $j$ and distinguish them into two types; fundamental ones and the others. The fundamental trace relations at level $j$ 
cannot be written as linear combinations of the trace relations at lower levels $j' (< j)$, multiplied by the gravitons at level $j-j'$. 
All trace relations of gravitons can be expressed as linear combinations of the fundamental 
trace relations with the coefficients being graviton cohomologies. 
We explicitly constructed the fundamental trace relations till certain levels, 
which will be presented below.

The single-trace generators of the $SU(3)$ BMN gravitons are given by
\begin{equation}\label{BMN-gen}
    \begin{aligned}
        u^{ij} \equiv & \; \textrm{tr}  \left(\phi^{(i} \phi^{j)}\right)\ ,\ \   
        u^{ijk} \equiv \; \textrm{tr} \left(\phi^{(i} \phi^{j} \phi^{k)}\right)\ , \\
        {v^{i}}_j \equiv & \; \textrm{tr} \left( \phi^i \psi_{j}\right) - {\textstyle \frac{1}{3} }
        \delta^i_j\textrm{tr}\left(\phi^a \psi_{a}\right)\ , \ \ 
        {v^{ij}}_k \equiv \; \textrm{tr} \left(\phi^{(i} \phi^{j)} \psi_{k} \right) 
        - {\textstyle \frac{1}{4}} \delta^{i}_{k}  
        \textrm{tr} \left( \phi^{(j} \phi^{a)} \psi_{a} \right) 
        - {\textstyle \frac{1}{4}} \delta^{j}_{k}  \textrm{tr} \left( \phi^{(i} \phi^{a)} \psi_{a} \right)\ , \\
        w^i \equiv & \; \textrm{tr} \left(f \phi^i + {\textstyle \frac{1}{2}} \epsilon^{ia_1a_2} \psi_{a_1} \psi_{a_2}\right)\ ,\ \ 
        w^{ij} \equiv \; \textrm{tr}\left(f \phi^{(i} \phi^{j)} +\epsilon^{a_1a_2(i}\phi^{j)} \psi_{a_1} \psi_{a_2}\right)\ ,
    \end{aligned}
\end{equation}
where we suppressed the subscript of $u_n, v_n, w_n$ since it can be easily read off from the number of the indices. Note that the $Q$-actions on $\phi, \psi, f$ are given by 
\begin{equation}
  Q \phi^m=0\ ,\quad Q\psi_{m}=-\frac{i}{2}\epsilon_{mnp}[\phi^n,\phi^p]\ ,\quad Qf=-i[\phi^m,\psi_{m}]\ .
\end{equation}
We would like to find the fundamental trace relations of \eqref{BMN-gen}.

It is helpful to start from the Gr\"obner basis for the trace relations. 
The Gr\"obner basis contains all fundamental trace relations. In general, the Gr\"obner basis 
also contains some non-fundamental trace relations. 
We shall obtain the fundamental trace relations from the Gr\"obner basis by induction.

At the lowest level of the trace relations, all of them are fundamental. Namely, every generator of the Gr\"obner basis at such level are the fundamental relations. For the $SU(3)$ theory, the lowest level is $j=10$. 
In order to organize them into covariant forms in the $SU(3)$ global symmetry, 
we use the following computational strategy (which also proves useful at higher orders).
We list the polynomials of \eqref{BMN-gen} which have the same representations 
as the lowest fundamental trace relations at $j=10$.
Among them, we should find particular linear combinations which vanish when all off-diagonal elements of $\phi^m,\psi_m, f$ are turned off, since the graviton trace relations vanish 
with diagonal fields. 
Once such combinations are identified, keeping $\phi,\psi,f$ general in this combination 
will yield the $Q$-exact operators for the lowest fundamental trace relations.
This way, we can find the fundamental trace relations at the lowest level.\footnote{There 
can be linear combinations which vanish even when the off-diagonal elements are turned on. 
In principle, they can also be the trace relations but most of them are just the identities that 
hold at arbitrary $N$. In practice, we only find them as mesonic identities between 
\eqref{BMN-gen} rather than the trace relations.}

Now, suppose that we found all fundamental trace relations until the level $j$. We can construct the fundamental trace relations at $j+2$ as follows. We first construct all non-fundamental trace relations at level $j+2$ by multiplying suitable graviton cohomologies to the fundamental ones below the level $j$. Not all of them are linearly independent so we should extract a linearly independent set among them. This lets us to compute the $SU(3)$ character of the non-fundamental trace relations at level $j+2$. Next, we consider a union of the non-fundamental trace relations and the Gr\"obner bases at level $j+2$. Note that the Gr\"obner basis will contain all fundamental trace relations and some non-fundamental ones. We extract a linearly independent set among such union, which contains all fundamental and non-fundamental relations. We also compute the $SU(3)$ character over them. Finally, we subtract the former character from the latter, which yields the $SU(3)$ character of the fundamental trace relations at level $j+2$. Then we list the multi-trace operators using \eqref{BMN-gen} which can account for it as before. 
Among them, we find particular linear combinations which vanish when all off-diagonal elements of $\phi,\psi, f$ are turned off, and which are linearly independent from the non-fundamental trace relations we constructed above. The final results are the fundamental trace relations at level $j+2$. In this way, one can construct the fundamental trace relations inductively.

In principle, one can obtain all fundamental trace relations of gravitons from the above induction. For the $SU(2)$ theory, it can be easily done. We found a 66-dimensional Gr\"obner basis, and there exist 48 fundamental trace relations among them. However, for the $SU(3)$ theory, we could not do a similar 
calculation since the construction of the Gr\"obner basis is time-consuming. We constructed it only in two subsectors: 
(1) all trace relations between $u_2, u_3, v_2, v_3$, and 
(2) trace relations between $u_2, u_3, v_2, v_3, w_2, w_3$ until $j \leq 20$. 
From the subsector (1), which has 1170 generators, we obtained all fundamental trace relations between $u_2, u_3, v_2, v_3$, i.e. the relations which do not involve $f$'s. There are in total 287 relations whose lowest level is $j=10$ and the highest level is $j=30$.
On the other hand, from the subsector (2),
we could generate the fundamental trace relations involving $f$'s until $j \leq 20$. 
There are in total 130 relations involving $f$'s between $14 \leq j \leq 20$.
These are enough to construct relations of relations at $j=24$.

Before presenting their explicit forms, we first explain our notation. When we write down certain operator in the irreducible representation $\mathbf{R}$ under $SU(3) \subset SU(4)_R$ as $O^{i_1i_2i_3...}_{j_1j_2j_3...}$, the actual form of such an operator should be understood as $O^{i_1i_2i_3...}_{j_1j_2j_3...}$ subtracted by its trace part to make it traceless, like 
\begin{equation}
    \begin{aligned}
        &[n,0] : O^{i_1i_2i_3\cdots i_n} \to O^{i_1i_2i_3\cdots i_n}\ , \qquad [0,n] : O_{i_1i_2i_3\cdots i_n} \to O_{i_1i_2i_3\cdots i_n}\ , \\
        &[1,1]: O^i_j \to O^i_j - {\textstyle \frac{1}{3}} \delta^i_j O^a_a\ , \\
        &[2,1]: O^{ij}_k \to O^{ij}_k -{\textstyle \frac{1}{2}} \delta^{(i}_kO^{j)a}_a 
        \ , \qquad [1,2]: O^{i}_{jk} \to 
        O^{i}_{jk} -{\textstyle \frac{1}{2}} \delta^{i}_{(j}O^{a}_{k)a} \ , \\
        &[3,1]: O^{ijk}_l \to O^{ijk}_l -{\textstyle \frac{3}{5}} \delta^{(i}_lO^{jk)a}_a \ , \qquad [1,3]: O^{i}_{jkl} \to O^{i}_{jkl} -{\textstyle \frac{3}{5}} \delta^{i}_{(j}O^{a}_{kl)a}\ , \\
        &[2,2]: O^{ij}_{kl} \to O^{ij}_{kl} - {\textstyle \frac{4}{5}} \delta^{(i}_{(k} O^{j)a}_{l)a} 
        + {\textstyle \frac{1}{10}} \delta^{(i}_{(k} \delta^{j)}_{l)} O^{a_1a_2}_{a_1a_2}\ , 
    \end{aligned}
\end{equation}
and so on. Here, $[\cdot ,\cdot ]$ are the Dynkin labels for $SU(3)$.

Below, we list the explicit forms of the fundamental trace relations according to their level $j$ and representation under $SU(3) \subset SU(4)_R$ as $t^j [R_1',R_2']$. The relations which do not involve $f$'s are given as follows\footnote{{\bf v3}: three of these, namely
$(R_{24}^{(0,3)})$, $(R_{26}^{(0,4)})^i$ and $(R_{30}^{(0,5)})$,
vanish on the Coulomb branch (each gauge-adjoint letter restricted to diagonal)
but are not $Q$-exact \cite{Kim:2026}.}:
{\allowdisplaybreaks
    \begin{align}\label{tr-rel}
        &t^{10} [1,2](u_2u_3): (R_{10}^{(0,0)})^i_{jk} = \epsilon_{a_1 a_2 (j} \epsilon_{k) b_1 b_2} u^{a_1 b_1} u^{i a_2 b_2} \nonumber\\
        &t^{12} [0,0](u_2u_2u_2) : R_{12}^{(0,0)} =  \epsilon_{a_1 a_2 a_3} \epsilon_{b_1 b_2 b_3} u^{a_1 b_1}u^{a_2 b_2}u^{a_3 b_3} \nonumber\\
        &t^{12} [2,2](u_2u_2u_2,u_3u_3) : (R_{12}^{(0,0)})^{ij}_{kl} = \epsilon_{a_1 a_2 (k} \epsilon_{l) b_1 b_2} \left( u^{a_1 b_1} u^{a_2 b_2} u^{ij} + 6 u^{a_1 b_1 (i} u^{j) a_2 b_2} \right)  \nonumber\\
        &t^{12} [0,3](u_2v_3): (R_{12}^{(0,1)})_{ijk} = \epsilon_{(i|a_1a_2}\epsilon_{|j|b_1b_2} u^{a_1b_1} {v^{a_2 b_2}}_{|k)} \nonumber\\
        &t^{12} [1,1](u_2v_3, u_3v_2) : (R_{12}^{(0,1)})^i_j =  \epsilon_{j a_1 a_2} \left( 4 u^{a_1 b} {v^{i a_2}}_{b} + 3u^{i a_1b} {v^{a_2}}_{b}\right) \nonumber\\
        &t^{12} [2,2](u_2v_3, u_3v_2) : (R_{12}^{(0,1)})^{ij}_{kl} =\epsilon_{a_1 a_2 (k}\left(u^{a_1(i}{v^{j)a_2}}_{l)}+
        u^{ij a_1}{v^{a_2}}_{l)}\right) \nonumber\\
        &t^{14} [1,0](u_2u_2v_2) : (R_{14}^{(0,1)})^i = \epsilon_{a_1a_2a_3} u^{i a_1} u^{b a_2} {v^{a_3}}_{b}  \nonumber\\
        &t^{14} [0,2](u_2u_2v_2, u_3v_3) : (R_{14}^{(0,1)})_{ij} = \epsilon_{a_1a_2 (i|}\left(\epsilon_{b_1b_2b_3} u^{a_1b_1} u^{a_2b_2}{v^{b_3}}_{|j)} -2 \epsilon_{|j) b_1b_2 } u^{a_1 b_1 c} {v^{a_2b_2}}_c \right) \nonumber\\
        &t^{14} [2,1](u_2u_2v_2, u_3v_3) : (R_{14}^{(0,1)})^{ij}_k =  \epsilon_{k a_1a_2} \left(3 u^{(a_1 b} u^{ij)} {v^{a_2}}_{b} + 4u^{a_1b} u^{a_2 (i} {v^{j)}}_b +24 u^{a_1 b (i} {v^{j) a_2}}_b \right)  \nonumber\\
        &t^{14} [1,3](u_2u_2v_2, u_3v_3) : (R_{14}^{(0,1)})^i_{jkl} =  \epsilon_{(j|a_1 a_2} \epsilon_{|k| b_1 b_2} \left(u^{a_1 b_1} u^{a_2 b_2} {v^i}_{|l)} + 6 u^{ia_1b_1} {v^{a_2b_2}}_{|l)}\right) \nonumber\\
        &t^{14} [3,2] (u_2u_2v_2, u_3v_3) : (R_{14}^{(0,1)})^{ijk}_{lm} = \epsilon_{a_1a_2(l} \left( u^{(a_1 i} u^{jk)} {v^{a_2}}_{m)} + 6 u^{a_1 (ij} {v^{k)a_2}}_{m)}  \right)  \nonumber\\
        &t^{14} [1,3](v_2v_3) : (R_{14}^{(0,2)})^{i}_{jkl} = \epsilon_{a_1 a_2 (j} {v^{a_1}}_{k} {v^{i a_2}}_{l)} \nonumber\\
        &t^{16}[0,1](u_2v_2v_2, v_3v_3):(R_{16}^{(0,2)})_i = \epsilon_{i a_1 a_2} \left(12u^{bc}{v^{a_1}}_b{v^{a_2}}_c +13u^{a_1 b}{v^{a_2}}_c{v^c}_b + 12{{v}^{a_1b}}_c{v^{a_2c}}_b \right) \nonumber\\
        &t^{16}[1,2] (u_2v_2v_2, v_3v_3): (R_{16}^{(0,2)})^i_{jk} = 
        \epsilon_{a_1a_2(j}\!\left(3u^{ib}{v^{a_1}}_{k)}{v^{a_2}}_b \!-\! 
        7u^{ia_1}{v^b}_{k)}{v^{a_2}}_b \!+\! 6u^{a_1b}{v^i}_{k)}{v^{a_2}}_b 
        \!+\! 24{v^{a_1b}}_{k)}{v^{ia_2}}_b \right) \nonumber\\
        &t^{16}[2,3] (u_2v_2v_2,v_3v_3): (R_{16}^{(0,2)})^{ij}_{klm} = \epsilon_{a_1a_2(k}\left({u^{a_1(i}{v^{j)}}_{l}v^{a_2}}_{m)}+ 3{v^{a_1(i}}_{l}{v^{j)a_2}}_{m)}\right) \nonumber\\
        &t^{18}[0,0](u_3v_2v_2): R_{18}^{(0,2)} = \epsilon_{a_1a_2a_3} u^{a_1bc} {v^{a_2}}_b {v^{a_3}}_c \nonumber\\
        &t^{20}[1,0](v_2v_2v_3) : (R_{20}^{(0,3)})^i = 2{v^a}_c{v^b}_a{v^{ic}}_b - 3{v^i}_a{v^c}_b{v^{ab}}_c \nonumber\\
        &t^{22}[2,0](u_2v_2v_2v_2) : (R_{22}^{(0,3)})^{ij} = u^{ij}{v^a}_b{v^b}_c{v^c}_a -3 u^{a(i}{v^{j)}}_b{v^b}_c{v^c}_a +3 u^{ab}{v^{(i}}_a{v^{j)}}_c{v^c}_b \nonumber\\
        &t^{24}[0,0] (u_2v_2v_2v_3) : R_{24}^{(0,3)} =  \epsilon_{a_1a_2a_3}u^{a_1b}{v^{a_2}}_b{v^{a_3c}}_{d}{v^{d}}_c \nonumber\\
        &t^{26}[1,0] (v_2v_2v_2v_3): (R_{26}^{(0,4)})^i= {v^{i}}_a {v^a}_b{v^d}_c{v^{bc}}_d 
        \nonumber\\
        &t^{30}[0,0] (v_2v_2v_2v_2v_2) : R_{30}^{(0,5)} =  {v^{a}}_b{v^{b}}_c{v^{c}}_d{v^{d}}_e{v^{e}}_a \nonumber\\
        &t^{30}[3,0] (v_2v_2v_2v_2v_2) : (R_{30}^{(0,5)})^{ijk}= \epsilon^{a_1a_2(i}{v^{j}}_{a_1}{v^{k)}}_{a_2} {v^{b}}_c{v^{c}}_d{v^{d}}_b\ .
\end{align}
}
Here, the superscripts of $R$ denote $(n_f, n_\psi)$ of the terms with maximal $n_f$ in the trace relations and the subscripts denote their $j$. Their $SU(3)$ representations can be read off from the number of upper and lower indices. The listed trace relations vanish up to $Q$-exact operators whose explicit form will be discussed below. As explained before, this is the exhaustive set of the fundamental trace relations of gravitons which do not involve $f$'s. The fundamental trace relations involving $f$'s until $j \leq 20$ are given by
{\allowdisplaybreaks
    \begin{align}\label{tr-rel-f} 
        &t^{14}[0,2](v_2v_3,u_2w_3) : (R_{14}^{(1,0)})_{ij}  = \epsilon_{a_1a_2(i} \left(8{v^{a_1 b}}_{j)} {v^{a_2}}_b + 5\epsilon_{j)b_1b_2}u^{a_1b_1}w^{a_2b_2}\right) \nonumber\\
        &t^{14}[2,1](v_2v_3,u_2w_3,u_3w_2): (R_{14}^{(1,0)})^{ij}_k =  2 {v^{(i}}_a{v^{j)a}}_k -5{v^{a}}_k{v^{ij}}_a +3\epsilon_{ka_1a_2}u^{a_1(i}w^{j)a_2} + 3\epsilon_{ka_1a_2}u^{ija_1}w^{a_2} \nonumber\\
        &t^{16}[0,1] (v_3v_3, u_2v_2v_2, u_2u_2w_2) : \nonumber\\
        &\hspace{.5cm} (R_{16}^{(1,0)})_i = 
        \epsilon_{ia_1a_2}\left(48 {v^{a_1b_1}}_{b_2} {v^{a_2b_2}}_{b_1} 
        + 9 u^{b_1b_2} {v^{a_1}}_{b_1} {v^{a_2}}_{b_2} 
        - 13 \epsilon_{b_1b_2b_3} u^{a_1b_1} u^{a_2b_2} w^{b_3}\right) \nonumber\\
        &t^{16}[1,2] (v_3v_3, u_2v_2v_2, u_3w_3, u_2u_2w_2) :\nonumber\\
        &\hspace{.5cm}
        (R_{16}^{(1,0)})^i_{jk}= 
        \epsilon_{a_1a_2(j|} \left(24{v^{i a_1}}_b {v^{ba_2}}_{|k)} +
        2u^{i a_1} {v^{a_2}}_b {v^{b}}_{|k)} \!-\! 6u^{a_1 b} {v^{a_2}}_b {v^{i}}_{|k)}
        \right.\nonumber\\ 
        &\hspace{2.5cm}
        \left.+6\epsilon_{|k)b_1b_2}u^{ia_1b_1} w^{a_2b_2} +\epsilon_{|k)b_1b_2}u^{a_1b_1} u^{a_2b_2} w^i \right)  \nonumber\\
        &t^{16}[3,1] (v_3v_3, u_2v_2v_2, u_3w_3, u_2u_2w_2) :\nonumber\\
        &\hspace{.5cm}
        (R_{16}^{(1,0)})^{ijk}_l = 24 {v^{(ij}}_a {v^{k)a}}_l + 7 u^{(ij}{v^{k)}}_a {v^{a}}_l -6u^{a(i}{v^{j}}_a {v^{k)}}_l + 18 \epsilon_{la_1a_2} u^{a_1(ij} w^{k)a_2} +3\epsilon_{la_1a_2} u^{(ij} u^{k)a_1} w^{a_2} \nonumber\\
        &t^{16}[1,2](v_2w_3, v_3w_2) : (R_{16}^{(1,1)})^i_{jk} = \epsilon_{a_1a_2(j}\left({v^{a_1}}_{k)}w^{a_2i} + {v^{a_1i}}_{k)}w^{a_2} \right) \nonumber\\
        &t^{18}[0,0](v_2v_2v_2,u_2v_2w_2) : R_{18}^{(1,1)} = {v^{a_1}}_{a_2}{v^{a_2}}_{a_3}{v^{a_3}}_{a_1} -3\epsilon_{a_1a_2a_3} u^{a_1 b}{v^{a_2}}_{b} w^{a_3} \nonumber\\
        &t^{18}[1,1](v_2v_2v_2,v_3w_3,u_2v_2w_2) : \nonumber\\
        &\hspace{.5cm}
        (R_{18}^{(1,1)})^i_j = 9{v^{i}}_{a_1} {v^{a_1}}_{a_2} {v^{a_2}}_{j}-24\epsilon_{ja_1a_2} {v^{ia_1}}_{b} w^{b a_2} \nonumber\\
        &\hspace{2.5cm} -13 \epsilon_{ja_1a_2} u^{ia_1} {v^{a_2}}_{b} w^b -16\epsilon_{ja_1a_2} u^{ib} {v^{a_1}}_{b} w^{a_2} +5\epsilon_{ja_1a_2} u^{a_1b} {v^{i}}_{b} w^{a_2}  \nonumber\\
        &t^{18}[0,3](v_2v_2v_2,v_3w_3,u_2v_2w_2) : \nonumber\\
        &\hspace{.5cm} 
        (R_{18}^{(1,1)})_{ijk} =  
        \epsilon_{a_1a_2(i|}\! \left(3{v^{a_1}}_{|j|}{v^{a_2}}_{b} {v^{b}}_{|k)} 
        \!-\! 3\epsilon_{b_1b_2|j}{v^{a_1b_1}}_{k)} w^{a_2b_2} 
        \!-\! \epsilon_{b_1b_2|j} u^{a_1b_1} {v^{a_2}}_{k)} w^{b_2}\right) \nonumber\\
        &t^{18}[2,2](v_2v_2v_2,v_3w_3,u_2v_2w_2) :(R_{18}^{(1,1)})_{ij}^{kl} = 2{v^{(i}}_{a}{v^{j)}}_{(k} {v^{a}}_{l)} -6 \epsilon_{a_1a_2(k} {v^{a_1 (i}}_{l)}w^{j)a_2}-\epsilon_{a_1a_2(k} u^{ij}{v^{a_1}}_{l)}w^{a_2}  \nonumber\\
        &t^{20}[0,2](v_2v_2w_2,u_2w_2w_2,w_3w_3): (R_{20}^{(2,0)})_{ij} = 2\epsilon_{a_1a_2(i|}{v^{a_1}}_{b}{v^{b}}_{|j)} w^{a_2} - 3 \epsilon_{a_1a_2a_3} {v^{a_1}}_{i}{v^{a_2}}_{j}w^{a_3} 
        \nonumber\\
        &\qquad \qquad \qquad \qquad \qquad \qquad \qquad \qquad \qquad \quad + \epsilon_{ia_1a_2}\epsilon_{jb_1b_2}u^{a_1b_1}w^{a_2} w^{b_2} + 3 \epsilon_{ia_1a_2}\epsilon_{jb_1b_2}w^{a_1b_1}w^{a_2b_2} \ .
    \end{align}
}
The relations involving one $f$ appear from $j=14$ and those involving two $f$'s appear from $j=20$. We do not find any relations involving three $f$'s until $j \leq 20$.

As explained before, the trace relations \eqref{tr-rel}, \eqref{tr-rel-f} vanish up to $Q$-exact operators, which we now construct explicitly. In principle, one should first construct the complete basis of the $Q$-exact operators, which have the same level $j$ and $SU(3)$ representation with the target trace relation. (The $Q$-action does not change $j$ and $SU(3)$ representation.) However, in practice, we can make some ans\"atze for the $Q$-exact form to reduce the dimension of $Q$-exact basis. One of our working assumptions is that the maximal number of $f$'s appearing before the $Q$-action is the same as that of the trace relation. There is a priori no reason to assume that but it turns out to be true for our examples. After imposing this assumption (and a couple of extra practical assumptions), we find a particular linear combination of the $Q$-exact operators in our basis for the target trace relation. In general, when we write $R_I\sim Qr_I$ for a trace relation 
$R_I$, there exist ambiguities of $r_I$ since we can add arbitrary $Q$-closed operators 
to $r_I$. We partly fix them by requiring $r_I$ to vanish when $\phi,\psi,f$ are restricted 
to diagonal matrices. We do not know whether such a requirement can be satisfied in general, but it does work for our examples.
The purpose of this requirement will be explained later. The other 
ambiguities are fixed by hand to get compact expressions.

Below, we list the operators $r_j^{(n_f,n_\psi)}$ related to the fundamental trace relations
$R_j^{(n_f,n_\psi-1)}$ by $i \, Q r_j^{(n_f,n_\psi)} = R_j^{(n_f,n_\psi-1)}$. We will not list
all $r_j^{(n_f,n_\psi)}$'s, but only those which are used in section 4.1.
For the relations without $f$'s in \eqref{tr-rel}, we obtain\footnote{{\bf v3:}
the last term in $(r_{14}^{(0,2)})_{ij}$ has changed from v2.
Both expressions give the same $(R_{14}^{(0,1)})_{ij}$ as the difference between
the two is $Q$-closed.}
{\allowdisplaybreaks
\begin{align}\label{tr-r}
        &(r_{10}^{(0,1)})^i_{jk} =  -2\, \epsilon_{a_1a_2(j} \tr \left( \phi^{a_1}\phi^{a_2}\phi^{i}\psi_{k)} \right)\ ,  \nonumber\\
        &r_{12}^{(0,1)} = \epsilon_{a_1a_2a_3} \left[ 6 \tr\left( \psi_b \phi^{a_1} \right) \tr \left( \phi^b \phi^{a_2} \phi^{a_3}\right) - \tr\left( \psi_b \phi^{a_1} \phi^{a_2}\right) \tr \left( \phi^b  \phi^{a_3}\right)  \right] \nonumber\\
        &\qquad \quad  -3\epsilon_{a_1a_2a_3} \left[ \tr\left(\psi_b \phi^b \phi^{a_1} \phi^{a_2} \phi^{a_3}\right)
        \!+\!\tr\left(\psi_b \phi^{a_1} \phi^b \phi^{a_2} \phi^{a_3}\right)
        \!+\!\tr\left(\psi_b \phi^{a_1} \phi^{a_2}\phi^b  \phi^{a_3}\right)
        \!+\!\tr\left(\psi_b\phi^{a_1} \phi^{a_2} \phi^{a_3} \phi^b \right)\right] \ , 
        \nonumber\\
        &(r_{12}^{(0,1)})^{ij}_{kl} =  -2\epsilon_{a_1a_2(k}\left[ \tr\left(\psi_{l)}\phi^{(i}\phi^{j)}\phi^{a_1}\phi^{a_2}\right) +7\tr\left(\psi_{l)}\phi^{(i|}\phi^{a_1}\phi^{|j)}\phi^{a_2}\right) \right]\ , 
        \nonumber\\
        &(r_{12}^{(0,2)})_{ijk} = \frac{1}{2} \epsilon_{a_1 a_2 (i} \tr \left( \phi^{a_1} \psi_{j}\phi^{a_2} \psi_{k)} \right) \ , \nonumber\\
        &(r_{12}^{(0,2)})^{i}_{j} =  6 \tr \left( \phi^{(i} \phi^{a)} \psi_{(a} \psi_{j)}\right) - 5 \tr\left(\phi^{[i} \psi_{a}\phi^{a]} \psi_{j}\right)\ ,
        \nonumber\\
        &(r_{12}^{(0,2)})^{ij}_{kl} =   \tr \left(\phi^{(i}\phi^{j)} \psi_{(k}\psi_{l)}\right)\ , \nonumber\\
        &(r_{14}^{(0,2)})^i  =3\, \tr \left( \phi^i \psi_{a_1} \phi^{a_1} \phi^{a_2} \psi_{a_2} \right) + 2\, \tr \left( \phi^i \phi^{a_1} \right) \tr\left( \phi^{a_2} \psi_{(a_1} \psi_{a_2)} \right) \nonumber\\ 
        & \qquad \qquad -6\, \tr\left( \phi^i \psi_{a_1} \right) \tr \left( \phi^{[a_1} \phi^{a_2]} \psi_{a_2} \right) -\, \tr \left(\phi^i \psi_{a_1}\psi_{a_2} \right) \tr\left(\phi^{a_1} \phi^{a_2}\right)\ , \nonumber\\
        &(r_{14}^{(0,2)})_{ij}  =   \frac{5}{9}\epsilon_{a_1a_2a_3}\left[2 \tr\left(\psi_{(i}\psi_{j)} \phi^{a_1} \phi^{a_2}\phi^{a_3}\right) + \tr\left(\psi_{(i} \phi^{a_1}\psi_{j)}\phi^{a_2}\phi^{a_3}\right)\right] \nn\\
        &\qquad \quad +\epsilon_{a_1a_2(i}\left[\tr\left(\psi_{j)} \psi_{a_3}\phi^{a_1}\phi^{a_2}\phi^{a_3}\right)+\tr\left(\psi_{j)} \psi_{a_3}\phi^{a_1}\phi^{a_3}\phi^{a_2}\right)+\tr\left(\psi_{j)} \psi_{a_3}\phi^{a_3}\phi^{a_1}\phi^{a_2}\right)\right] \nn\\
        &\qquad \quad +\epsilon_{a_1a_2(i}\left[\tr\left(\psi_{j)} \phi^{a_1}\psi_{a_3}\phi^{a_2}\phi^{a_3}\right)+\tr\left(\psi_{j)} \phi^{a_1}\psi_{a_3}\phi^{a_3}\phi^{a_2}\right)+\tr\left(\psi_{j)} \phi^{a_3}\psi_{a_3}\phi^{a_1}\phi^{a_2}\right)\right] \nn\\
        &\qquad \quad +\epsilon_{a_1a_2(i}\left[\tr\left(\psi_{j)} \phi^{a_1}\phi^{a_2}\psi_{a_3}\phi^{a_3}\right)+\tr\left(\psi_{j)} \phi^{a_1}\phi^{a_3}\psi_{a_3}\phi^{a_2}\right)+\tr\left(\psi_{j)} \phi^{a_3}\phi^{a_1}\psi_{a_3}\phi^{a_2}\right)\right] \nn\\
        &\qquad \quad +\epsilon_{a_1a_2(i}\left[\tr\left(\psi_{j)} \phi^{a_1}\phi^{a_2}\phi^{a_3}\psi_{a_3}\right)+\tr\left(\psi_{j)} \phi^{a_1}\phi^{a_3}\phi^{a_2}\psi_{a_3}\right)+\tr\left(\psi_{j)} \phi^{a_3}\phi^{a_1}\phi^{a_2}\psi_{a_3}\right)\right] \nn\\
        &\qquad \quad -\frac{5}{3}\epsilon_{a_1a_2(i}\left[\tr\left(\psi_{j)}\phi^{a_1}\phi^{a_2}\right)\tr\left(\psi_{a_3}\phi^{a_3}\right)+\tr\left(\psi_{j)}\phi^{a_1}\phi^{a_3}\right)\tr\left(\psi_{a_3}\phi^{a_2}\right)+\tr\left(\psi_{j)}\phi^{a_3}\phi^{a_1}\right)\tr\left(\psi_{a_3}\phi^{a_2}\right)\right]\ , \nonumber\\
        &(r_{14}^{(0,2)})^{ij}_{k}  =12\tr\left(\phi^{(i} \phi^{a} \phi^{j)} \psi_{(a} \psi_{k)}\right)+12\tr\left(\phi^{(i|} \phi^{a} \phi^{|j)} \psi_{(a} \psi_{k)}\right) +54 \tr\left( \phi^{(i} \phi^{j} \psi_{(a} \phi^{a)} \psi_{k)} \right)-36 \tr\left( \phi^{(i} \phi^{j)} \psi_{(a} \phi^a \psi_{k)} \right) \ ,\nonumber\\
        &(r_{14}^{(0,2)})^i_{jkl}  = 2\epsilon_{a_1a_2(j}\left[\tr\left(\phi^i \phi^{a_1} \phi^{a_2}\psi_k \psi_{l)} \right) +3\tr\left(\phi^i \phi^{a_1} \psi_k \phi^{a_2} \psi_{l)} \right)-2\tr\left(\phi^i  \psi_k \phi^{a_1} \phi^{a_2} \psi_{l)}\right)\right]\ , 
        \nonumber\\
        &(r_{14}^{(0,3)})^i_{jkl}  =  -\frac{1}{2} \tr \left(\phi^i \psi_{(j}\psi_{k}\psi_{l)} \right)\ , \nonumber\\
        &(r_{16}^{(0,3)})_i = 
        \frac{39}{4} \tr \left( \psi_i \{\psi_{b_1}\psi_{b_2}, \phi^{b_1}\phi^{b_2}\} \right) +2\tr \left( \psi_i \psi_{b_1} \phi^{b_1}\psi_{b_2}\phi^{b_2} \right) - \frac{61}{4}\tr \left( \psi_i \psi_{b_1} \phi^{b_2}\psi_{b_2}\phi^{b_1} \right) + \frac{97}{4}\tr \left( \psi_i  \phi^{b_1}\psi_{b_1}\psi_{b_2}\phi^{b_2} \right)  
        \nonumber\\
        &\qquad \quad-\frac{41}{4} \tr \left( \psi_i  \phi^{b_2}\psi_{b_1}\psi_{b_2}\phi^{b_1} \right) -5 \tr \left( \psi_i  \psi_{b_1}\phi^{b_1}\phi^{b_2}\psi_{b_2} \right)-\frac{25}{2} \tr \left( \psi_i  \psi_{b_1}\phi^{b_2}\phi^{b_1}\psi_{b_2} \right)  +2\tr \left( \psi_i  \phi^{b_1}\psi_{b_1}\phi^{b_2}\psi_{b_2} \right)  \nonumber\\
        &\qquad \quad- \frac{61}{4}\tr \left( \psi_i  \phi^{b_2}\psi_{b_1}\phi^{b_1}\psi_{b_2} \right) - \frac{11}{4} \tr \left(  \phi^{b_1}\phi^{b_2}\right) \tr \left(\psi_i \psi_{b_1}\psi_{b_2} \right) - \frac{27}{2} \tr \left( \psi_{b_1}\psi_{b_2} \right) \tr \left(\psi_i \phi^{b_1}\phi^{b_2} \right)  \nonumber\\
        &\qquad \quad+ \frac{29}{4} \tr \left(\phi^{b_2} \psi_{b_2} \right) \tr \left(\psi_i [\psi_{b_1},\phi^{b_1}] \right)\ , \nonumber\\
        &(r_{16}^{(0,3)})^i_{jk} = 2 \tr \left( \psi_{(j} \psi_{k)} \psi_b \phi^b \phi^i \right) -4\tr \left( \psi_{(j} \psi_{k)} \psi_b \phi^i \phi^b \right) - \tr \left( \psi_{(j|} \psi_b \psi_{|k)} \{\phi^b ,\phi^i\} \right) -4 \tr \left( \psi_{(j} \psi_{k)} \phi^{(b}\psi_b  \phi^{i)} \right)  \nonumber\\
        & \qquad \qquad +7\tr \left( \psi_{(j|} \{\psi_b, \phi^{b}\}\psi_{|k)}  \phi^{i} \right) -11\tr \left( \psi_{(j|} \{\psi_b, \phi^{i}\}\psi_{|k)}  \phi^{b} \right) -4\tr \left( \psi_{(j} \psi_{k)}  \phi^b \phi^i \psi_b\right)+2\tr \left( \psi_{(j} \psi_{k)}  \phi^i \phi^b \psi_b\right) \nonumber\\
        &\qquad \qquad +3 \tr \left( \psi_{(j|} \psi_b \right) \tr \left( \psi_{|k)} [\phi^b, \phi^i] \right) +6 \tr \left( \psi_{(j} \phi^{[b} \right) \tr \left( \{\psi_{k)},\psi_b\} \phi^{i]} \right) \ .
\end{align}
}
For the relations involving $f$'s in \eqref{tr-rel-f}, we find
{\allowdisplaybreaks
    \begin{align}\label{tr-r-f}
        &(r_{14}^{(1,1)})_{ij} = 5 \epsilon_{a_1a_2(i} \tr \left(f \phi^{a_1}\psi_{j)}\phi^{a_2}\right) + \tr \left( \phi^a \left\{\psi_a , \psi_{(i} \psi_{j)}\right\} \right) -4 \, \tr \left(\phi^a \psi_{(i|} \psi_{a} \psi_{|j)}\right) \ , \nn \\
        &(r_{14}^{(1,1)})^{ij}_k = 3 \, \tr \left(f\phi^{(i} \phi^{j)}\psi_k\right) - 3 \, \tr \left(f\psi_k\phi^{(i} \phi^{j)}\right) + \epsilon^{a_1 a_2 (i} \tr \left( \phi^{j)} \psi_k \psi_{a_1} \psi_{a_2}\right) - \epsilon^{a_1 a_2 (i} \tr \left( \phi^{j)} \psi_{a_1} \psi_{a_2} \psi_k \right) \ , \nn \\
        &(r_{16}^{(1,1)})_i = 13 \epsilon_{a_1a_2a_3} \tr \left(f \psi_i\right) \tr \left(\phi^{a_1}\phi^{a_2}\phi^{a_3}\right) + \frac{10}{3} \epsilon_{a_1a_2a_3} \tr \left(f \phi^{a_1}\right) \tr \left(\psi_i\phi^{a_2}\phi^{a_3}\right) + \frac{10}{3} \epsilon_{a_1a_2a_3} \tr \left(f \phi^{a_1}\phi^{a_2}\right) \tr \left(\psi_i\phi^{a_3}\right) \nn \\ 
        &+46 \epsilon_{ia_1a_2} \tr \left(f \phi^{b}\right) \tr \left(\psi_b\phi^{a_1}\phi^{a_2}\right) - 7\epsilon_{ia_1a_2} \tr \left(f \phi^{a_1}\right) \tr \left(\psi_b\phi^{a_2}\phi^{b}\right) -7\epsilon_{ia_1a_2} \tr \left(f \phi^{b}\phi^{a_1}\right) \tr \left(\psi_b\phi^{a_2}\right) \nn \\
        &+6\epsilon_{ia_1a_2} \tr \left(f \phi^{a_1}\phi^{a_2}\right) \tr \left(\psi_b\phi^{b}\right) - \frac{115}{3} \epsilon_{a_1a_2a_3} \tr\left(f\psi_i \phi^{a_1} \phi^{a_2} \phi^{a_3}\right) - \frac{95}{3}\epsilon_{a_1a_2a_3} \tr\left(f\phi^{a_1}\psi_i  \phi^{a_2} \phi^{a_3}\right) \nn \\
        &+5 \epsilon_{a_1a_2a_3} \tr\left(f\phi^{a_1} \phi^{a_2} \psi_i \phi^{a_3}\right) +36 \epsilon_{ia_1a_2} \tr\left(f\psi_{b} \phi^{a_1} \phi^{a_2} \phi^{b}\right) -43\epsilon_{ia_1a_2} \tr\left(f\psi_{b} \phi^{a_1} \phi^{b} \phi^{a_2}\right) \nn \\
        &+39\epsilon_{ia_1a_2} \tr\left(f \phi^{a_1}\psi_{b} \phi^{a_2} \phi^{b}\right) -68\epsilon_{ia_1a_2} \tr\left(f \phi^{a_1}\phi^{a_2}\psi_{b}  \phi^{b}\right) + 39\epsilon_{ia_1a_2}\tr\left(f \phi^{a_1}\phi^{b}\psi_{b}  \phi^{a_2}\right)  \nn \\
        &+ 13\tr \left( \psi_i\{\psi_{b_1}\psi_{b_2},\phi^{b_1}\phi^{b_2}\} \right) -31\tr \left( \psi_i\{\psi_{b_1}\psi_{b_2},\phi^{b_2}\phi^{b_1}\} \right) + 14\tr \left( \psi_i\psi_{b_1}\phi^{b_1}\psi_{b_2}\phi^{b_2} \right) \nn \\
        &-22\tr \left( \psi_i\psi_{b_1}\phi^{b_2}\phi^{b_1}\psi_{b_2} \right)+ 14\tr \left( \psi_i\phi^{b_1}\psi_{b_1}\phi^{b_2}\psi_{b_2} \right) \ , \nn \\
        &(r_{16}^{(1,1)})^i_{jk} = \epsilon_{a_1a_2(j}\left[-4 \tr \left( f \phi^i\right) \tr \left( \psi_{k)} \phi^{a_1} \phi^{a_2}\right)  - \tr \left( \phi^i \phi^{a_2} \right) \tr \left( f [\psi_{k)}, \phi^{a_1}]\right) \right] \nn \\
        &+\epsilon_{a_1a_2(j} \left[3 \tr \left( f\phi^{a_1}\{ \psi_{k)} ,\phi^i\} \phi^{a_2} \right) +5 \tr \left(f \{\psi_{k)}, \phi^{a_1} \phi^i \phi^{a_2}\}\right) -4 \tr\left( f \psi_{k)} \phi^i \phi^{a_1} \phi^{a_2}\right)-4 \tr\left( f \phi^{a_1} \phi^{a_2} \phi^i \psi_{k)}\right)\right] \nn \\
        &+2\tr\left( \psi_{(j} \psi_{k)} \psi_b [\phi^b, \phi^i] \right) -3 \tr \left( \psi_{(j|} \psi_b \psi_{|k)} \{\phi^b, \phi^i\} \right) +6 \tr \left( \psi_{(j|}\{\psi_b, \phi^b\} \psi_{|k)} \phi^i\right) -9 \tr \left(\psi_{(j|}\{\psi_b, \phi^i\} \psi_{|k)} \phi^b  \right) \nn \\
        &-2\tr \left( \psi_{(j} \psi_{k)} \left[\phi^b, \phi^i\right] \psi_b \right) + \tr \left( \psi_{(j|} \psi_b\right) \tr \left( \psi_{|k)} [\phi^b, \phi^i] \right) + \tr \left(\psi_{(j|}\phi^b\right)\tr\left(\{\psi_{|k)},\psi_b\}\phi^i\right)\ , \nn \\
        &(r_{16}^{(1,2)})^i_{jk} = -\frac{1}{2} \tr\left(f\phi^i\psi_{(j} \psi_{k)} \right)-\frac{1}{2} \tr\left(f\psi_{(j}\phi^i \psi_{k)} \right)-\frac{1}{2} \tr\left(f\psi_{(j} \psi_{k)}\phi^i \right) -\frac{1}{4} \epsilon^{ia_1a_2}\tr \left(\psi_{a_1}\psi_{a_2}\psi_{(j}\psi_{k)}\right)  \ , \nn \\
        &(r_{18}^{(1,2)})^i_j = -4\tr\left(f \phi^i \phi^a\right)\tr\left(\psi_j\psi_a\right) -5 \tr\left(f \phi^a \phi^i\right)\tr\left(\psi_j\psi_a\right) \nn \\
        &-\frac{53}{2}\tr\left(f \phi^i \psi_j\right)\tr\left(\phi^a\psi_a\right) + 7\tr\left(f \phi^i \psi_a\right)\tr\left(\phi^a\psi_j\right) + \frac{15}{2}\tr\left(f \phi^a \psi_j\right)\tr\left(\phi^i\psi_a\right) +12 \tr\left(f \phi^a \psi_a\right)\tr\left(\phi^i\psi_j\right)\nn \\
        &+2  \tr\left(f \psi_j\phi^i \right)\tr\left(\phi^a\psi_a\right) -13\tr\left(f \psi_a\phi^i \right)\tr\left(\phi^a\psi_j\right) +4\tr\left(f \psi_j\phi^a \right)\tr\left(\phi^i\psi_a\right) \nn \\
        &+6 \tr\left(f  \psi_j\psi_a\right)\tr\left(\phi^i\phi^a\right) + \frac{13}{2}\tr\left(f  \psi_a\psi_j\right)\tr\left(\phi^i\phi^a\right) \nn \\
        &-4 \tr\left(f \phi^i \right)\tr\left(\phi^a\psi_j\psi_a\right) + 14\tr\left(f \phi^i \right)\tr\left(\phi^a\psi_a\psi_j\right)  -8\tr\left(f \phi^a \right)\tr\left(\phi^i\psi_j\psi_a\right) -8 \tr\left(f \phi^a \right)\tr\left(\phi^i\psi_a\psi_j\right)\nn \\
        &-4 \tr\left(f \psi_j \right)\tr\left(\psi_a\phi^i\phi^a\right) -9\tr\left(f \psi_a \right)\tr\left(\psi_j\phi^i\phi^a\right) +6 \tr\left(f \psi_a \right)\tr\left(\psi_j\phi^a\phi^i\right)\nn \\
        &+3 \tr\left(f \phi^i \phi^a\psi_j\psi_a\right) - \frac{31}{2} \tr\left(f \phi^i \phi^a\psi_a\psi_j\right) +3\tr\left(f \phi^a \phi^i\psi_j\psi_a\right)  +\frac{5}{2}\tr\left(f \phi^a \phi^i\psi_a\psi_j\right) \nn \\
        &+12\tr\left(f \phi^i \psi_j\phi^a\psi_a\right) -\frac{13}{2}\tr\left(f \phi^i \psi_a\phi^a\psi_j\right) -6\tr\left(f \phi^a \psi_j\phi^i\psi_a\right) -\frac{13}{2}\tr\left(f \phi^a \psi_a\phi^i\psi_j\right) \nn \\
        &+18 \tr\left(f \phi^i \psi_j\psi_a\phi^a\right)\nn \\
        &-12 \tr\left(f \psi_j\phi^i \phi^a \psi_a\right) +\frac{17}{2}\tr\left(f \psi_a\phi^i \phi^a \psi_j\right) -\frac{43}{2} \tr\left(f \psi_a\phi^a \phi^i \psi_j\right)\nn \\
        &+\frac{1}{3} \epsilon^{a_1a_2a_3}\tr\left( \phi^{i} \psi_{j} \right) \tr\left( \psi_{a_1}\psi_{a_2}\psi_{a_3} \right)-2\epsilon^{a_1a_2i}\tr\left( \phi^{b} \psi_{a_1} \right) \tr\left( \psi_{b}\psi_{j}\psi_{a_2} \right) \nn \\
        &- 10\epsilon^{a_1a_2a_3} \tr\left( \phi^i \psi_j \psi_{a_1}\psi_{a_2}\psi_{a_3}\right) + 8\epsilon^{a_1a_2a_3} \tr\left( \phi^i  \psi_{a_1} \psi_j \psi_{a_2}\psi_{a_3}\right)-2\epsilon^{a_1a_2a_3} \tr\left( \phi^i  \psi_{a_1}  \psi_{a_2}\psi_j\psi_{a_3}\right) \ , \nn \\
        &(r_{18}^{(1,2)})_{ijk}\nn \\
        &=-\epsilon_{a_1a_2(i} \left[ \tr \left(f \phi^{a_1}\right) \tr\left( \phi^{a_2} \psi_j \psi_{k)}\right) -\frac{3}{2} \tr \left(f \psi_j \right) \tr \left(\psi_{k)} \phi^{a_1} \phi^{a_2}\right) +3  \tr \left(f \phi^{a_1} \psi_j \phi^{a_2} \psi_{k)} \right) -3 \tr \left(f \psi_j \phi^{a_1}  \psi_{k)} \phi^{a_2} \right) \right] \nn \\
        &-\frac{1}{2} \tr \left(\phi^a \psi_a\right) \tr \left(\psi_{(i}\psi_j\psi_{k)}\right)+\frac{3}{2} \tr \left(\phi^a \psi_{(i|}\right) \tr \left(\psi_{a}\psi_{|j}\psi_{k)}\right) + \frac{1}{2} \tr \left(\phi^a \psi_{(i} \psi_{j|}\right) \tr \left(\psi_a \psi_{|k)}\right) \nn \\
        &+\frac{3}{2} \tr \left(\phi^a \psi_{(i|}\psi_a\psi_{|j}\psi_{k)}\right)- \frac{3}{2} \tr \left(\phi^a \psi_{(i}\psi_{j|}\psi_a\psi_{|k)}\right)\ ,  \nn \\
        &(r_{20}^{(2,1)})_{ij} \nn \\
        &= -\epsilon_{a_1a_2(i} \left[\tr\left( ff \right) \tr \left( \phi^{a_1} \phi^{a_2} \psi_{j)} \right) +\frac{1}{2}\tr\left( f \psi_{j)}\right) \tr \left( f\phi^{a_1} \phi^{a_2}  \right)+2\tr\left( f \phi^{a_1} \right) \tr \left( f[\phi^{a_2}, \psi_{j)} ]\right)\right] \nn \\
        &+\epsilon_{a_1a_2(i} \left[ 4 \tr\left( ff  \phi^{a_1} \phi^{a_2} \psi_{j)} \right) -\tr\left( f  \phi^{a_1} \phi^{a_2}f \psi_{j)} \right) \right] \nn \\
        &+2\tr\left( f \phi^a \psi_{(i} \right) \tr \left( \psi_{j)} \psi_a  \right) -4\tr\left( f  \psi_{(i} \phi^a \right) \tr \left( \psi_{j)} \psi_a  \right)\nn \\
        &-\frac{1}{2} \tr\left(f \psi_{(i}\right) \left( \phi^a \psi_{j)} \psi_a \right) - \frac{5}{2}  \tr\left(f \psi_{(i|}\right) \left( \phi^a \psi_{a} \psi_{|j)} \right) + 2\tr\left(f \psi_a\right) \left( \phi^a \psi_{(i} \psi_{j)} \right) \nn \\
        &-4\tr\left(f \psi_{(i|}\psi_a\right) \left( \phi^a  \psi_{|j)} \right) \nn \\
        &+2\tr\left( f \phi^a \psi_{(i} [\psi_{j)}, \psi_a]  \right) +4\tr\left( f \phi^a \psi_a  \psi_{(i} \psi_{j)} \right) \nn \\
        &+4 \tr\left( f \psi_{(i} \phi^a  \psi_{j)} \psi_a  \right)-3\tr\left( f \psi_{(i|} \phi^a  \psi_a \psi_{|j)}   \right) -2 \tr\left( f \psi_{a} \phi^a  \psi_{(i} \psi_{j)}  \right) \nn \\
        &- \tr\left( f \psi_{(i|}   \psi_a \phi^a \psi_{|j)}  \right) +4 \tr\left( f    \psi_a \psi_{(i} \phi^a \psi_{j)}  \right) \nn \\
        &+\frac{2}{5} \epsilon^{a_1a_2a_3}\left[2 \tr\left(\psi_{a_1} \psi_{a_2}\right)\tr\left(\psi_{a_3}\psi_{(i}\psi_{j)}\right) -3\tr\left(\psi_{(i|}\psi_{a_1} \psi_{|j)}\psi_{a_2}\psi_{a_3}\right)\right]\ .
    \end{align}
}

Finally, we construct relations of these trace relations. Consider a linear combination of the trace relations with coefficients being the graviton cohomologies. If it vanishes identically, we call it a relation of relations. While the trace relations are identities that can be seen at the 
level of `gluons' $\phi,\psi,f$, the relations of relations are the identities 
of mesons $u_2, u_3, v_2, v_3, w_2, w_3$. We do not need to know how $u_2, u_3, v_2, v_3, w_2, w_3$ are made of $\phi,\psi,f$ to obtain the relations of relations. 
After constructing relations of relations, one can write them as the $Q$-action on certain operators using \eqref{tr-r}, \eqref{tr-r-f}. They are the $Q$-closed operators since their $Q$-actions vanish due to the relations of relations. This is the way we obtain the $Q$-closed operators in section 4.1. They can be either $Q$-exact or not and there is no trivial way to judge it easily. 
If they are not $Q$-exact, they are the non-graviton cohomologies since they are made of the linear combinations of $r_I$'s, which vanish with diagonal $\phi,\psi,f$. For the check of the (non-)$Q$-exactness, refer to section 4.2.

Now we will construct relations of relations at the threshold level $j=24$ which are singlets under $SU(3) \subset SU(4)_R$, from the trace relations \eqref{tr-rel}, \eqref{tr-rel-f}. 
There are 5 choices of $(R,J)$ in this sector in which relations of relations exist. 

\paragraph{i) $(R,J) = (2,2)$} Let us first enumerate all $SU(3) \subset SU(4)_R$ singlets in this sector made by the product of the trace relations in \eqref{tr-rel}, \eqref{tr-rel-f} and the graviton cohomologies. There are following 6 singlets:
\begin{equation}
    \begin{aligned}
         &s_1^{(2,0)} = u^{ij}  (R_{20}^{(2,0)})_{ij}\ , \ \ 
         s_2^{(2,0)} = w^{ij}  (R_{14}^{(1,0)})_{ij}\ , \ \ 
         s_3^{(2,0)}= w^i  (R_{16}^{(1,0)})_i \ , \\
        &s_1^{(1,2)} = {v^{jk}}_i  (R_{16}^{(1,1)})^i_{jk}\ , \ \ 
        s_2^{(1,2)}= {v^j}_i   (R_{18}^{(1,1)})^i_j\ , \ \
        s_3^{(1,2)}= w^i  (R_{16}^{(0,2)})_i\ .
    \end{aligned}
\end{equation}
The superscripts denote ($n_f, n_\psi$) of the terms with maximal $n_f$ in the operator, as before.
There is one relation of these relations given by
\begin{equation}
    i \, QO^{(2,1)} \equiv 65s_1^{(2,0)} -39s_2^{(2,0)} +5s_3^{(2,0)} -312s_1^{(1,2)} -26s_2^{(1,2)} +6s_3^{(1,2)} = 0\ .
\end{equation}
This is the $Q$-action on the $Q$-closed operator \eqref{O1}.

\paragraph{ii) $(R,J) = (\frac{5}{2},\frac{3}{2})$} There exist 12 $SU(3)$ singlets in this sector given by
\begin{equation}
    \begin{aligned}
        &s_1^{(1,1)} = u^{a(i} {v^{j)}}_{a}  (R_{14}^{(1,0)})_{ij} \ , \ \
        s_2^{(1,1)}= \epsilon_{a_1a_2(i}u^{a_1k} {v^{a_2}}_{j)}  (R_{14}^{(1,0)})^{ij}_k\ , \ \
        s_3^{(1,1)} = {v^{jk}}_{i}  (R_{16}^{(1,0)})^i_{jk}\ , \\
        &s_4^{(1,1)} = u^{ijk}  (R_{18}^{(1,1)})_{ijk}\ , \ \
        s_5^{(1,1)} = {v^{(j}}_{i} w^{k)}  (R_{10}^{(0,0)})^i_{jk}\ , \ \
        s_6^{(1,1)} = u^{(ij}w^{k)}  (R_{12}^{(0,1)})_{ijk}\ , \\
        &s_7^{(1,1)} = \epsilon_{a_1a_2i} u^{a_1j} w^{a_2}  (R_{12}^{(0,1)})^i_j\ , \ \
        s_8^{(1,1)} = w^{ij}  (R_{14}^{(0,1)})_{ij}\ , \\
        &s_{1}^{(0,3)} = \epsilon^{a_1a_2(i} {v^{j}}_{a_1} {v^{k)}}_{a_2} (R_{12}^{(0,1)})_{ijk}\ , 
        \ \
        s_{2}^{(0,3)} = {v^j}_a {v^a}_i  (R_{12}^{(0,1)})^i_j\ , \\
        &s_{3}^{(0,3)} = u^{(jk} {v^{k)}}_i  (R_{14}^{(0,2)})^i_{jkl}\ , \ \
        s_{4}^{(0,3)} = {v^{jk}}_i  (R_{16}^{(0,2)})^i_{jk}\ .
\end{aligned}
\end{equation}
There are 4 relations of these relations, given by
\begin{equation}
    \begin{aligned}
        &i \, Q O_1^{(1,2)} \equiv 3s_5^{(1,1)} -3s_6^{(1,1)} +s_7^{(1,1)}= 0\ , \\
        &i \, Q O_2^{(1,2)} \equiv 9s_1^{(1,1)} -10s_2^{(1,1)} - 30 s_5^{(1,1)}-60s_{3}^{(0,3)}= 0\ , \\
        &i \, Q O_3^{(1,2)} \equiv 3s_1^{(1,1)} -6s_2^{(1,1)}+4s_4^{(1,1)} -14s_5^{(1,1)}-6s_8^{(1,1)} -12s_1^{(0,3)} -4s_{2}^{(0,3)}= 0\ , \\
        &i \, Q O_4^{(1,2)} \equiv 3s_1^{(1,1)} -14s_2^{(1,1)}-8s_3^{(1,1)} -42s_5^{(1,1)} +12 s_6^{(1,1)} -24 s_8^{(1,1)} -36s_1^{(0,3)} + 8 s_4^{(0,3)}= 0\ .
    \end{aligned}
\end{equation}
They are the $Q$-action on \eqref{O2}.

\paragraph{iii) $(R,J) = (3,1)$} There exist 16 $SU(3)$ singlets in this sector given by
\begin{equation}
    \begin{aligned}
        &s_1^{(1,0)} = \epsilon_{a_1a_2i}\epsilon_{b_1b_2j} u^{a_1b_1}u^{a_2b_2k}   (R_{14}^{(1,0)})^{ij}_k\ , \ \
        s_2^{(1,0)} = \epsilon_{a_1a_2 i } u^{a_1 (j} w^{k) a_2} (R_{10}^{(0,0)})^i_{jk}\ , \\
        &s_3^{(1,0)}  = \epsilon_{a_1a_2 i } u^{a_1 jk} w^{a_2}  (R_{10}^{(0,0)})^i_{jk}\ , \\
        &s_1^{(0,2)}  = {v^a}_i {v^{jk}}_a  (R_{10}^{(0,0)})^i_{jk} \ ,\ \
        s_2^{(0,2)}  = {v^{(j}}_a {v^{k)a}}_i  (R_{10}^{(0,0)})^i_{jk}\ ,\ \
        s_3^{(0,2)}  = u^{a(i}{v^{jk)}}_a   (R_{12}^{(0,1)})_{ijk}\ ,\\
        &s_4^{(0,2)}  = u^{a(ij}{v^{k)}}_a   (R_{12}^{(0,1)})_{ijk} \ ,\ \
        s_5^{(0,2)}  = \epsilon_{a_1a_2 i} u^{a_1 b}{v^{a_2 j}}_{b}   (R_{12}^{(0,1)})^i_j \ ,\ \
        s_6^{(0,2)}  = \epsilon_{a_1a_2 i} u^{a_1 bj}{v^{a_2}}_{b}  (R_{12}^{(0,1)})^i_j \ ,\\
        &s_7^{(0,2)}  = \epsilon_{a_1a_2(i} u^{a_1 (k}{v^{l)a_2}}_{j)} (R_{12}^{(0,1)})^{ij}_{kl} 
        \ ,\ \
        s_8^{(0,2)}  = \epsilon_{a_1a_2(i} u^{a_1 kl}{v^{a_2}}_{j)}  (R_{12}^{(0,1)})^{ij}_{kl} 
        \ ,\\
        &s_9^{(0,2)}  = \epsilon_{a_1 a_2 i} u^{a_1 (j}u^{kl) a_2}  (R_{14}^{(0,2)})^{i}_{jkl} \ ,\ \
        s_{10}^{(0,2)}  = \epsilon_{a_1 a_2 i} u^{a_1 b} {v^{a_2}}_b  (R_{14}^{(0,1)})^i\ ,\ \
        s_{11}^{(0,2)}  = u^{a(i} {v^{j)}}_{a}   (R_{14}^{(0,1)})_{ij} \ ,\\
        &s_{12}^{(0,2)}  = \epsilon_{a_1a_2(i} u^{a_1 k} {v^{a_2}}_{j)} (R_{14}^{(0,1)})^{ij}_k \ ,\\
        &s_{13}^{(0,2)}  = u^{(jk}{v^{l)}}_i  (R_{14}^{(0,1)})^i_{jkl} \ .
    \end{aligned}
\end{equation}
There are 13 relations of these relations, given by
{\allowdisplaybreaks
    \begin{align}
    &i \, Q O_1^{(1,1)} \equiv s_2^{(1,0)} =0\ , \nonumber\\
    &i \, Q O_2^{(1,1)} \equiv s_3^{(1,0)} =0\ , \nonumber\\
    &i \, Q O_3^{(1,1)} \equiv s_1^{(1,0)} +5s_1^{(0,2)} -2s_2^{(0,2)} =0\ , \nonumber\\
    &i\, QO_1^{(0,3)} \equiv4s_5^{(0,2)}+3s_6^{(0,2)} = (R_{12}^{(0,1)})^i_j  
    (R_{12}^{(0,1)})^j_i = i\, Q\left[\frac{1}{2} i\, Q ((r_{12}^{(0,2)})^i_j (r_{12}^{(0,2)})^j_i) \right] = 0\ , \nonumber\\
    &i\, QO_2^{(0,3)} \equiv  s_7^{(0,2)}+s_8^{(0,2)}= (R_{12}^{(0,1)})^{ij}_{kl} 
    (R_{12}^{(0,1)})^{kl}_{ij}= i\, Q\left[\frac{1}{2} i\, Q ((r_{12}^{(0,2)})^{ij}_{kl} (r_{12}^{(0,2)})^{kl}_{ij}) \right]= 0\ , \nonumber\\
    &i\, QO_3^{(0,3)} \equiv s_3^{(0,2)} = 0\ , \nonumber\\
    &i\, QO_4^{(0,3)} \equiv s_{10}^{(0,2)}= 0\ , \nonumber\\
    &i\, QO_5^{(0,3)} \equiv 6s_1^{(0,2)}-6s_4^{(0,2)}-s_6^{(0,2)}= 0\ ,\nonumber\\
    &i\, QO_6^{(0,3)} \equiv 24 s_2^{(0,2)} -6 s_{11}^{(0,2)} + s_{12}^{(0,2)}= 0\ ,\nonumber\\
    &i\, QO_7^{(0,3)} \equiv s_1^{(0,2)} -10s_2^{(0,2)} -6s_4^{(0,2)} -10 s_8^{(0,2)} = 0\ ,\nonumber\\
    &i\, QO_8^{(0,3)} \equiv  5s_1^{(0,2)} -2s_2^{(0,2)} -9s_4^{(0,2)} +6s_9^{(0,2)}= 0\ ,\nonumber\\
    &i\, QO_9^{(0,3)} \equiv 6s_1^{(0,2)} +12s_2^{(0,2)} -18s_4^{(0,2)} +s_{12}^{(0,2)}= 0\ ,\nonumber\\
    &i\, QO_{10}^{(0,3)} \equiv 38s_1^{(0,2)} +4s_2^{(0,2)} -24s_4^{(0,2)} -5s_{13}^{(0,2)}  = 0\ .
\end{align}
}
They are the $Q$-action on \eqref{O3}. 
Here, $O^{(0,3)}_1$ and $O^{(0,3)}_2$ are explicitly shown to be $Q$-exact.

\paragraph{iv) $(R,J) = (\frac{7}{2},\frac{1}{2})$} There exist 8 $SU(3)$ singlets in this sector given by
\begin{equation}
    \begin{aligned}
        &s^{(0,1)}_1  =  \epsilon_{a_1a_2i} u^{a_1 b} u^{jk} {v^{a_2}}_{b}   
        (R_{10}^{(0,0)})^i_{jk}\ , \ \
        s^{(0,1)}_2  = \epsilon_{a_1a_2 i}u^{a_1 b} u^{a_2 (j} {v^{k)}}_{b} 
        (R_{10}^{(0,0)})^i_{jk}\ , \\
        &s^{(0,1)}_3  = \epsilon_{a_1a_2i} u^{a_1b(j} {v^{k)a_2}}_{b} 
        (R_{10}^{(0,0)})^i_{jk}\ , \ \
        s^{(0,1)}_4  = \epsilon_{a_1a_2(i} u^{a_1(k} {v^{l)a_2}}_{j)}  
        (R_{12}^{(0,0)})^{ij}_{kl} \ , \\
        &s^{(0,1)}_5  = \epsilon_{a_1a_2(i} u^{a_1kl} {v^{a_2}}_{j)} 
        (R_{12}^{(0,0)})^{ij}_{kl}\ , \ \
        s^{(0,1)}_6  = \epsilon_{a_1a_2(i} \epsilon_{j)b_1b_2} u^{a_1b_1} u^{a_2b_2} u^{kl} 
        (R_{12}^{(0,1)})^{ij}_{kl} \ , \\
        &s^{(0,1)}_7  = \epsilon_{a_1a_2(i} \epsilon_{j)b_1b_2} u^{a_1b_1} u^{a_2b_2k}   (R_{14}^{(0,1)})^{ij}_k \ , \ \
        s^{(0,1)}_8  = \epsilon_{a_1a_2i} u^{a_1(j} u^{kl)a_2}  (R_{14}^{(0,1)})^i_{jkl} \ .
    \end{aligned}
\end{equation}
There are 6 relations of these relations, given by
\begin{equation}
    \begin{aligned}
        &i\, QO^{(0,2)}_1 \equiv s_1^{(0,1)}-2s_2^{(0,1)} = 0\ , \\
        &i\, QO^{(0,2)}_2 \equiv 6s_3^{(0,1)} + s_4^{(0,1)} = 0\ , \\
        &i\, QO^{(0,2)}_3 \equiv s_1^{(0,1)} + s_5^{(0,1)} = 0\ , \\
        &i\, QO^{(0,2)}_4 \equiv s_1^{(0,1)} + s_6^{(0,1)} = 0\ , \\
        &i\, QO^{(0,2)}_5 \equiv 4s_1^{(0,1)} + 24s_3^{(0,1)} - s_7^{(0,1)} = 0\ , \\
        &i\, QO^{(0,2)}_6 \equiv s_1^{(0,1)} - 12s_3^{(0,1)} + 3s_8^{(0,1)} = 0\ .
    \end{aligned}
\end{equation}
They are the $Q$-action on \eqref{O4}.

\paragraph{v) $(R,J) = (4,0)$} There exist 4 $SU(3)$ singlets in this sector given by
\begin{equation}
    \begin{aligned}
        &s^{(0,0)}_1  = \epsilon_{a_1a_2a_3} \epsilon_{b_1b_2 i} u^{a_1b_1} u^{a_2b_2} u^{a_3jk} 
        (R_{10}^{(0,0)})^i_{jk}\ , \\
        &s^{(0,0)}_2  = R_{12}^{(0,0)}   R_{12}^{(0,0)}\ , \\
        &s^{(0,0)}_3  = \epsilon_{a_1a_2(i} \epsilon_{j)b_1b_2} u^{a_1b_1} u^{a_2b_2} u^{kl} 
        (R_{12}^{(0,0)})^{ij}_{kl}\ ,\\
        &s^{(0,0)}_4  = \epsilon_{a_1 a_2 (i} \epsilon_{j) b_1 b_2}  u^{a_1 b_1 (k} u^{l) a_2 b_2} 
        (R_{12}^{(0,0)})^{ij}_{kl}\ .
    \end{aligned}
\end{equation}
There is 1 relation of these relations, given by
\begin{equation}
    i \, QO^{(0,1)} \equiv 36s_1^{(0,0)} +5s_2^{(0,0)} -6s_3^{(0,0)} = 0\ .
\end{equation}
This is the $Q$-action on \eqref{O5}.


It is straightforward to generate relations of relations in other charge sectors. 
We present relations of relations at $j=30$ and $(R,J) = (3,2)$ 
which are singlets under $SU(3)$ global symmetry and do not involve $f$'s. They will yield the fermionic $Q$-closed operators with $(R,J) = \left(\frac{5}{2},\frac{5}{2}\right)$, which 
may be good ans\"atze for the non-graviton cohomology detected by the index 
at $j=30$. The $SU(3)$ singlets in this sector are given by
\begin{equation}
\begin{aligned}
  p_1 &= \epsilon^{a_1a_2a_3} {v^{(i}}_{a_1}{v^j}_{a_2}{v^{k)}}_{a_3}   (R_{12}^{(0,1)})_{ijk} \;,  \ \
  p_2 = {v^j}_a{v^a}_b{v^b}_i   (R_{12}^{(0,1)})^i_j \;,  \ \
  p_3 = {v^{(k}}_a{v^{l)}}_{(i}{v^a}_{j)}   (R_{12}^{(0,1)})^{ij}_{kl} \;,  \\
  p_4 &= u^{(jk}{v^{l)}}_a{v^a}_i   (R_{14}^{(0,2)})^i_{jkl} \;,  \ \
  p_5 = u^{a(j}{v^{k}}_a{v^{l)}}_i   (R_{14}^{(0,2)})^i_{jkl}\;,  \ \
  p_6 = {v^{(jk}}_a{v^{l)a}}_i   (R_{14}^{(0,2)})^i_{jkl} \;,  \\
  p_7 &= {v^{(k}}_{(i}{v^{lm)}}_{j)}  (R_{16}^{(0,2)})^{ij}_{klm} \;,  \ \
  p_8 = {v^{(j}}_a{v^{k)a}}_i   (R_{16}^{(0,2)})^i_{jk}\;,  \ \
  p_9 = {v^a}_i{v^{jk}}_a   (R_{16}^{(0,2)})^i_{jk}\;,  \\ 
  p_{10} &= {v^{ia}}_b{v^b}_a   (R_{16}^{(0,2)})_i \;,  \ \
  p_{11} = \epsilon_{ia_1a_2}u^{a_1b}{v^{a_2}}_b   (R_{20}^{(0,3)})^i\;.
\end{aligned}
\end{equation}
The relations of relations of the above singlets are given as follows:
\begin{equation}
\begin{aligned}
  & 5 p_1 - 10 p_2 - 30 p_4 + 8 p_{11} = 0\ , \\
  & 15 p_1 + 6 p_2 - 40 p_3 - 30 p_4 + 90 p_5 = 0\ ,  \\
  & 105 p_1 - 336 p_2 + 140 p_3 - 1050 p_4 + 10080 p_6 - 900 p_7 = 0\ ,  \\
  & 15 p_1 - 138 p_2 - 160 p_3 - 570 p_4 + 864 p_6 - 48 p_8 = 0\ ,  \\
  & 375 p_1 + 6 p_2 + 1760 p_3 + 150 p_4 + 4320 p_6 - 120 p_9 = 0\ ,  \\
  & 55 p_1 - 266 p_2 - 160 p_3 - 1050 p_4 + 2880 p_6 + 40 p_{10} = 0\ .
\end{aligned}
\end{equation}

\end{document}